\documentclass[12pt]{article} 
\usepackage{amsfonts}
\usepackage{amsmath}
\usepackage{amssymb}
\usepackage{array}
\usepackage{bigints}
\usepackage{booktabs}
\usepackage[nosort]{cite}
\usepackage{comment}
\usepackage{caption}
\usepackage{color}
\usepackage{dsfont}
\usepackage{float}
\usepackage{framed}
\usepackage{graphicx}
\usepackage{indentfirst}
\usepackage{mathrsfs}
\usepackage{multirow}
\usepackage{setspace}
\usepackage{subdepth}
\usepackage{subfig}
\usepackage{titlesec}
\usepackage[dotinlabels]{titletoc}
\usepackage{wrapfig}
\usepackage[all]{xy}
\usepackage{young}
\usepackage{slashed}
\usepackage[vcentermath]{youngtab}

\usepackage{hyperref}
\hypersetup{colorlinks=true}
\hypersetup{linkcolor=black}
\hypersetup{citecolor=red}
\hypersetup{urlcolor=blue}

\numberwithin{equation}{section}

\usepackage[left=2.5cm,right=2.5cm,top=2.5cm,bottom=3cm]{geometry}
\newcommand{\be}{\begin{equation}}
\newcommand{\ee}{\end{equation}}
\newcommand{\bea}{\begin{eqnarray}}
\newcommand{\eea}{\end{eqnarray}}
\def\la{\label}
\def\nref#1{(\ref{#1})}
\def\half{{1 \over 2 }}

\def\tilde{\widetilde}
\def\t{\tilde}
\def\hat{\widehat}
\def\bar{\overline}
\def\b{\bar}
\def\d{\partial}
\def\grad{\nabla}
\def\ep{\varepsilon}
\def\1{{\mathds 1}}
\def\Re{\mathop{\rm Re}}
\def\Im{\mathop{\rm Im}}
\DeclareMathOperator{\tr}{\mathrm{tr}}

\newcommand{\Z}{{\mathbb Z}}
\newcommand{\C}{{\mathbb C}}
\newcommand{\R}{{\mathbb R}}
\def\P{\hbox{$\mathbb P$}}

\def\CA{{\mathcal A}}

\def\CC{{\mathcal C}}

\def\CG{{\mathcal G}}
\def\CH{{\mathcal H}}

\def\CM{{\mathcal M}}

\def\CO{{\mathcal O}}

\def\CR{{\mathcal R}}

\def\CT{{\mathcal T}}
\def\CU{{\mathcal U}}
\def\CV{{\mathcal V}}

\def\sfT{{\mathsf T}}

\def\sfC{{\mathsf C}}

\DeclareFontShape{OT1}{cmr}{mx}{n}%
    {<->cmr10}{}
\newcommand{\mytitlefont}{\fontseries{mx}\selectfont}
\DeclareMathAlphabet{\titlemath}{OT1}{cmr}{mx}{n}

\begin{document}


\begin{titlepage}

\begin{center}

~\\[20pt]

{\fontsize{26pt}{0pt} \mytitlefont Comments on QED$_3$ in a Magnetic Field} 

\vskip30pt

Thomas T.~Dumitrescu$^1$ and Juan Maldacena$^2$

\vskip20pt

$^1$
{\it Mani L.\,Bhaumik Institute for Theoretical Physics,\\[-2pt] \it Department of Physics and Astronomy,}\\[-2pt]
       {\it University of California, Los Angeles, CA 90095, USA}\\[6pt]
       
$^2$
{\it Institute for Advanced Study,  Princeton, NJ 08540, USA }

\bigskip
\bigskip
\bigskip

\end{center}

\noindent  We discuss the low-energy dynamics of massless Dirac fermions interacting with a propagating, relativistic photon in 2+1 spacetime dimensions, when we turn on a uniform magnetic field. This problem can be solved when the magnetic field is sufficiently strong. As observed previously, we find that the vacuum spontaneously breaks some of the global symmetries. We also determine the spectrum of excitations around this vacuum, and compute the resulting low-energy effective action. We use techniques that were previously developed for quantum Hall ferromagnets in condensed matter physics. 
\vfill

\begin{flushleft}
August 2025
\end{flushleft}

\end{titlepage}

\tableofcontents

\section{Introduction }

In this paper,  we consider the problem of massless QED in 2+1 dimensions in the presence of a magnetic field. We consider $N_f$ massless Dirac fermions interacting with a dynamical $U(1)$ gauge field around a configuration with a constant magnetic field. We consider only the case with time-reversal symmetry, i.e.~there are no Chern-Simons terms and~$N_f$ is even. We would like to understand the low-energy dynamics of the system. It was argued in~\cite{Klimenko,Gusynin:1994re,Gusynin:1995nb,Semenoff:1999xv,Shovkovy:2012zn} that the magnetic field triggers the spontaneous breaking of some of the symmetries of the problem, and this phenomenon was called ``magnetic catalysis.'' 
   
   Some motivations to study this problem are the following:

\begin{itemize}
    \item It is a simpler version of a similar problem in 3+1 dimensions, where symmetry breaking was also argued to be catalyzed by the presence of a magnetic field \cite{Gusynin:1994xp,Gusynin:1995gt,Gusynin:1997kj,Gusynin:1998zq,Shovkovy:2012zn,Hattori:2017qio}. 
    \item It is a problem where, for weak~$U(1)$ gauge coupling, $e^2 \ll \sqrt{B}$, we have many light states, with a degeneracy that is exponential in the spatial area.  

    \item  It is very similar to quantum Hall ferromagnets \cite{PhysRevB.30.5655,sondhi1993skyrmions,girvin1999quantumhalleffectnovel}. In these systems, fermions in a magnetic field are restricted to a two-dimensional transverse plane, and one can neglect the magnetic dipole (or Zeeman) coupling. Other similar problems include graphene in a magnetic field (see e.g.~\cite{Semenoff:2011ya,Miransky:2015ava}). We will see that we can use ideas developed in those contexts to analyze QED$_3$ in a magnetic field. In essence, we just need to replace the three-dimensional Coulomb potential $v(r) \sim 1/r$ that is relevant there by the two-dimensional one $ v(r) \sim \log r $ that is relevant to QED$_3$. 

\item When $N_f\gg 1$, the theory flows to a conformal field theory (CFT) in the IR. We can study its spectrum of monopole operators, which carry magnetic charge~$N$. Following~\cite{Borokhov:2002ib}, this requires understanding the energies of states on $S^2 \times$(time) with magnetic flux~$2 \pi N$ on $S^2$. The analysis in the present paper makes sharp statements about these operators in the large-flux limit, $N \gg 1$, because the effects we take into account lift degeneracies that are present at leading order in the large-$N_f$ expansion~\cite{Pufu:2013vpa,Chester:2017vdh}. In particular, we find that the lowest-dimension operators with a given, large magnetic charge~$N \gg 1$ are in a specific, non-trivial representation of the $SU(N_f)$ flavor symmetry (which we determine), and that they do not carry any spacetime spin. Our conclusions are consistent with~\cite{Dyer:2013fja}.

\end{itemize}

The physics of QED$_3$ in a magnetic field is essentially the following: let us discuss here the simplest case~$N_f=2$. 
   Consider a large magnetic field, $B \gg e^4$.  At energy scales lower than $\sqrt{B}$ we can integrate out all higher Landau levels, leaving only the lowest one where the fermions have exactly zero energy when we set the gauge coupling~$e^2 \to 0$. At lower energies,  the effects of non-vanishing~$e^2$ become important. In particular, the Coulomb interaction selects a vacuum that spontaneously breaks the flavor symmetry $SU(2) \to U(1)$.   This leads to a Nambu-Goldstone boson (NGB) with a quadratic dispersion relation $\omega \sim e^2 k^2 / B$ for small momentum~$k$. There is also another light mode, the massless photon, which has a linear, relativistic dispersion relation $\omega \sim k$. It can be viewed as an NGB for the magnetic~$U(1)$ global symmetry under which the monopole operators are charged.

   In section~\ref{sec:freedirac}, we review some aspects of free Dirac fermions in a magnetic field, including a discussion of anomalies and Chern-Simons terms. 

   In section~\ref{qed3sec}, we analyze QED$_3$ in a strong magnetic field, $\sqrt{B} \gg e^2$, in the simplest case of~$N_f=2$ flavors.  We discuss how to organize the theory according to energy scales, following a Wilsonian procedure that leaves us with the fermion zero-modes in the lowest Landau level and the low-energy modes of the photon. We explain how to solve this low-energy theory and find the correct vacuum, which breaks the $SU(2)$ flavor symmetry.
   
   We also  determine the spectrum of the simplest excitations around this vacuum. Their low-energy limit describes the NGBs associated with the broken symmetries. We discuss the effective non-linear sigma model that describes all the NGBs mentioned above (including the dual photon) at low energies. An interesting feature is that the coefficients of this sigma model are calculable from the microscopic theory. We also show that the low-energy theory matches all 't Hooft anomalies of the full UV theory. Finally, we compare with the case of a very weak magnetic field, $B \ll e^4$, using the conjectured low energy dynamics in \cite{Chester:2024waw, Dumitrescu:2024jko}. 

   In section~\ref{LargeNsec}, we discuss the generalization to $N_f \geq 4$ flavors. Here, the strong magnetic field regime~$B \gg ( e^2 N_f)^2$ is very similar to the $N_f=2$ case. A new feature arises when $N_f \gg 1$,  in which case the theory without a magnetic field flows to a weakly-coupled CFT. We can then add a small magnetic field $B \ll ( e^2 N_f)^2$ and show that both the conformal symmetry and the non-spacetime symmetries are spontaneously broken. For large $N_f$ it is possible to compute the parameters of the resulting low-energy sigma model at any value of the magnetic field~$B$, interpolating between the large- and small-field regimes. 
    
Several appendices contain background material and computational details.

    \section{Free Dirac Fermion in a Constant Magnetic Field} \label{sec:freedirac} 
    
In this section,  we set up the notation and review some aspects of free Dirac fermions in a constant  magnetic background field. The effects of making the electromagnetic field dynamical will be analyzed in section~\ref{qed3sec}.  Readers who are familiar with 2+1 dimensional free fermions in a magnetic field can skim this section on a first reading.  

\subsection{The Free Dirac Lagrangian} \label{dirlagsec}

Unless stated otherwise we will work in Lorentzian signature, i.e.~in 2+1 dimensional Minkowski spacetime.\footnote{~The metric is $\eta_{\mu\nu} = -++$ and the totally antisymmetric Levi-Civita symbol is normalized as $\ep^{012} = 1$, so that~$\int A \wedge dA = \int d^3 x \, A_0 F_{12} + (\text{permutations})$ in Lorentzian signature.  The path integral weight is $e^{iS}$, where the action $S$ is real.} Since the magnetic field breaks Lorentz invariance, we will often split the spacetime coordinates $x^\mu$ (sometimes abbreviated as $x$) into time $x^0 =t$ and space coordinates $x^a~(a = 1,2)$,
\begin{equation}
x^\mu = \left(t, x^a\right)~, \qquad \mu = 0, 1, 2~, \qquad a = 1, 2~.
\end{equation}
We will also refer to the spatial coordinates as follows,
\begin{equation}
\vec x = (x^1, x^2) = \left(x, y\right)~.
\end{equation}
For the purpose of integration, we denote the spacetime volume element by $d^3 x = d t \, d^2 x$ and its spatial analogue, the area element, by~$d^2 x = dx dy$. 

In 2+1 dimensions, a single Dirac fermion is described by a complex 2-component Grassmann field $\Psi(x)$, whose spinor indices we suppress. Let us first consider the case where $\Psi$ is free and massless, described by the following action,
\begin{equation} \la{MzerDir}
S_\text{Dirac}(m = 0) = \int d^3x ~\, \big(-i \b \Psi \gamma^\mu \d_\mu \Psi\big)~.
\end{equation}
The Dirac gamma matrices $\gamma^\mu$ satisfy\footnote{~When needed, we choose the following explicit parametrization of the gamma matrices: 
\begin{equation}\label{explicitG}
\gamma^0 = - i \gamma^3 = - i \sigma_3~, \qquad \gamma^1 = \sigma_1~, \qquad \gamma^2 =  \sigma_2~,
\end{equation}
where $\vec \sigma = (\sigma_1, \sigma_2, \sigma_3)$ are the standard Pauli matrices. 
}
\begin{equation}
\{\gamma^\mu, \gamma^\nu\} = 2 \eta^{\mu\nu} \1_{2 \times 2}~,
\end{equation}
so that $(\gamma^0)^\dagger = - \gamma^0$ and $(\gamma^a)^\dagger = \gamma^a$. We define the Dirac bar $\b \Psi$ of $\Psi$ as follows,\footnote{~Throughout, $\dagger$ denotes order-reversing Hermitian conjugation.} 
\begin{equation}\label{dirbar}
\b \Psi \equiv \Psi^\dagger \gamma^0~.
\end{equation}
With this definition $i \b \Psi \Psi$ is Hermitian and can be added to the action, where it multiplies the real Dirac mass $m$, 
\begin{equation}\label{mDirac}
S_\text{Dirac} (m) = \int d^3x \, \big( - i \b \Psi \gamma^\mu \d_\mu \Psi + i m \b \Psi \Psi \big)~, \qquad m \in \R~.
\end{equation}

\subsection{Symmetries, Anomalies, and Chern-Simons Terms}\label{dirsymsec}

In addition to Lorentz invariance, the massless Dirac Lagrangian~\eqref{MzerDir} has the following ordinary (zero-from) global symmetries: 
\begin{itemize}
\item A $U(1)$ flavor symmetry under which $\Psi$ has charge~$-1$,\footnote{~By this we mean that~$[Q, \Psi(x)] = - \Psi(x)$, where~$Q$ is the~$U(1)$ Noether charge.} leading to the conserved Noether current $j_\mu = -\b \Psi \gamma_\mu \Psi$ and charge~$Q \equiv \int d^2x \, j^0 = \int d^2 x \, \Psi^\dagger \Psi$. The corresponding $U(1)$ background gauge field $A_\mu(x)$ couples to $\Psi$ by replacing $\d_\mu \to D_\mu \equiv \d_\mu - i A_\mu$ in~\eqref{MzerDir},\footnote{~Background $U(1)$ gauge transformations act via $\Psi(x) \to e^{i \lambda(x)} \Psi(x)$, $A_\mu(x) \to A_\mu(x) + \d_\mu \lambda(x)$, where $\lambda(x) \sim \lambda(x) + 2\pi$ is an angle-valued $U(1)$ gauge parameter.} which amounts to adding the following term to the action,
\begin{equation}\label{jAcoup}
    \Delta S[A] = \int d^3 x \, A_\mu j^\mu~, \qquad j_\mu = -\b \Psi \gamma_\mu \Psi~, 
\end{equation}
so that
\begin{equation}\la{DiracmA}
    S_\text{Dirac}[m, A] = \int d^3 x \, \left(- i \b \Psi \gamma^\mu (\d_\mu - iA_\mu) \Psi + i m\b \Psi \Psi\right)~.
\end{equation}

Even though $\Psi$ is a fermion, it is possible to make sense of it on spacetime manifolds $\CM_3$ that do not admit a spin structure by extending $A_\mu$ to a Spin$^c$ connection. On such manifolds, the flux quantization condition for $A_\mu$ on closed surfaces $\Sigma_2$ is modified,\footnote{~For the purpose of describing these geometric structures, it is convenient to work in Euclidean signature and take $\CM_3$ to be an oriented Riemannian three-manifold. The second Stiefel-Whitney class of the tangent bundle of the manifold~$w_2(T\CM_3) \in H^2(\CM_3, \Z_2)$  vanishes if and only if $\CM_3$ admits a spin structure.} 
\begin{equation}\label{spinc}
{1 \over 2\pi} \int_{\Sigma_2} dA = \half \int_{\Sigma_2} w_2(T\CM_3) \;\; \text{mod}  \;\; \Z~.
\end{equation}

\item A unitary charge-conjugation symmetry $\CC$ under which\footnote{~Here $*$ denotes Hermitian conjugation of the field operator without transposition of its spinor indices.} 
\begin{equation}\label{Cdef}
\CC : \Psi \to C \Psi^*~,~~ 
~~~~~ \qquad A_\mu \to - A_\mu~,
\end{equation}
where the matrix $C = \gamma^1$ in the parametrization~\nref{explicitG}. 

\item An anti-unitary time-reversal symmetry~$\CT$ under which\footnote{~Here we use the shorthand $\CO(\pm t) \equiv \CO(\pm t, \vec x)$, for every local field $\CO(x^\mu)$, to indicate the action of~$\CT$.} 
\begin{equation}\label{Tdef}
\CT : \Psi(t) \to T \Psi(-t)~, \qquad A_0(t) \to A_0(-t)~, \qquad A_a (t) \to - A_a(-t)~. 
\end{equation}
where the matrix $T = \gamma^2$ in the parametrization of~\nref{explicitG}. This symmetry is explicitly broken by the mass~$m$ in~\eqref{mDirac}, which gets a minus sign under~$\CT: m \to -m$. 
\end{itemize}

The theory is also symmetric under spatial reflections~$\CR$, but since the whole theory is unitary and relativistic, it possesses~$\CC\CR\CT$ symmetry. Thus, all consequences of spatial reflections are accounted for by considering~$\CC$ and~$\CT$.

Famously, there is a mixed 't Hooft anomaly, referred to as the parity anomaly~\cite{Redlich:1983kn,Redlich:1983dv,Niemi:1983rq},  involving the $U(1)$ flavor symmetry and any orientation-reversing symmetry, which we can take to be~$\CT$ or $\CC\CT$. This anomaly is conveniently summarized by the following invertible 3+1 dimensional anomaly inflow action -- or symmetry protected topological (SPT) phase --  for the background fields, extended to a four-manifold $\CM_4$ with boundary $\CM_3$,
\begin{equation}\label{paranom}
S_\text{Dirac Inflow}[A] = {\pi \over 8 \pi^2} \int_{\CM_4} dA \wedge dA + \cdots
\end{equation}
This is a conventional 3+1 dimensional~$U(1)$ $\theta$-term with $\theta = \pi$, which respects time-reversal.\footnote{~\label{nograv}The ellipsis in~\eqref{paranom} indicates an analogous gravitational $\theta$-term $\sim \int_{\CM_4} \tr (R \wedge R)$. This term is needed to ensure full background gauge invariance when $\CM_4$ does not admit a spin structure, and it recently played an important role in the study of Higgs-confinement transitions in QCD$_4$~\cite{Dumitrescu:2023hbe}. However, such purely gravitational terms will ultimately cancel out when we study QED$_3$, and hence we indicate them by ellipses (more details appear in~\cite{Seiberg:2016rsg} for the free Dirac fermion, and in~\cite{Dumitrescu:2024jko} for QED$_3$).  We will do the same for gravitational Chern-Simons terms in 2+1 dimensions.} 

We choose to regularize the boundary theory in a way that preserves full $U(1)$ gauge invariance, at the expense of violating time reversal. Following standard practice, we express this in terms of (effective) Chern-Simons terms. A conventionally normalized Chern-Simons term for the Spin$^c$ gauge field $A_\mu$ takes the form\footnote{~See footnote~\ref{nograv} for a discussion of the ellipsis.}
\begin{equation}\label{CSdef}
S_\text{CS}[A] = {k \over 4 \pi} \int_{\CM_3} A \wedge dA  + \cdots = {k \over 4 \pi} \int d^3 x \, \ep^{\mu\nu\rho} A_\mu \d_\nu A_\rho + \cdots~.
\end{equation}
Gauge invariance quantizes the Chern-Simons level $k \in \Z$ to be an integer, and thus integrating out massive fields can only lead to quantized~$k$. However, integrating out gapless or topological degrees of freedom can lead to an effective Chern-Simons level $k_\text{eff}$ that may be fractional or even irrational. A useful definition of~$k_\text{eff}$ involves the two-point function of the current $j_\mu$ in the vacuum (where~$A = 0$), see e.g.~\cite{Closset:2012vp} for a detailed discussion,
\begin{equation}\la{jjcorr}
    \langle j_\mu(x) j_\nu(0)\rangle\big|_{A = 0} = (\CT\text{-even}) + {ik_\text{eff} \over 4 \pi} \ep^{\mu\nu\rho} \d_\rho\delta^{(3)}(x)~.
\end{equation}
Since the massless Dirac fermion in the vacuum preserves~$\CT$-symmetry, the correlator~\eqref{jjcorr} must be~$\CT$-preserving at separated points~$x \neq 0$. However, it has a~$\CT$-violating contact term when~$k_\text{eff} \neq 0$. If~$k_\text{eff} \in \Z$, then it can be set to zero by adding a properly quantized Chern-Simons (counter-) term with level~$k_\text{bare} = - k_\text{eff}$. Thus, the integer part of~$k_\text{eff}$ depends on the regularization scheme used to define the theory.

Comparing~\eqref{paranom} and~\eqref{CSdef}, we see that the parity anomaly implies that the effective Chern-Simons level of a free Dirac fermion of unit charge is
\begin{equation}
k_\text{eff} =  \half + k_\text{bare}~, \quad k_\text{bare} \in \Z~.
\end{equation}
We will follow~\cite{Cordova:2017kue} and work in a scheme where $k_\text{bare} = 0$ and~$k_\text{eff} = \half$ for every Dirac fermion. 

Turning on the real Dirac mass $m$ in~\eqref{mDirac} preserves all symmetries of the theory except for $\CT$ (and reflections~$\CR$). Since $\Psi$ is massive, the low-energy theory is trivially gapped; integrating it out shifts the effective Chern-Simons level by $\Delta k_\text{eff} = \half \text{sign}(m)$, so that 
\begin{equation}\label{keffdir}
k_\text{eff} = \half \left(1 + \text{sign}(m)\right) = \begin{cases} 0 \quad (m < 0) \\ \half \quad (m = 0) \\ 1 \quad (m > 0)
\end{cases}
\end{equation}
As required, $k_\text{eff} \in \Z$ in the trivially gapped $m \neq 0$ phases.

It is instructive to define a running~$k_\text{eff}(p^2)$, by examining~\eqref{jjcorr} in momentum space,
\be
\int d^3 x \, e^{-i p \cdot x} \, \langle j_\mu(x) j_\nu(0)\rangle\big|_{A = 0} = - {k_\text{eff}(p^2) \over 4 \pi} \ep_{\mu\nu\rho} p^\rho~,
\ee
This is no longer a contact term, because the mass~$m \neq 0$ explicitly breaks~$\CT$-symmetry. Instead, the function~$k_\text{eff}(p^2)$ (which was explicitly computed in appendix~A of~\cite{Closset:2012vp}) interpolates between~$k_\text{eff}(\infty) = \half$ in the massless UV  theory at large spacelike (equivalently, Euclidean) momenta $p^2 \gg m^2$, and~$k_\text{eff}(0) = \half (1 + \text{sign}(m))$ in the gapped IR theory at small momenta~$p^2 \ll m^2$. In particular, the shift~$\Delta k_\text{eff} = \half \text{sign}(m)$ does not depend on the regularization scheme, because it comes from the current two-point function at separated points. 

The parity anomaly~\eqref{paranom} implies that the partition function~$Z[m = 0, A]$ of the massless theory, regularized in a gauge invariant fashion, is not invariant under~$\CT$ or~$\CC\CT$. This statement can be extended to all values of the mass~$m$ by promoting it to~$\CT$-odd background field (or spurion), 
\be\label{zmanom}
Z[-m, \CT(A)] =  Z[m, A]\exp\left(-iS_\text{CS}[A]\right)~.
\ee
By contrast, the partition function is fully~$\CC$-invariant if we take~$m$ to be~$\CC$-even.

\subsection{Landau Levels and Zero Modes in a Constant Magnetic Field} 
  
Let us turn on a constant magnetic field $B > 0$.\footnote{~Here we have used charge-conjugation~$\CC$ to fix~$B > 0$ without loss of generality.} We will work in Landau gauge, where
\begin{equation}\label{Landg}
A_x = 0~, \qquad A_y = B x~, \qquad F_{xy} = (dA)_{xy} = B~.
\end{equation}
Note that this background magnetic field explicitly breaks charge conjugation $\CC$ in~\eqref{Cdef} and time-reversal~$\CT$ in~\eqref{Tdef}, but it preserves $\CC\CT$, i.e.~there is a preserved notion of time-reversal in the magnetic field, which exchanges particles and anti-particles. The magnetic field strength $B = F_{xy} \neq 0$ also explicitly breaks Lorentz boosts, while superficially preserving all translations and spatial rotations. However, only translations in the $y$-direction are manifest in Landau gauge for the vector potential~\eqref{Landg}. As we will review below, these symmetries are indeed present, but their algebra is deformed. 

We are interested in studying the effect of the constant magnetic field~\eqref{Landg} on the massless Dirac field $\Psi$ in~\eqref{MzerDir}. To this end, we must mode-expand $\Psi$ in solutions of the massless Dirac equation,
\begin{equation}\label{direqA}
\slashed D \Psi =  (\slashed \d - i \slashed A) \Psi = 0~.
\end{equation}
This equation can be solved exactly, see e.g.~\cite{berestetskii2012quantum,Miransky:2015ava}. The energy spectrum is given by relativistic Landau levels,
\begin{equation}\label{LLenergy}
p^0 =  \pm \sqrt{2 B n}~, \qquad n \in \Z_{\geq 0}~.
\end{equation}
The Landau levels are degenerate, with a total degeneracy~$ {B \over 2\pi }  $(Area) that scales like the area. 

We are particularly interested in the zero modes with $n = 0$, which we also refer to as the lowest Landau level (with $n\geq 1$ describing the higher Landau levels).  The fact that the energy of these zero modes exactly vanishes can be thought of as a cancellation between the orbital Landau levels and the Zeeman energy arising from the magnetic moment of the electron (with gyromagnetic ratio $g= 2$). 

The lowest Landau level can be described in terms of  Grassmann-even mode functions~$u_q(\vec x)$ that are $t$-independent solutions of~\eqref{direqA}.  Distinct degenerate solutions are labeled by a single real variable~$q \in \R$,  
   \be \la{DirSol}
   u_q(\vec x) = \left(B\over \pi \right)^{1/4} \exp\left[ i q y  - { B \over 2 }  \left( x - { q \over B}\right)^2  \right] \zeta ~, \qquad  \gamma^x \gamma^y \zeta = i  \zeta~.
   \ee 
Let us make some comments:
\begin{itemize}
\item[(i)] We unit-normalize the constant, Grassmann-even  Dirac spinor~$\zeta$ in~\eqref{DirSol}, i.e.~$\zeta^\dagger \zeta = 1$.\footnote{~Noting that~$\zeta$ is unique up to a phase, we use the explicit gamma matrices~\eqref{explicitG} to fix
\be\la{expZeta}
\zeta =   \begin{pmatrix}1 \\ 0 \end{pmatrix}~.
\ee} It follows that the wave functions $u_q(\vec x)$ satisfy the following orthonormality relation,
\begin{equation}\label{unorm}
\int d^2 x \, u_{q'}^\dagger(\vec x) u_q(\vec x) = 2 \pi \delta(q-q')~.
\end{equation}

\item[(ii)] The degeneracy label $q \in \R$ denotes the momentum in the $y$-direction, which is conserved in Landau gauge~\eqref{Landg}. By contrast, translations in the $x$-direction and spatial rotations must be combined with background gauge transformations to leave $A_\mu$ invariant. This has the effect of centrally extending the algebra of spatial translations $\vec P$ by the global~$U(1)$ charge~$Q$ to the so-called magnetic translation algebra~\cite{PhysRev.134.A1602},\footnote{~See \cite{Seiberg:2024yig} for a recent discussion through the lens of anomalies, which also discusses magnetic translations from a second-quantized point of view.}
\begin{equation}\label{magtrans}
[P_x, P_y] = -i B Q~, ~~~~~{\rm with }~~~~~~P_x = - i \partial_x - B y ~,~~~~~~~~ P_y = - i \partial_y~, 
\end{equation}
where we have indicated the action of~$P_x, P_y$ on the eigenfunctions~\nref{DirSol} in Landau gauge. While these operators are not gauge invariant (e.g.~$P_y$ is the canonical~$y$-momentum of the single-particle problem), they commute with the Hamiltonian and keep us within the lowest Landau level (e.g.~$P_x$ shifts~$q$ by a constant). Moreover, their commutator in~\eqref{magtrans} is gauge invariant, and it implies that we can only diagonalize one spatial momentum at a time (e.g.~$P_y = q$ in Landau gauge).    

\item[(iii)] The wavefunction~\eqref{DirSol} is a plane wave of momentum $P_y = q$ multiplied by a Gaussian centered at~$x = q/ B$, so increasing $q$ shifts the wavefunction in the $x$-direction. This characteristic feature of Landau level wavefunctions is related to the following fact about particles of unit charge in a magnetic field: a constant external force $f_y = {d P_y / d t }$ in the $y$-direction makes the particle drift in the $x$-direction with velocity $v_x = f_y/B$. If $f_y$ is due to an applied electric field, this phenomenon is the familiar Hall effect.

\item[(iv)] Note that~\eqref{DirSol} is the wavefunction for a {\it positively} charged particle, as opposed to the electron in the real world, which is negatively charged.\footnote{~This is consistent with the fact that~$u_q(\vec x)$ multiplies an annihilation operator~$\psi_q$ of charge~$-1$.} 
\end{itemize}

Upon quantization, each Grassmann-even zero mode wavefunction~$u_q(\vec x)$ in~\eqref{DirSol} gives rise to one complex, Grassmann-odd quantum mechanical fermion~$\psi_q(t)$, whose adjoint is~$\psi_q^\dagger(t)$. The full Dirac field can then be expanded as follows,\footnote{~Note that the mode expansion of $\Psi^*(t, \vec x)$ involves $u_q^*(\vec x)$, which is not the same as~$u_{-q}(\vec x)$.}
\be \la{Pmodexp}
\Psi(t, \vec x) = \Psi_0(t, \vec x) + \Psi_H(t, \vec x)~, \qquad \Psi_0(t, \vec x) = \int { d q \over 2 \pi } \,   u_q(\vec x) \, \psi_q(t)~.    
\ee 
Here~$\Psi_0(x)$ is the projection of the Dirac field onto the zero-modes in the lowest Landau level; its complement~$\Psi_H(x)$ describes all non-zero modes in the higher Landau levels, with Landau level index~$n \geq 1$ and energies $|p^0| \geq \sqrt{2B}$ in~\eqref{LLenergy}. 

Substituting the mode expansion~\eqref{Pmodexp} into the Dirac Lagrangian~\eqref{mDirac} leads to\footnote{~Here we use~\eqref{explicitG} and~\eqref{expZeta} to compute~$\zeta^\dagger \gamma^0 \zeta = -i$ and~$\zeta^\dagger \vec \gamma \zeta = 0$.} 
\begin{equation}\la{SzmQM}
S_\text{Dirac}(m) = \int dt \int {dq \over 2\pi} \left(i  \, \psi_q^\dagger \d_t \psi_q + m \psi^\dagger_q \psi_q\right) + \left(\text{higher Landau levels}\right)~.
\end{equation}
When~$m = 0$ the zero-mode sector enjoys a scaling symmetry under $t \to \lambda t$, under which all~$\psi_q(t)$ have scaling dimension~$0$. Consequently, the kinetic term for the zero modes in~\eqref{SzmQM} is a marginal operator of scaling dimension 1, while the mass term is a relevant operator of scaling dimension~$0$. 

The~$\psi_q(t)$ inherit their transformations under global symmetries (discussed in section~\ref{dirsymsec} above) from~$\Psi(x)$: they have~$U(1)$ charge~$-1$, (so that~$[Q, \psi_q] = -\psi_q$), and they transform under the anti-unitary~$\CC\CT$ symmetry that is present when~$m = 0$ and preserved by the magnetic field as follows,  
\be\la{CTonzm}
\CC\CT : \psi_q(t) \to -i \psi_q^\dagger(-t)~, \qquad m \to -m~.
\ee

\subsection{Compactifying the Spatial Directions}\label{sec:compact} 

It is interesting to generalize the preceding discussion to the case where the spatial manifold~$\Sigma_2$ is compact, the most common choices being the sphere~$\Sigma_2 = S^2$ and the torus $\Sigma_2 = T^2$. In this case, the number~$N$ of magnetic flux quanta must be an integer,
\begin{equation}\label{fluxquant}
  {1 \over 2\pi}  \int_{\Sigma_2} B = N \in \Z~,
\end{equation}
which we take to be positive ($N > 0$). Note that the flat-space limit~$\Sigma_2 = \R^2$ discussed above is a large-flux limit: $N \to \infty$ and~$\text{Area}(\Sigma_2) \to \infty$, with fixed~$B$. The Atiyah-Singer index theorem implies that the Dirac equation~\eqref{direqA} has~$N$ complex zero modes on such a compact~$\Sigma_2$.

Let us focus on the case where~$\Sigma_2 = S^2$ is a sphere, which will make an appearance in our discussion. A monopole field of flux~$N$ with constant~$B$ preserves the~$SU(2)_R$ rotation symmetry of the sphere. In this case the zero modes~$\psi_m$ transform under a single, irreducible~spin-$j$ representation of that~$SU(2)_R$ symmetry, with~$N = 2j +1$ (see e.g.~\cite{Borokhov:2002ib}).\footnote{~If we write the gauge potential and metric on~$S^2$ as 
 \be 
 A  = -{ N \over 2 }   \cos \theta   d\phi  ~, \qquad ds^2 = d\theta^2 + \sin^2{\theta} \, d\phi^2 ~,
  \ee 
  and choose gamma matrices $\gamma^{\hat \theta} = \sigma^x$ and $\gamma^{\hat \varphi}  = \sigma^y$,  
  then the zero-mode wavefunctions are given by the following Dirac spinors on~$S^2$,
  \be \la{WFSph}
  u_m(\theta, \phi) = \left( \sin { \theta \over 2 } \right)^{j-m} \left(\cos{ \theta \over 2 } \right)^{j+m} e^{ i m \phi } \left( \begin{array}{c} 1 \\ 0 \end{array} \right) ~,~~~~~~ 2 j + 1 = N ~,~~~{ -j \leq m \leq j~.} 
  \ee  } As usual $m = -j, -j+1, \ldots, j$ labels the~$J_3$ eigenvalue of the spin-$j$ representation. The analogue of~\eqref{SzmQM} for the zero modes of a massless Dirac fermion on~$S^2$ thus takes the form
   \be \la{s2zm}
   S_{S^2 \text{zero modes}} = \sum_{m =-j}^j  \int dt  \, i \psi_m^\dagger \d_t \psi_{m}~, \qquad 2j +1 = N = {1 \over 2\pi}  \int_{S^2} B~. 
   \ee   

We could also consider a torus~$\Sigma_2 = T^2$, with $x \sim x + L_x$, $y \sim y + L_y$, and constant magnetic field~$B = N/(2\pi L_x L_y)$. In this case,  the translation symmetries of the torus are broken to a copy of~$\Z_N$ for each cycle; these discrete translations form a non-Abelian algebra~\cite{PhysRev.134.A1602} that can be thought of as an exponentiated version of~\eqref{magtrans}. The wavefunctions on the torus involve Jacobi theta functions, and we will not discuss them.

\subsection{Effective Field Theory of the Lowest Landau Level} \la{LLLEFT}

\subsubsection{Massless Case}

In a constant magnetic field, where~$A_\mu$ (in Landau gauge) and its field strength~$F =dA$ are given by~\eqref{Landg}, the zero modes in the lowest Landau level and the higher Landau levels in~\eqref{SzmQM} are decoupled. Thus, the effective theory at energies~$E \ll \sqrt{B}$ is correctly described by only retaining the zero-mode term in~\eqref{SzmQM}.

This is no longer the case if we consider variations~$a_\mu$ of the total~$U(1)$ background gauge field, which is now given by~$A_{\mu \, , \text{total}} = A_\mu + a_\mu$, with field strength~$F_\text{total} = dA_\text{total} = F + f$. Here~$a_\mu$ is also a~$U(1)$ background gauge field, with field strength~$f = da$, that describes electromagnetic fields on top of the constant magnetic field described by~$A_\mu$. We will assume that the magnetic flux is fixed by~$A$, so that~$\int f_{xy} = 0$, i.e.~$f_{xy}$ only describes spacetime gradients of the magnetic field. 

\begin{figure}[h!]
   \begin{center}
    \includegraphics[scale=.35]{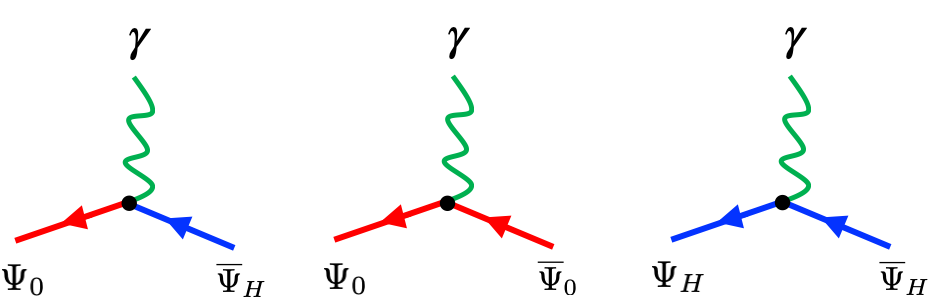}
    \end{center}
\caption{Three-point couplings of the background photon~$a_\mu$ (shown in green) to the zero-modes~$\Psi_0$ in the lowest Landau level (shown in red), and the heavy modes~$\Psi_H$ in the higher Landau levels (shown in blue).}
    \label{EFTvert}
\end{figure}

In section~\ref{qed3sec},  we will gauge~$a_\mu$, but for now it is also a background field. Consider the massless Dirac action~$S_\text{Dirac}[m = 0, A_\text{total} = A + a]$ in~\eqref{DiracmA} with~$U(1)$ background field~$A_\text{total} =  A + a$. Expanding~$\Psi = \Psi_0 + \Psi_H$ in Landau levels of the constant~$B$-field described by~$A_\mu$, as in~\eqref{Pmodexp}, leads to three-point couplings (shown in figure~\ref{EFTvert}) of a background photon~$a_\mu$ (shown in green) to an incoming and outgoing charged fermion, either of which could be a zero-mode~$\Psi_0$ (shown in red) or a heavy mode~$\Psi_H$ (shown in blue). 

Let us assume that~$a_\mu$ has energies and momenta~$\ll \sqrt{B}$. Then we can integrate out the higher Landau levels to obtain a local, Wilsonian effective action for the zero modes~$\Psi_0$ and the photon~$a_\mu$. In a standard gradient expansion, the leading contribution~$S^{(0)}_{\text{eff}}$ to this effective action is obtained by projecting~$S_\text{Dirac}[m = 0, A + a]$ onto the lowest Landau level, 
\begin{equation}\label{eq:seff0}
    S^{(0)}_{\text{eff}}[\Psi_0, a_\mu] = \int d^3 x \, \left( - i \b \Psi_0 \gamma^\mu (\d_\mu - i a_\mu) \Psi_0 \right) = \int dt \, \int {dq \over 2\pi} \, \left( i \psi_q^\dagger \d_t \psi_q\right) + \int d^3 x \, a_0 j^0_\text{eff}~.
\end{equation}
Note that~$\Psi_0$ has mass dimension~$1$, as befits a Dirac field in 2+1 dimensions; the scalings of the~$\psi_q$ were discussed below~\eqref{SzmQM}. 

It follows from~\eqref{eq:seff0} that the effective current~$j^\mu_\text{eff} = - \b \Psi_0 \gamma^\mu \Psi_0$ in the low-energy theory has vanishing spatial components~$\vec j_\text{eff} = 0$. This is the familiar statement that there are only densities, no spatial currents, in the lowest Landau level (at least at leading order). Below, it will be useful to use a mixed representation of~$j^0_\text{eff}$ that is Fourier-transformed in the spatial coordinates; using the wavefunctions in~\eqref{DirSol},\footnote{~With these conventions, the Fourier-transformed operator~$\CO(\vec p) = \int d^2 x \, e^{- i \vec p \cdot \vec x} \CO(\vec x)$ carries momentum~$-\vec p$, so that~$[\vec P, \CO(\vec p)] = - \vec p \, \CO(\vec p)$.} 
\begin{equation}\label{jeffk}
    j^0_\text{eff}(t, \vec p) = \int d^2 x \, e^{- i \vec p \cdot  \vec x} j^0_\text{eff}(x) = \int {dq dq' \over (2 \pi)^2} \, 2 \pi \delta(q-q'-p_y) \exp\left(-{ {\vec p}^{\,2} \over 4 B} - {i p_x(q + q') \over 2B}\right) \psi^\dagger_{q'} \psi_q~.
\end{equation}
The phase factor in the integrand is natural, given that~$(q + q')/2B$ is the~$x$-component of the center of mass for the two fermions.

Let us make several comments about the leading tree-level effective action~$S^{(0)}_\text{eff}$ in~\eqref{eq:seff0}:
\begin{itemize}
\item[(i)] Consider a~$U(1)$ gauge transformation~$a_\mu \to a_\mu + \d_\mu \lambda$ (with fixed~$A$). This requires a compensating phase rotation~$\Psi_0 \to e^{i \lambda} \Psi_0$ of the zero-mode field, but that field is constrained to satisfy~$\vec \gamma \cdot ( \vec \d - i \vec A) \Psi_0(t, \vec x) = 0$, as can be verified directly from~\eqref{Pmodexp}. This implies that only~$\vec x$-independent gauge transformations~$\lambda(t)$ leave~\eqref{eq:seff0} invariant.  Thus already classically, i.e.~at tree level, \eqref{eq:seff0} cannot be the full effective action and must be corrected to restore gauge invariance under arbitrary~$\lambda(x)$, as we explain below. 

\item[(ii)] To quantize~\eqref{eq:seff0}, we use the following canonical equal time commutators,
\be \la{CanCom}
\big\{\psi_q^\dagger, \psi_{q'}\big\} = 2 \pi \delta(q-q')~, \qquad \big\{\psi_q, \psi_{q'}\big\} = 0 = \big\{\psi^\dagger_q, \psi_{q'}^\dagger\big\}~.
\ee
The Fock vacuum $|0\rangle$ is defined to be the state annihilated by all $\psi_q$, which we view as lowering operators. We should emphasize that, at this stage, there are many other states with zero energy, which are obtained by acting with $\psi^\dagger_q$  on the Fock vacuum.   

\item[(iii)] If we put the system in finite volume, and we restrict~$a_0(t)$ to be an~$\vec x$-independent   background gauge field (so that~\eqref{eq:seff0} is gauge invariant, see point (i)~above), then we can view the problem as a 0+1 dimensional one.   Then integrating out the~ $N$ fermions $\psi_q(t)$ generates a 0+1 dimensional Chern-Simons term~$\hat k \int dt \, a_0(t)$ in the (non-Wilsonian) quantum effective action for~$a_0$, with effective level~$\hat k_\text{eff} = \half$ (see appendix~\ref{sec:degqm}). More precisely, as discussed around~\eqref{effCSferm}, this means that a background charge~$+\half$ must be added for every zero mode, so that the expectation value of the current~$j^0_\text{eff}$ (which superficially vanishes in the Fock vacuum, see~\eqref{jeffk}) is given by the following expression in the quantum theory, 
\begin{equation}\label{bgcharge}
    \langle 0 | j^0_\text{eff} | 0 \rangle\big|_{a = 0} = \hat k_\text{eff} = {N\over 2}~. 
\end{equation}
Here~$N$ is the number of zero modes. This precisely matches the effective Chern-Simons level~$k_\text{eff} = \half$ of the massless 2+1 dimensional Dirac fermion in~\eqref{keffdir}, and hence also the parity anomaly~\eqref{paranom}, upon substituting~$A_\text{total} = A+ a$ into~\eqref{CSdef} and using~$\int (dA)_{xy} = 2 \pi N$. 
\end{itemize}

We have already established that the leading-order Wilsonian effective action~\eqref{eq:seff0} cannot be the whole story and must be corrected. The corrections come from two kinds of Feynman diagrams (see~\cite{Hong:1997uw} for a closely related discussion in 3+1 dimensions):

  \begin{figure}[t!]
   \begin{center}
    \includegraphics[scale=.400]{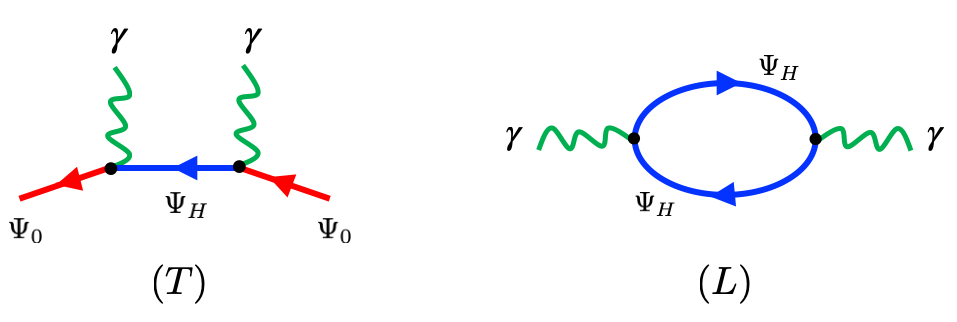}
    \end{center}
\caption{Diagrams that contribute to the Wilsonian effective action of the zero modes~$\Psi_0$ (red) and the photon~$a_\mu$ (green): tree diagrams (T) with a single fermion line connecting two zero modes with internal heavy $\Psi_H$-propagators (blue) and an arbitrary number of external photons; and loop diagrams (L) with an arbitrary number of external photons and heavy~$\Psi_H$ particles in the loop. In both diagrams, we show the case of two external photon lines.}
    \label{EFTtreeloop}
\end{figure}

\begin{itemize}
\item[(T)] Tree-level diagrams involving exactly two zero modes~$\Psi_0$ (one incoming and one outgoing) and~$m \geq 2$ external photons~$a_\mu$, connected by~$m-1$ tree-level propagators for the heavy fields~$\Psi_H$. The case~$m =2$ is shown in the left panel of figure~\ref{EFTtreeloop}. The~$\Psi_H$ have~$|p^0| \geq \sqrt{2 B}$ (see~\eqref{LLenergy}), so that their propagators can be Taylor expanded in non-negative integer powers of~$p_0^2 / B \ll 1$ and~${\vec p}^{\, 2} / B \ll 1$, times a single overall factor with scaling~$\sim p_\mu/B$.\footnote{~See e.g.~\cite{Miransky:2015ava, Hong:1997uw} for a detailed presentation of these propagators.} These tree diagrams lead to local terms of the schematic form
    \begin{equation}\la{SeffT}
        S_\text{eff}^{(T)}[\Psi_0, a] \sim \sum_{\substack{m = 2 \\ n =0}}^\infty {c_{mn} \over B^{m-1+n}} \;  \b \Psi_0 (a_\mu)^{m} (\d_\nu)^{m-1 + 2n} \Psi_0~.
    \end{equation}
Since~$\Psi_0, a_\mu,$ and~$\d_\nu$ all have mass dimension~$1$, the~$c_{mn}$ are dimensionless Wilson coefficients, and the fields in~\eqref{SeffT} are suitably contracted (in general also with various gamma-matrices that we do not show explicitly) to ensure invariance under translations and spatial rotations.\footnote{~A much more stringent requirement, to be discussed below, is that the path integral over~$\Psi_0$ leads to a Lorentz- and (modulo some divergent counterterms) also a conformally-invariant functional of~$A_\text{total} = A+a$, since these symmetries are respected by the massless Dirac action~$S[m =0, A_\text{total} = A + a]$ in~\eqref{DiracmA}.} Since~$\CC\CT$ symmetry is not anomalous at tree level, $S_\text{eff}^{(T)}$ must  respect this symmetry, as is the case for~$S^{(0)}_\text{eff}$ in~\eqref{eq:seff0}. However, only the full tree-level effective action~$S^{(0)}_\text{eff} + S_\text{eff}^{(T)}$ is invariant under arbitrary $a_\mu$ gauge transformations. 
    
    \item[(L)] Loop diagrams involving no zero modes, an arbitrary number of external photons~$a_\mu$, and heavy~$\Psi_H$ particles running in the loop. An example is shown in the right panel of figure~\ref{EFTtreeloop}. They induce~$\CC\CT$-preserving, gauge-invariant local terms that can be Taylor-expanded in integer powers of~$f/B$. The leading contribution is computed in appendix~\ref{App1Loop}; it can be compactly written in terms of~$F_\text{total} = dA + da = F + f$,
    \begin{equation}\label{Fconfkin}
        S^{(L)}_\text{eff}[A_\text{total}] = - \gamma \int d^3 x \left(\half \left(F_\text{total}^{\mu\nu}\right)^2\right)^{3/4} + \cdots ~, \qquad \gamma = {\zeta(3/2) \over 4 \pi^2 \sqrt 2}~.
    \end{equation}
    Expanding to second order in~$f$ leads to the local, conformally-invariant kinetic terms in~\eqref{FinAc}; the ellipses indicate terms involving gradients of~$F_\text{total}$. Note that loop diagrams involving only the heavy field~$\Psi_H$ do not give rise to any~$\CC\CT$-violating terms, including Chern-Simons terms, as we explicitly confirm in appendix~\ref{Toddapp}. 
\end{itemize}

The total Wilsonian effective action at energies and momenta~$\ll \sqrt B$ is given by adding the three contributions in~\eqref{eq:seff0}
\begin{equation}\label{totalWilson}
    S_\text{eff}[\Psi_0, a] = S^{(0)}_\text{eff}[\Psi_0, a] + S^{(T)}_\text{eff}[\Psi_0, a] + S^{(L)}_\text{eff}[A + a]~.
\end{equation}
This action can be varied with respect to~$a_\mu$ to determine the full effective current~$j_\text{eff}^\mu$, which includes corrections to~\eqref{jeffk}. In particular, these corrections ensure that the full~$j_\text{eff}^\mu$ is gauge invariant and transforms like a Lorentz vector.

The discussion above raises a puzzle: recall from around~\eqref{bgcharge} that integrating out the~$N$ massless zero modes~$\psi_q$ in the leading effective action~\eqref{eq:seff0} generates a 0+1 dimensional Chern-Simons term~$\hat k \int dt \, a_0$ with effective level~$\hat k_\text{eff} = N/2$. However, given our preceding observations about the Lorentz invariance of the effective action, we expect this term to arise from the relativistic Chern-Simons term \eqref{CSdef} upon substituting~$A_\text{total} = A + a$,
\be\la{CSsplit}
S_\text{CS}[A_\text{total} = A + a] = {k N} \int dt\, a_0 + {k \over 4\pi} \int a \wedge da~, \qquad N = {B  \over 2\pi}(\text{Area})~.
\ee
We indeed recover~$\hat k_\text{eff} = N/2$  from~$k = 1/2$, but so far we have not explained the origin of the second term~$\sim a \wedge da$ in~\eqref{CSsplit}. Note that this term can be obtained from the first one by replacing~$B \to B + da$, so that it makes the background charge~\eqref{bgcharge} compatible with locality.  

We claim that the second term in~\eqref{CSsplit}, with~$k_\text{eff} = 1/2$, is also generated by a quantum effect in the Wilsonian effective theory: it arises by integrating out the zero modes~$\Psi_0$ in the leading~$m = 2, n=0$ term in~\eqref{SeffT}, 
\begin{equation}\la{specialST}
    S^{(T)}_\text{eff}[\Psi_0, a]\Big|_{m = 2, n = 0} \sim {c_{2, 0} \over B} \; \b \Psi_0 (a_\mu)^2 (\d_\nu) \Psi_0~.
\end{equation}
This is the only term in~\eqref{SeffT} that scales like~$1 / B$, and we already know from the first term in~\eqref{CSsplit} that integrating out the~$N$ zero modes can generate a factor of~$B$. Integrating out~$\Psi_0$ in~\eqref{specialST}, which amounts to closing the red~$\Psi_0$ lines in figure~\ref{EFTtreeloop} into a loop, we obtain a contribution to the two-point function of the current~$j_\mu$ that is $B$-independent and contains a single derivative. It must therefore be a Chern-Simons contact term of the form~\eqref{jjcorr}. We will now verify this, and the fact that~$k_\text{eff} = 1/2$, by deforming the theory with small Dirac masses~$|m| \ll \sqrt{B}$ of either sign and computing the resulting~$k_\text{eff}$. 

 \subsubsection{Adding a Small Mass} 

It was shown in~\eqref{SzmQM} that a Dirac mass~$m$ in~\eqref{mDirac} changes~$S^{(0)}_\text{eff}$ in~\eqref{eq:seff0} by
\begin{equation}\label{deltam}
    \Delta S^{(0)}_\text{eff}[\Psi_0] = \int dt \int {dq \over 2\pi} \, m \psi_q^\dagger \psi_q~.
\end{equation}
As long as~$m \ll \sqrt{B}$, this is the leading~$m$-dependence of the Wilsonian effective action. When~$m \neq 0$, it leads to a fully gapped theory for the background field~$a_\mu$ at low energies. Since~$\CC\CT$ symmetry is broken, the low-energy theory may contain Chern-Simons terms for~$a_\mu$, but they must be properly quantized, with~$k_\text{eff} \in \Z$. In appendix~\ref{Toddapp} we verify through an explicit one-loop calculation that~$k_\text{eff} = \half (1 + \text{sign}(m))$ is indeed generated by the Feynman diagram described below~\eqref{specialST}, with massive propagators. This implies that the same diagram generates~$k_\text{eff} = \half$ in the massless theory. 

In canonical quantization, the contribution of~\eqref{deltam} to the Hamiltonian is given by
\begin{equation}
\Delta H =\int {dq \over 2 \pi} \, {m \over 2} \left(- \psi^\dagger_q \psi_q + \psi_q  \psi^\dagger_q   \right) = \int {dq \over 2 \pi} \, m \left(- \psi^\dagger_q \psi_q\right) + {N \over 2} m ~, 
\end{equation}
where~$N$ is the number of zero modes. As shown in appendix~\ref{FermApp}, this operator ordering gives the Hamiltonian and the partition function the correct spurious transformations under the broken~$\CC\CT$ symmetry, which sends~$m \to -m$. By comparing with~\eqref{mDirac}, we can deduce expectation values of the mass operator~$i \b \Psi \Psi$, evaluated in the massless theory and in translationally invariant states~$|\chi\rangle$, 
\begin{equation}
    \langle \chi|  i \b \Psi \Psi | \chi \rangle = {1 \over \text{Area}} \int {dq \over 2 \pi} \, {1 \over 2} \langle \chi | \left(\psi^\dagger_q \psi_q - \psi_q  \psi^\dagger_q   \right)| \chi \rangle~.
\end{equation}
For instance, if~$|\chi\rangle = |0\rangle$ is the Fock vacuum, then~$\langle 0 | i \b \Psi \Psi|0\rangle = -{N / (2 \text{Area}}) = -B / (4 \pi)$.

\section{QED$_3$ in a Strong Magnetic Field} \label{qed3sec}

In this section, we will discuss QED$_3$ (or simply QED), i.e.~$U(1)$ gauge theory in 2+1 spacetime dimensions with $N_f = 2$ two-component Dirac fermions of unit charge, vanishing~$U(1)$ Chern-Simons level, and time-reversal symmetry.  We will study the effect of a strong, constant magnetic field~$B \gg e^4$ (here~$e^2$ is the dimensionful~$U(1)$ gauge coupling with units of mass), which can be analyzed reliably to leading order in~$e^2/\sqrt{B}$.

\subsection{Lagrangian, Symmetries, and Anomalies} 

Here we review basic aspects of QED$_3$ that we will need in our analysis. See~\cite{Dumitrescu:2024jko} for a detailed recent discussion with references. In the absence a magnetic field, the QED action involves~$N_f$ massless flavors~$\Psi^i~(i = 1, \ldots, N_f)$ of two-component Dirac fermions with unit charge, that couple to a dynamical~$U(1)$ gauge field~$a_\mu$,\footnote{~More precisely, as discussed around~\eqref{spinc}, $a_\mu$ is a dynamical Spin$^c$ connection.}  
   \be \la{LagraNor}
   S_\text{QED} =  \int d^3 x \, \left\{  - { 1 \over 4 e^2 } f_{\mu \nu} f^{\mu \nu } - i   \bar \Psi_i \gamma^\mu ( \partial_\mu   - i  a_\mu ) \Psi^i   + { k_{\rm bare}  \over 4 \pi } \ep^{\mu \nu \rho } a_\mu \d_{\nu} a_\rho \right\}~, \quad k_\text{bare} = -{N_f \over 2} \in \Z~. 
   \ee 
Here~$e^2$ is the~$U(1)$ gauge coupling, which has dimensions of energy and serves as a strong-coupling scale: the theory is weakly coupled in the UV, at energies~$E \gg e^2$, and strongly coupled in the IR at energies~$E \lesssim e^2$.  

Note that we have included in the Lagrangian~\eqref{LagraNor} a bare Chern-Simons term with properly quantized level~$k_\text{bare}$. We will follow the regularization scheme discussed around~\eqref{keffdir}, according to which each massless Dirac flavor contributes~$\Delta k_\text{eff} = 1/2$ to the effective Chern-Simons level. Since we are interested in studying a theory that is invariant under a time-reversal symmetry~$\CT$ (and hence also under spatial reflections~$\CR$),  we must ensure that the total effective level vanishes,
\begin{equation}\la{nokeff}
    k_\text{eff} = {N_f \over 2} + k_\text{bare} = 0~.
\end{equation}
This fixes the bare Chern-Simons level in~\eqref{LagraNor}, and the requirement that it is quantized implies that~$N_f \in 2 \Z$ must be even.

The presentation above has the virtue of treating all fermions equally, which makes the~$SU(N_f)$ flavor symmetry of~\eqref{LagraNor} manifest, but it obscures its time-reversal symmetry: the bare Chern-Simons term is not~$\CT$-invariant, but it is canceled by an anomalous~$\CT$-violating one-loop contribution from the fermions. Once this has happened, the theory is fully~$\CT$-invariant at the quantum level.

Alternatively, we could use a regularization scheme that manifestly preserves a certain notion~$\t \CT$ of time-reversal symmetry (specified in~\eqref{CTilde} below). In this scheme, half of the fermions contribute~$\Delta k_\text{eff} = 1/2$, and the other half~$\Delta k_\text{eff} = -1/2$.  Then no~$k_\text{bare}$ is needed and the classical action is manifestly~$\t \CT$-invariant, but it obscures the~$SU(N_f)$ flavor symmetry.\footnote{~This clash between manifest~$\CT$ and~$SU(N_f)$ symmetries anticipates the 't Hooft anomaly in~\eqref{U2infl}.} This scheme is natural if one groups the fermions into~$N_f/2$ four-component Dirac fermions, each of which contains a pair of two-component fermions with electric charges~$\pm1$.

With these comments in mind, we note that~\eqref{LagraNor} has a unitary charge-conjugation symmetry~$\CC$, which acts as in~\eqref{Cdef} on every fermion flavor, and an anti-unitary time-reversal symmetry~$\CT$, which acts on every flavor as in~\eqref{Tdef}. Consequently, there is also a spatial reflection symmetry~$\CR$, which is implied by the~$\CC\CR\CT$ theorem. 

In the remainder of this section, we will focus on the minimal case of~$N_f = 2$ massless two-component flavors. The case of higher even~$N_f > 2$ will be discussed in section~\ref{LargeNsec} below. The~$N_f = 2$ theory has a unitary zero-form symmetry given by
\begin{equation}\la{u2symm}
    U(2) = {SU(2)_f \times U(1)_m \over \Z_2}~.
\end{equation}
We refer to~$SU(2)_f$ and~$U(1)_m$ as the flavor and magnetic symmetries, respectively. An important point is that the global symmetry is defined to act faithfully, and hence non-projectively, on gauge-invariant local operators -- including the monopole operators discussed below. In QED, all of these operators are bosonic, even though the theory is formulated in terms of fermion fields. This is because fermion parity~$(-1)^F$ coincides with the element $-1 \in U(1)$ of the gauge group, and hence it is gauged.\footnote{~Here it is important that all fermions have odd (in our case unit) electric charge. It is this fact that enables us to extend~$a_\mu$ to a Spin$^c$ connection on non-spin manifolds.} Thus the Lorentz symmetry of the theory is~$SO(2,1)$, not its covering group~$\text{Spin}(2,1)$. Note that the fermions~$\Psi^i$ in~\eqref{LagraNor} transform projectively under the~$SO(2,1) \times U(2)$ symmetry, because they are doublets of both~$\text{Spin}(2,1)$ and~$SU(2)_f$, while being~$U(1)_m$ neutral; this is possible because they are not gauge-invariant local fields. 

In order to describe the action of the~$U(2)$ symmetry~\eqref{u2symm} on gauge-invariant operators, it is useful to make the following distinction:
\begin{itemize}
    \item {\bf Non-Monopole Operators} are by definition neutral under the magnetic~$U(1)_m$ symmetry. They are the standard gauge-invariant local operators constructed out of the fields. Important examples are the field strength~$f_{\mu\nu}$, whose dual
    \begin{equation}\label{u1mcurr}
j^\mu_m = {1 \over 4\pi} \ep^{\mu\nu\rho} f_{\nu\rho}
    \end{equation}
   serves as the conserved~$U(1)_m$ current, and the Hermitian fermion bilinears 
    \begin{equation}\la{Opmdef}
    \CO = i \b \Psi_i \Psi^i~, \qquad  \CO^A = i \bar \Psi_i {(\sigma^A)^i}_j \Psi^j~. 
    \end{equation}
    Here~$\CO$ and~$\CO^A$ transform as an~$SU(2)_f$ singlet and triplet, respectively.\footnote{~All non-monopole operators transform in faithful representations of~$SO(3)_f = SU(2)_f/\Z_2$.} When they are added to the action, these operators give rise to fermion masses,
    \begin{equation}\la{masses}
        \Delta S_\text{masses} = \int d^3 x \, \left(m \CO +   m_A   \CO^A \right)~. 
    \end{equation}
    Both operators are~$\CC\CT$-odd, 
    \begin{equation}\la{OpmCT}
        \CC\CT : \CO, ~ \CO^A \quad \to \quad - \CO, ~ -  \CO^A~.
    \end{equation}
For some purposes, it is convenient to introduce a notion of time-reversal symmetry~$\t \CT$ that is obtained by combining~$\CT$ with a certain~$SU(2)_f$ rotation, 
 \be \la{CTilde}
 \tilde \CT = \CT {\cal U}_f~, \qquad \CU_f = - i \sigma^2 \in SU(2)_f~.
 \ee 
Note that~$\CU_f$ acts by exchanging the~$i = 1,2$ flavor indices of the fermions~$\Psi^i$ (up to a sign). Consequently, $\t \CT$ preserves the~$A = 3$ component of the triplet operator~$\CO^A$ in~\eqref{Opmdef}, but it sends  the singlet operator~$\CO \to - \CO$. Since both of these operators are~$\CC$-even, it follows that
\begin{equation}\label{cttoo3}
    \CC \t \CT : \CO(t) \to - \CO(-t)~, \qquad \CO^3(t) \to \CO^3(-t)~.
\end{equation}
Since~$\t \CT$ and~$\CC\t \CT$ are both compatible with the~$m_3 \CO^3$ Dirac mass term in~\eqref{masses}, there is a regularization scheme (e.g.~of Pauli-Villars type) in which these symmetries are manifestly preserved. This is precisely the scheme described below~\eqref{nokeff}, in which no bare Chern-Simons term is needed in~\eqref{LagraNor}.

\item {\bf Monopole operators} are gauge-invariant local operators of disorder type that create~$N \neq 0$ units of flux for the dynamical~$U(1)$ gauge field~$f$ around a spacetime point; hence they also carry global~$U(1)_m$ charge~$N$. Monopole operators can be defined by studying the theory on a small spatial~$S^2$, with~$N$ units of flux on the~$S^2$ (see~\cite{Borokhov:2002ib}). In QED, the monopole operators can transform under the~$SU(2)_f$ flavor symmetry because they must be dressed with fermion zero modes to make them gauge invariant. For instance, it was shown in~\cite{Borokhov:2002ib} the minimal monopole operator~$\CM^i$ of flux~$N = 1$ transforms in the doublet representation of~$SU(2)_f$. Note that in Lorentzian signature, the total~$U(1)_m$ charge is nothing but the magnetic flux in the~$xy$-plane,
\begin{equation}\la{nfluxjm}
    N = \int d^2 x \, j_m^0 = {1 \over 2 \pi} \int d^2 x \, f_{xy}~.
\end{equation}
\end{itemize}

Finally, the global~$U(2)$ symmetry has a mixed 't Hooft anomaly of parity type with any orientation-reversing symmetry, such as~$\CT$ or~$\CC\CT$. The anomaly inflow action is given by
\begin{equation}\la{U2infl}
    S_{U(2) \, \text{inflow}} = \pi \int_{\CM_4} c_2(U(2))~.
\end{equation}
The second Chern class~$c_2(U(2))$ of the~$U(2)$ background gauge fields involves both the~$SU(2)_f$ and the~$U(1)_m$ instanton densities, i.e.~suitable~$\theta$-angles with~$\theta = \pi$ for both symmetries.

\subsection{Turning on a Strong Magnetic Field}

\subsubsection{Basic Setup}

We are interested in QED, with action~\eqref{LagraNor} and~$N_f = 2$ massless flavors, in a uniform magnetic field~$B$. To this end, we make the following replacement in the QED action,
\begin{equation}\label{qeda}
a \quad \to \quad A_\text{total} = A + a~, 
\end{equation}
where~$A$ (in Landau-gauge~\eqref{Landg}) describes a constant magnetic field~$(dA)_{xy} = B$, and~$a$ is the dynamical~$U(1)$ gauge field describing the fluctuating photon of QED around this background. Note that we fix~$\int (da)_{xy} =0$, so that the magnetic flux is entirely due to the background field~$A$. Recall from~\eqref{nfluxjm} that turning on such a constant magnetic field is tantamount to working in a sector of fixed~$U(1)_m$ charge density~$N/\text{Area} = {B \over  2\pi}$. 

The constant magnetic field~$F_{xy} = B$ breaks Lorentz boosts, and also~$\CC$ and~$\CT$, but it preserves~$\CC\CT$. Since the gauge field is dynamical, the broken symmetries are only spontaneously (not explicitly) broken. This will be important below. 

The magnetic field sets an energy scale $\sqrt{B}$. Let us consider the strong-field regime,
\be \la{strongB}
    e^2 \ll \sqrt{B}~.
\ee 
We are interested in understanding the dynamics at low energies~$E \ll \sqrt{B}$. Due to~\eqref{strongB} we might hope that the problem is weakly coupled, but this is not the case; nevertheless, it can be solved to leading order in the small ratio~$e^2 / \sqrt{B}$.

As a first approximation, we ignore the higher Landau levels (with energies~$E \gtrsim \sqrt{B}$) and focus on the zero-modes in the lowest Landau level. This description is incomplete because the fluctuating QED gauge field~$a_\mu$ in~\eqref{qeda} mediates interactions between the fermions. We must now understand these interactions, which break the enormous degeneracy among the gapless states in the lowest Landau level.

\subsubsection{Intuitive Picture of the Zero-Mode Dynamics}

    Let us quickly sketch the picture that we will subsequently develop in detail below: each Dirac fermion flavor~$\Psi^i~(i = 1,2)$ can be expanded in zero modes~$\psi_q^i$ (as in~\eqref{Pmodexp}). For each Landau orbital, labeled by~$q \in \R$, we have four states, which can be thought of as filled or unfilled states for both flavors. As discussed below~\eqref{eq:seff0}, the zero-modes contribute to the electric charge density~$j^0$, but they do not sustain a spatial current, $\vec j = 0$. Thus they only couple to the~$a_0$ component of the photon. 

One of the effects of path-integrating over~$a_\mu$ is the imposition of the Gauss law. As we will see below, this implies that -- on average -- we must have one filled fermion per Landau orbital~$q$; in fact, we will see that the states of lowest energy are translationally invariant, with precisely one filled state for every~$q$.  However, each filled state could harbor either of the two flavors, which translates into their~$SU(2)_f$ spins pointing up or down.  In the absence of interactions, all of these spin configurations have equal energy.

In our problem -- viewed in Coulomb gauge -- the spatial components~$\vec a$ describe a decoupled massless photon, while~$a_0$ mediates the Coulomb interaction between the zero-modes. It is a repulsive four-fermion interaction, whose ground states are such that all~$SU(2)_f$ spins align. Thus, the~$SU(2)_f$ symmetry is spontaneously broken. This is a manifestation of the famous exchange mechanism for the origin of ferromagnetism~\cite{Heisenberg}: the exclusion principle prevents fermions from being on top of each other if their spins are aligned, but not otherwise, so when the spins are aligned, the repulsive Coulomb energy is minimized. 
 
   \subsection{The Coulomb Interaction in the Lowest Landau Level}

   We will now describe the leading tree-level Coulomb interaction between the fermions, which is mediated by a single photon, and explain in the language of Wilsonian effective field theory why this interaction is sufficient to accurately determine the low-energy physics in the strong-magnetic-field regime~$e^2 \ll \sqrt{B}$. 

\subsubsection{The Electromagnetic Current}\label{emcurr}

In the leading tree-level approximation, the effective field theory of the zero modes is given by a copy of~\eqref{eq:seff0} for each of the zero-mode flavors~$\Psi_0^i~(i = 1,2)$, together with the Maxwell and Chern-Simons terms in~\eqref{LagraNor}. In this approximation, the electromagnetic current~$j^\mu$ (obtained by varying the effective action with respect to~$a_\mu$) has no spatial part, $\vec j = 0$. The electric charge density~$j^0$ is given by~\eqref{jeffk}, summed over the flavors, with an additional correction due to the Chern-Simons term, 
\be \la{CurrExp}
j^0 = j^0_\Psi + { k_{\rm bare} \over 2 \pi }  (B +  f_{xy} )  ~, \qquad j^0_\Psi \equiv  - \bar \Psi_{0 i} \gamma^0 \Psi_0^i  ~, \qquad k_{\rm bare }=-1~.
\ee 
Here the spatial Fourier transform of the zero-mode contribution~$j_\Psi^0$ is (see~\eqref{jeffk}) 
  \be 
  \la{CurInv} 
    { j}^0_\Psi({ \vec p }) = \int {dq dq' \over (2 \pi)^2} \, 2 \pi \delta(q-q'-p_y) \exp\left(-{ {\vec p}^{\,2} \over 4 B} - {i p_x(q + q') \over 2B}\right) (\psi^i_{q'})^\dagger \psi^i_q~. 
\ee  
Each flavor~$\psi_q^i$ obeys the canonical anticommutation relations~\nref{CanCom}, and the two flavors anticommute,~$\{\psi^i_q, \psi^j_{q'}\}=0$.

Even though~$j^0$ is a~$\CC\CT$-odd operator, this is not manifest in~\eqref{CurrExp} because of the bare Chern-Simons term. Recall that this term cancels an equal and opposite quantum-mechanical contribution that is generated by the fermion zero modes. This cancellation can be made explicit by reordering one of the fermion flavors (but not the other, at the expense of manifest~$SU(2)_f$ symmetry),  
\be \la{CurrNew}
    { j}^0({ \vec p }) = \int {dq dq' \over (2 \pi)^2} \, 2 \pi \delta(q-q'-p_y) \exp\left(-{ {\vec p}^{\,2} \over 4 B} - {i p_x(q + q') \over 2B}\right) \left[ (\psi^1_{q'})^\dagger \psi^1_q - \psi^2_q (\psi^2_{q'})^\dagger \right]~. 
\ee 
This expression is manifestly~$\CC \tilde \CT$-odd, see \nref{CTilde}, and since it is~$SU(2)_f$ invariant, is is also odd under~$\CC\CT$. 

Let us summarize the (somewhat subtle) cancellations leading to~\eqref{CurrNew}:  
\begin{itemize}
    \item Naively, the normal ordering constant that arises in going from \nref{CurInv} to \nref{CurrNew} only cancels the constant background~$\sim B$ contributed by the Chern-Simons term in~\eqref{CurrExp}.
\item This naive answer is incompatible with the (non-linearly realized) Lorentz symmetry of the full problem, which requires the entire Chern-Simons contribution~$\sim (B + f_{xy})$ in~\eqref{CurrExp} to cancel, rather than just the term~$\sim B$. However, this symmetry is not preserved if we naively truncate to the lowest Landau level, rather than correctly integrating out the higher levels.  Thus,  the term~$\sim f_{xy}$ is not canceled in this naive truncation. 

\item As discussed around~\eqref{specialST}, the term~$\sim f_{xy}$ in~\eqref{CurrExp} is canceled by a one-loop quantum effect in the low-energy theory of the zero modes, once we take into account the irrelevant operator~\eqref{specialST} that arises from integrating out the higher Landau levels. 

\item Alternatively, we can use a manifestly~$\CC \tilde \CT$-preserving regularization, with a single four-component Dirac fermion containing a pair of two-component fermions with electric charges~$\pm 1$. This explains the relative sign in the current~\eqref{CurrNew}. In this regularization~$k_\text{bare}$ vanishes and the current has no Chern-Simons contribution, as in \nref{CurrNew}.  
\end{itemize}
With this understanding, we will use~\eqref{CurrNew} as our leading~$U(1)$ gauge current.

Of course,  the full current~$j^\mu$ differs from~\eqref{CurrNew} by the variation of higher-order terms in the effective action with respect to~$a_\mu$ (see sections~\ref{LLLEFT} and~\ref{QEDEFT} for a discussion of such terms). These corrections render the full~$j^\mu$ Lorentz-covariant, as well as invariant under arbitrary~$U(1)$ gauge transformations (as discussed below~\eqref{jeffk}, this is not the case for~\eqref{CurrNew}). For instance, the explicit dependence of~\eqref{CurrNew} on~$B$ implies that if we have a slowly varying fluctuation of the magnetic field, we should replace~$B \to B + f_{xy}$. This induces some higher-order couplings of the lowest Landau level to the spatial components of~$a_\mu$; these will make an appearance in section~\ref{photanom} below.

Let us discuss the action of the leading-order current~\eqref{CurrNew} on the Fock space of the zero modes. For instance, the Fock vacuum~$|0\rangle$ has constant negative charge density, 
\be \la{VacZer}
\psi^i_q |0 \rangle =0~, \qquad j^0({\vec x})|0\rangle = -{B \over 2\pi}~.
\ee 
It follows that the Fock vacuum has total electric charge~$-N$, and hence it is a bosonic state for even~$N$ and a fermionic state for odd~$N$. Note that~$|0\rangle$ is also invariant under the~$SU(2)_f$ symmetry.\footnote{~The~$SU(2)_f$ generators are represented on the zero modes as~$ S^A_f = \half \int {dq \over 2\pi} \, (\psi^i_q)^\dagger {( \sigma^A)^i}_j \psi^j_q$, see section~\ref{exSol}.} 

One interesting state of vanishing total~$U(1)$ charge can be obtained by filling all the orbitals~$q \in \R$ with the second fermion flavor, 
\be \la{VacChi3}
|\downarrow~\rangle \equiv C_\downarrow \prod_{q \in \R}  (\psi_q^{2})^\dagger |0 \rangle 
~, \qquad \psi^1_q |\downarrow~\rangle =0 = (\psi^2_q)^\dagger |\downarrow~\rangle~, \qquad \langle \downarrow | \downarrow \rangle = 1~.
\ee 
Here the notation~$|\downarrow~\rangle$ is to indicate that all~$SU(2)_f$ spins are pointing down in this state,\footnote{~We could write~$|\downarrow \downarrow \cdots \downarrow\rangle$, to indicate that all spins are pointing down, but this would be cumbersome.} see section~\ref{expval}, and we have included an (infinite) normalization factor~$C_\downarrow$ to ensure that the state~$|\downarrow~\rangle$ has unit norm. Note that this state is always bosonic, as required for all gauge-invariant states in QED, because it is obtained by acting on the Fock vacuum of charge~$-N$ with~$N$ creation operators~$(\psi_q^2)^\dagger$ of charge~$+1$. Interestingly, the state~\eqref{VacChi3} also has exactly zero charge density, 
\begin{equation}\label{jonchi3}
    j^0(\vec p \,)|\downarrow~\rangle = 0~,
\end{equation}
which is manifest if we use~\eqref{CurrNew}. 

At this point~\nref{VacZer} and \nref{VacChi3} are just two possible states among a large set of low-energy states in the zero-mode Fock space.  In order to find the true ground states, we will need to understand the leading interaction effects on this highly degenerate space. 

\subsubsection{Leading Coulomb Interaction}
\label{leadCoulInt}

In the leading approximation described above, the electromagnetic current~$j^\mu$ that couples to the dynamical QED photon~$a_\mu$ only has a~$j^0$ component given by~\eqref{CurrNew}. It thus does not couple to the spatial components of the photon field~$\vec a$. This makes it particularly convenient to analyze the problem in Coulomb gauge, $\vec \grad \cdot \vec a = 0$. In this gauge, $\vec a$ describes a decoupled, transverse photon, so that the~$U(1)_m$ symmetry is spontaneously broken, with the photon playing the role of the Nambu-Goldstone Boson. 

As usual, $a_0$ is non-propagating in Coulomb gauge; integrating it out leads to the instantaneous Coulomb interaction, which is bilinear in the charge density~$j^0$ and described by the following Hamiltonian,
   \be \la{VerFor}
    H_C =  { e^2 \over 2} \int { d^2 p  \over (2 \pi)^2 } \, \left[  {j^0 ({ \vec p}\, )} \right]^\dagger    { 1 \over {\vec p}^{\, 2} }  \, j^0({\vec p })~.
    \ee
Several comments are in order:
\begin{itemize}
    \item Since~$j^0$ contains two zero-modes~$\psi_q^i$, this is a four-fermion interaction that describes the Coulomb repulsion between electrons (as well as attraction between electrons and positrons). Given that the scaling dimension of~$\psi_q^i$ is zero, as discussed below~\eqref{SzmQM}, the four-fermion interaction~\eqref{VerFor} also has vanishing scaling dimension and is thus highly relevant. This leads to a strongly-coupled quantum mechanics problem for the zero modes, whose solution we present in section~\ref{exSol}.

\item The Gauss law, which states that the total electric charge~$j^0(\vec p = \vec 0) = 0$ should vanish, arises from integrating out the zero-momentum mode of~$a_0$. It is therefore an automatic consequence of~$\eqref{VerFor}$, which has an IR divergence unless the Gauss law holds. 

\item In Coulomb gauge, the fact that the intermediate photon in~\nref{VerFor} has zero frequency is automatic, because the kinetic term for~$a_0$ has no time derivatives. In fact, \nref{VerFor} also correctly describes the interaction mediated by a single tree-level photon in any gauge, even if the~$a_0$ propagator depends on the frequency of the photon, because the external zero-mode fermions have vanishing energy. 
\end{itemize}

   \subsubsection{Wilsonian Point of View}\la{QEDEFT}

   Let us comment on the nature of the approximation in which~\eqref{VerFor} is the leading interaction for the zero modes. We will adopt a Wilsonian point of view and organize the theory by {\it energy} scales (not momentum scales), with a floating energy cutoff~$\Lambda$. Let us first consider the case 
    \be \la{LamLE}
     e^2 \ll \Lambda \ll \sqrt{B}~.
     \ee 
    At such an energy scale~$\Lambda$, the degrees of freedom in the Wilsonian action are as follows:
\begin{itemize}
    \item Photons with an (approximately) relativistic dispersion relation~$E^2 = {\vec p}^{\,2}$. Thus, all photons with~$E, |\vec p| > \Lambda$ have been integrated out. 

\item The zero-mode fermions~$\psi_q^i$ describing the Lowest Landau level. All higher Landau levels with energies~$E > \Lambda$ have been integrated out. This part of the problem is similar to the one discussed in section~\ref{LLLEFT}.  

Note that even though the parameter~$q \in \R$ that labels the zero modes can be thought of as the momentum in the $y$-direction, it can be arbitrarily large while respecting the energy cutoff~\nref{LamLE}. We will therefore not impose a cutoff on~$q$. It is, however, true that the wavefunctions of the zero modes can only be localized to spatial distance scales of order the magnetic length~$\ell_B \sim 1/\sqrt{B}$. 

Thus, the zero-modes~$\psi^i_q$ have dimension zero under the scaling symmetry discussed below~\eqref{SzmQM}, which acts on time, but not on their momentum label~$q$. 
\end{itemize}

After integrating out all the photon and higher-Landau-level modes with energies~$E > \Lambda$,  we expect to generate (approximately) local interactions involving the zero-modes~$\psi_q^i$ and the low-energy modes of the photon. One way to organize these local terms, which was used in section~\ref{LLLEFT} (see also~\cite{Hong:1997uw}), is to work in terms of the projection~$\Psi_0^i(x)$ of the Dirac field onto the lowest Landau level, and only expand this field in terms of the quantum mechanical fermions~$\psi_q^i$ at the very end.  
      
An important fact is that the theory is weakly coupled at the scale $\Lambda$, since~$e^2 \ll \Lambda$, so that tree-level interactions dominate. Indeed, the leading tree-level interaction is precisely the Coulomb interaction~\nref{VerFor} that arises from single photon exchange. It gives interaction terms that are of order $e^2$ and involve four fermions. In addition, we also have terms that come from integrating out higher Landau levels. Such terms are suppressed by extra powers of $e^2/\sqrt{B}$ or $(\text{momentum})/ \sqrt{B}$, where the denominator comes from the energy $\sqrt{B}$ of the higher Landau level fermions. See e.g.~the discussion around~\eqref{SeffT}, as well as~\cite{Hong:1997uw,Hattori:2017qio} for a related discussion in 3+1 dimensions (where the details are different, because the fermions do not scale in the same way).  

One potential worry is the following: some of the higher-order terms described above (that we would like to neglect in our analysis) contain spatial derivatives~$\vec \grad$ of $\Psi_0$,\footnote{~Using the leading-order equations of motion, time derivatives of~$\Psi_0$ vanish.} and these can in principle give rise to large factors of $q$, which is the (unbounded) $y$-momentum of the zero-mode wavefunctions~\nref{DirSol}. Thankfully, invariance under background gauge transformations of~$A$ (taking us out of Landau gauge~\eqref{Landg}) ensures that any spatial derivatives acting on~$\Psi_0$ should come in the form of covariant derivatives~$\vec D = \vec \grad - i \vec A$. In Landau gauge~\eqref{Landg} this means that~$D_x = \d_x$; this is bounded by~$|\d_x| \lesssim \sqrt{B}$ when acting on the~$x$-dependent part of the zero-mode wavefunctions~\nref{DirSol}, which is a Gaussian of width~$\sim \ell_B = {1 / \sqrt{B}}$. By contrast, $|D_y|$ evaluates to~$|q- B x| \lesssim \sqrt{B}$ when acting on the zero-mode wavefunctions; the bound is again due to the width of the Gaussian. Note that the covariant gradient~$\vec D$ is analogous to the velocity of the single-particle problem, which (unlike the canonical momentum~$P_y = q$) is gauge invariant.

Let us comment on the range of momenta that contribute to the Coulomb Hamiltonian~\eqref{VerFor} of the zero-modes:
\begin{itemize}
\item If we impose the Wilsonian cutoff \nref{LamLE}, then we should strictly speaking only integrate over $|\vec p|> \Lambda$ in \nref{VerFor}. 
\item The integral is exponentially suppressed at momenta~$|\vec p| \gg \sqrt{B}$, due to the decaying Landau level form factor in~\eqref{CurrNew}. 
\item As we lower the Wilsonian energy cutoff~$\Lambda$, we will eventually reach the scale~$\Lambda \sim e^2$, integrating over all but the lowest photon momenta in  \nref{VerFor}. As we will see below, we must include low-momentum photons in the IR effective action, so we should not integrate them out. We thus implicitly have an IR cutoff at~$|\vec p| \sim e^2$ in \nref{VerFor}, which we will not specify precisely. 

\item Changing the IR cutoff is subleading in~$e^2$. At leading order in small~$e^2$, we can therefore use~\eqref{VerFor} at face value, without imposing any UV or IR restrictions on the momentum. This has the advantage of automatically imposing the Gauss law,  as discussed below~\eqref{VerFor}.    
\end{itemize}

The discussion above explains why the tree-level Coulomb Hamiltonian~\eqref{VerFor} is the leading interaction in the energy range~\eqref{LamLE}. As we will show below, this interaction gives rise to spontaneous symmetry breaking, with a weakly-coupled effective theory for the resulting Nambu-Goldstone bosons at suitably small energies and momenta. In order for this picture to be self-consistent, it is crucial that loop effects in the Wilsonian effective theory remain subleading, even if we lower the cutoff~$\Lambda$ below the scale~$e^2$. This is not the case for massless QED in the vacuum (with~$B = 0$), which becomes strongly coupled when~$\Lambda \lesssim e^2$. By contrast, in a sufficiently strong magnetic field, the zero modes -- though themselves strongly coupled -- do not drastically modify the IR behavior of the photon, see e.g.~the discussion around~\eqref{OneLLLL}.

\subsection{Solving the Coulomb Hamiltonian and Symmetry Breaking} \la{exSol}

\subsubsection{Finding the Vacua}\la{vacua}

The leading Coulomb Hamiltonian~\nref{VerFor} is a relevant four-fermion deformation in the low-energy theory, at energies less than $\sqrt{B}$, so that the problem becomes strongly coupled in the IR. However, the exchange potential in~\nref{VerFor} has some special features that make it possible to find both the ground states and the excitations exactly, as we will now explain. In fact, our problem is very similar to the problem of quantum Hall ferromagnetism~\cite{PhysRevB.30.5655,sondhi1993skyrmions,girvin1999quantumhalleffectnovel}. 

Roughly speaking  (see appendix~\ref{FMQHE}), quantum Hall ferromagnets arise when electrons with spin are confined to a two-dimensional plane and subjected to a magnetic field, but one can neglect their magnetic moment (or Zeeman) coupling. In that case, there is a global~$SU(2)$ spin-rotation symmetry. An important difference is that the Coulomb potential~$\sim 1/r$ that arises in those systems still lives in 3+1 dimensions. In our case, the Coulomb potential arises from 2+1 dimensional photon exchange and is~$\sim \log r$. This means that we can repeat the analysis of~\cite{PhysRevB.30.5655,sondhi1993skyrmions,girvin1999quantumhalleffectnovel} to find the ground state and its excitations, making just small changes.\footnote{~One other difference is that the quantum Hall ferromagnets studied in~\cite{sondhi1993skyrmions} break~${\cal C  T }$ symmetry and have a non-zero Hall conductance. In our case, the~${\cal CT}$ symmetry ensures that the Hall conductance vanishes in the vacuum. In that respect, our problem is more similar to graphene in a magnetic field, see e.g.~\cite{Semenoff:2011ya,Miransky:2015ava}.}

The first observation is that the Coulomb Hamiltonian~\nref{VerFor} for the zero-modes is non-negative,
   \be \la{VerForPos}
    H_C =  { e^2 \over 2} \int { d^2 p  \over (2 \pi)^2 } \, \left[  {j^0 ({ \vec p}\, )} \right]^\dagger    { 1 \over {\vec p}^{\, 2} }  \, j^0({\vec p }) \geq 0~,
    \ee
    because it is an integral over non-negative operators of the form~$j^\dagger j \geq 0$ with a positive measure provided by the~$a_0$ propagator, i.e.~the Fourier-transformed Coulomb potential. Clearly any state~$|\chi\rangle$ that saturates the bound~\eqref{VerForPos} must be a ground state of~$H_C$ with exactly zero energy, and this happens if and only if,\footnote{~\label{fn:SUSY}This is reminiscent of the zero-energy ground states in a supersymmetric theory, which are annihilated by the supercharges~$Q$, while the Hamiltonian~$H_\text{SUSY} \sim Q^\dagger Q \geq 0$ is quadratic in the~$Q$'s and non-negative.}
\be \la{CondOk}
j^0({\vec p}\, ) |\chi \rangle =0 ~, \quad \text{for all } \vec p~.
\ee 
Thus, the state~$|\chi\rangle$ must have vanishing charge density. 

It is possible that no states~$|\chi\rangle$ obeying~\eqref{CondOk} exist, in which case the ground states of~$H_C$ have strictly positive energy and must be determined by other means.\footnote{~In the analogy with supersymmetric theories (see footnote~\ref{fn:SUSY}), this would correspond to the case of spontaneous supersymmetry breaking, which is generally more difficult to analyze.} Luckily, we have already encountered a state of exactly zero charge density: it is the all-down~$SU(2)_f$ spin state in~\nref{VacChi3}, which we repeat here 
\be \la{VacChi3Bis}
|\downarrow~\rangle \equiv C_\downarrow \prod_{q \in \R}  (\psi_q^{2})^\dagger |0 \rangle 
~, \qquad \psi^1_q |\downarrow~\rangle =0 = (\psi^2_q)^\dagger |\downarrow~\rangle~, \qquad \langle \downarrow | \downarrow \rangle = 1~.
\ee 
By a global~$SU(2)_f$ rotation we can construct a ground state~$|n^A\rangle$ that points along any unit ~$SU(2)_f$ triplet vector~$n^A$, with~$|\downarrow~\rangle$ corresponding to~$n^A = - \delta^A_3$. Thus, the space of vacua forms a round~$S^2$, or equivalently~$\C\P^1$, parametrized by~$n^A$.

It is not difficult to check that these rotated~$|n^A\rangle$ constitute all states~$|\chi\rangle$ that obey~\nref{CondOk}. To this end, we examine the expression~\eqref{CurrNew} for~$j^0(\vec p) \sim (\psi^1_{q'})^\dagger \psi^1_q -  \psi^2_q (\psi^1_{q'})^\dagger$ in terms of the zero modes: dialing~$p_y$ in that expression allows us to fix any~$q - q'$, and suitably Fourier-transforming in~$p_x$ fixes arbitrary~$q + q'$, so that~\nref{CondOk} is equivalent to 
\be  \la{CondVa}
\left[ (\psi^1_q)^\dagger \psi^1_{q'} - \psi^2_{q'} (\psi^2_q)^\dagger  \right]|\chi \rangle =0~, \qquad \text{for all } q, q'~.
\ee 
When~$q = q'$, this constraint requires precisely one occupied fermion for every~$q$,\footnote{~It is helpful to imagine that~$q$ is a discrete label, so that the canonical anticommutation relations~\eqref{CanCom} have a Kronecker~$\delta_{q q'}$ on the right side, and we can reorder the expression in square brackets that appears in the constraint~\eqref{CondVa} as follows,
\begin{equation}
   (\psi^1_q)^\dagger \psi^1_{q'} - \psi^2_{q'} (\psi^2_q)^\dagger  =  (\psi^i_q)^\dagger \psi^i_{q'} - \delta_{q q'}~.
\end{equation}
Note that this is~$SU(2)_f$ invariant.} so that 
\begin{equation}
    |\chi\rangle = \prod_q \alpha_q^i ( \psi^i_q)^\dagger|0\rangle~, \qquad \psi_q^i|0\rangle = 0~.
\end{equation}
Here~$|0\rangle$ is the Fock vacuum. To see that all~$\alpha_q^i || \alpha_{q'}^i$ are parallel~$SU(2)_f$ spinors, it suffices to impose~\eqref{CondVa} for~$q \neq q'$. We can then use an~$SU(2)_f$ rotation to set all~$\alpha_q^i 
\sim \delta^i_2$, in which case we recover~$|\downarrow~\rangle$ in~\eqref{VacChi3Bis}.
Of course, this final state is a very simple product state with no entanglement.

\subsubsection{Expectation Values}\label{expval}

Let us compute the expectation operators of some simple operators in the vacua found above. It is a feature of the lowest-Landau-level approximation we are using that Lorentz scalars and vectors that are bilinear in the fermions become identical at leading order. For instance, consider the~$SU(2)_f$ triplet fermion mass operator in~\eqref{Opmdef},
\begin{equation}\la{Obis}
    \CO^A = i \b \Psi_i {( \sigma^A)^i}_j \Psi^j~,
\end{equation}
and the~$SU(2)_f$ flavor currents,  together with the associated charges, 
\begin{equation}
     j_\mu^A = - \half \b \Psi_i {(  \sigma^A)^i}_j  \gamma_\mu \Psi^j~,  \qquad    S^A_f = \int d^2 x \, {  j}^{\, 0 A}(\vec x)~.
\end{equation}
At leading order, their spatial components~$  j^A_{x,y}$ flow to zero in the deep IR, while the~$SU(2)_f$ charge density is proportional to~\eqref{Obis},
\begin{equation}\label{vOisvJ}
     {  j}^{\, 0 ,A }(\vec x) = \half  \CO^A(\vec x) = \half \int {dq dq' \over (2 \pi)^2} \, u_{q'}^\dagger(\vec x) u_{q}(\vec x) \, \psi^\dagger_{q'i} {(  \sigma^A)^i}_j \psi_{q}^j~.
\end{equation}
This is only possible because these two operators transform identically under the symmetries left unbroken by the magnetic field; in particular, they are both~$\CC\CT$-odd, as can be verified directly from~\eqref{CTonzm}.

We can now compute the expectation values~$\langle \cdots \rangle_{\downarrow}$ of these operators in the translationally-invariant vacuum state~$|\downarrow~\rangle$ in~\eqref{VacChi3} (see~\cite{Gusynin:1994re,Semenoff:1999xv} for related computations),
\begin{equation}\label{ChargDen}
    \langle  {  j}^{\, 0 , A}(\vec x) \rangle_{\downarrow} = \half   \langle  \CO^A(\vec x) \rangle_{\downarrow} = -{B \over 4 \pi} \delta^A_3 ~.
\end{equation}
Note that these vevs are all pointing in the direction of the south pole~$n^A = - \delta^A_3$ of the sphere (or~$\C\P^1$) of vacua~$|n^A\rangle$ found in section~\ref{vacua} above. (We can obtain all other vacua on~$\C\P^1$ by~$SU(2)_f$ rotations.) The non-zero~$SU(2)_f$ charge-density~\eqref{ChargDen} is exactly what is expected for a ferromagnet with~$N = {B \over 2\pi} (\text{Area})$ spins, all of which point down and contribute~$S^3_f$ spin~$-\half$. On its own, this background charge density does not spontaneously break the~$SU(2)_f$ Cartan generator~$S^3_f$ that preserves the spin axis, which is unbroken in a standard ferromagnet. However, we will see this symmetry is actually spontaneously broken in our problem, because of its interplay with the transverse photon that is not present in typical ferromagnets (see section~\ref{LowSigma}).  

An interesting fact about the charge density~\eqref{ChargDen} is that it is proportional to the magnetic field~$B$, which  according to~\eqref{nfluxjm} is the charge density of the~$U(1)_m$ symmetry,
\begin{equation}\la{magdens}
  \langle  { j}_m^{0}(\vec x) \rangle_{\downarrow}= {B \over 2 \pi} ~.
\end{equation}
Note that the dynamical, transverse photon~$f_{xy}$ does not contribute to this expectation value. Thus the~$U(1)_m$ symmetry is also spontaneously broken, with Nambu-Goldstone boson~$f_{xy}$, as already discussed in section~\ref{leadCoulInt}.

It follows from~\eqref{ChargDen} and~\eqref{magdens} that a linear combination of the~$SU(2)_f$ Cartan and~$U(1)_m$ has no background charge density, 
\begin{equation}\label{jmplus2j3}
    \langle  j_m^0(\vec x) + 2 j^{\, 0, A = 3}(\vec x) \rangle_{\downarrow} = 0~,
\end{equation}
and, in fact, this linear combination is unbroken even once the mixing with the transverse photon is taken into account (see section~\ref{LowSigma}).  We thus see that, as far as the internal symmetries are concerned,\footnote{~We already saw that Lorentz boosts are spontaneously broken. Below, we will also see that time translations mix with a spontaneously broken~$U(1)$ flavor symmetry, as in a superfluid.} the symmetry breaking pattern is
\begin{equation}\label{SSBpat}
U(2) = {SU(2)_f \times U(1)_m \over \Z_2} \to U(1)_\text{unbroken} = \half \left(U(1)_m + U(1)_f\right)~,
\end{equation}
where~$U(1)_f \subset SU(2)_f$ and~$U(1)_\text{unbroken}$ are conventionally normalized~$U(1)$ symmetries with integer charges.\footnote{~Thus the~$U(1)_f$ charge is~$2 S_f^3$.} As was explained in~\cite{Dumitrescu:2024jko}, this pattern is consistent with the Vafa-Witten theorem~\cite{Vafa:1983tf}, which continues to hold in a magnetic field. It follows from~\eqref{SSBpat} that the full space of vacua is in fact a (squashed) three sphere,
\begin{equation}\la{s3vacua}
    {\t S}^3 = U(2) / U(1)_\text{unbroken}~,
\end{equation}
which is Hopf-fibered over the~$\C\P^1$ base of vacua labeled by the states~$|n^A\rangle$ described in section~\ref{vacua} above. Here, the Hopf fiber is furnished by the transverse photon, as discussed in section ~\ref{LowSigma}.

It is instructive to imagine what would happen if we placed the theory on a very large spatial~$S^2$ -- much larger than~$1/e^2, 1/\sqrt{B}$ -- as described in section~\eqref{sec:compact}. In this case, there is no symmetry breaking. Instead, the ground states on the sphere transform in a single irreducible representation of the~$SU(2)_f$ flavor symmetry, whose total flavor spin is fixed by the magnetic flux, 
\be \la{TotSpin}
s_f = { N \over 2  } ~, \qquad  N = { 1 \over 2 \pi } \int_{S^2}  d^2x\,  B~.
\ee 
Then~$|\downarrow~\rangle$, with charge density~\nref{ChargDen} is the all-down spin state, with~$S_f^3 = - s_f$. Note that the ground states on the spatial sphere are singlets under the~$SU(2)_R$ geometric rotation symmetry of the sphere,\footnote{~We can argue this directly by considering the analogue of the state~$|\downarrow~\rangle$ in~\eqref{VacChi3} on~$S^2$, which must take the form~$|\downarrow~ \rangle_{S^2} = \prod_{m=-j}^j\psi^{2 \dagger}_{m} |0 \rangle$. Here, each flavor~$\psi_m^i$ gives a copy of the zero modes on~$S^2$ discussed around~\eqref{s2zm}. Clearly~$|\downarrow~ \rangle_{S^2}$ has~$J_3 = 0$. To see that it is a full~$SU(2)_R$ singlet, observe that the~$N$ zero-modes~$\psi_m^2$ transform in the fundamental of an~$SU(N)$ symmetry that contains~$SU(2)_R$. It is therefore sufficient to argue that the state~$|\downarrow~ \rangle_{S^2}$ is an~$SU(N)$ singlet, but this is manifest from Fermi statistics.} consistent with the fact that they become translationally invariant vacua in the flat-space limit. Even though the sphere is large, we can interpret these states in terms of monopole operators (which correspond to states on a small sphere): as far as their charges are concerned, they transform in the~$N$-fold symmetric power of the minimal~$N =1$, $SU(2)_f$ doublet monopole~$\CM^i$ discussed above~\eqref{nfluxjm},
\begin{equation}\label{SymPow}
    (\text{vacua on } S^2) \sim \text{Sym}^N \left(\CM^{i} \right)~.
\end{equation}
Note, however, that the symmetry-breaking pattern~\eqref{SSBpat} in flat space, and the resulting vacuum manifold~\eqref{s3vacua}, are not dictated by the stability group of the charge-$N$ monopole operator in~\eqref{SymPow}. Rather, they are consistent with the stability group of the minimal~$N = 1$ monopole operator~$\CM^i$, as discussed further in section~\ref{weakfield}. 

Let us examine the analogue of~\eqref{vOisvJ} in the~$SU(2)_f$ singlet sector, where we have the singlet mass~$\CO = i \b \Psi_i \Psi^i$ in~\eqref{Opmdef}, and for the electromagnetic~$U(1)$ charge density~$j^0$ in~\eqref{CurrNew}, 
\begin{equation}\label{Oisjem}
    \CO(\vec x) = { j}^0({ \vec x })~. 
\end{equation}
Recall that the operator ordering~\eqref{CurrNew} makes both operators manifestly odd under~$\CC\t \CT$, and since they are~$SU(2)_f$ singlets, also under~$\CC\CT$. We already known from~\eqref{CondOk} that the current~$j^0$ must annihilate the vacuum, and hence~$\CO$ does too; a fortiori, they both have vanishing vevs,
\be \la{OddOp}
\langle   \CO(\vec x)  \rangle_\downarrow = \langle j^0(\vec x)\rangle_\downarrow= 0~.
\ee 
The idea that such~$\CC\t \CT$-odd operators should not acquire symmetry-breaking vevs was advocated in~\cite{Vafa:1984xh}, by appealing to features of the Euclidean QED$_3$ path integral that continue to hold in a magnetic field.

\subsubsection{Adding Small Masses}

Using the discussion above, we can comment on what happens when we turn on the singlet and triplet masses~$m, m^A$ in~\eqref{masses}, as long as they are much smaller than~$\sqrt{B}$. At leading order, the zero-mode Hamiltonian changes as follows,
\begin{equation}\label{massH}
    \Delta H_\text{zero-modes} = - m \int d^2 x \, j^0 - m^A \int d^2 x \, \CO^A~.
\end{equation}
Here we have used the fact~\eqref{Oisjem} that~$\CO$ flows to the electromagnetic charge density, while~$\CO^A$ is the zero-mode operator in~\eqref{vOisvJ}.  Since the vacua~$|n^A\rangle$ in the massless theory are annihilated by~$j^0$, the~$\CC\CT$-odd mass~$m$ has no effect at leading order. By contrast, the triplet mass~$m^A$ pins~$n^A \sim m^A$, leading to a unique vacuum for the $S^2$ part of the sigma model. (Of course, we still have the vacua parametrized by the dual photon.)  Note that, via~\eqref{vOisvJ}, we can express the second term in~\eqref{massH} as a Zeeman coupling~$ \sim -  m^A S_f^A$. These properties of the vacua will be reflected in our discussion of the low-energy effective theory in section~\ref{LowSigma}.



 \subsection{Excitation Spectrum of Nambu-Goldstone Bosons}\label{excitations}

Having found the symmetry-breaking ground states~\eqref{s3vacua} of the Coulomb Hamiltonian~$H_C$ in~\eqref{VerFor} for the zero modes, we expect on general grounds that the system has gapless Nambu-Goldstone bosons (NGBs). This is indeed the case, as we will now confirm. Moreover, the special form of~$H_C$ makes it possible to determine the dispersion relation~$\omega(\vec k)$ of the NGBs exactly, for all momenta~$\vec k$, as has long been known in condensed matter physics~\cite{Bychkov1981, PhysRevB.30.5655}. The analysis in this section can be repeated on a spatial~$S^2$, which would give the excitations above the ground states described around~\eqref{TotSpin}, see e.g.~\cite{NAKAJIMA1994327}, but we will not do it here.

We will focus on excitations of the all-spins-down ground state~$|\downarrow~\rangle$ in~\eqref{VacChi3Bis}. Naively, the simplest ones are single spin flips, obtained by acting with either $( \psi^1_q)^\dagger$ or $\psi^2_q$ on~$|\downarrow~\rangle$. However, these are electrically charged excitations, which are logarithmically confined by the Coulomb force in 2+1 dimensions: their energy diverges 
as~$\log L$, with~$L \to \infty$ an IR cutoff, as we will explicitly verify below.

Thus, the lowest-energy excitations must be electrically neutral particle-hole pairs,\footnote{~In 3+1 dimensional QED such a pair is known as positronium; in condensed matter physics, it is referred to as an exciton.} 
\be \la{AlSta}
 |e^+ e^- ; {q',q} \rangle = (\psi^1_{q'})^\dagger \psi^2_{q} |\downarrow~\rangle~.
 \ee 
Note that this excitation has flavor spin~$\Delta S_f^3 =  +1$ relative to the all-down ground state~$|\downarrow~\rangle$, as befits a single spin flip,\footnote{~More precisely~\eqref{AlSta} has charge~$+1$ under the~$U(1)_\text{unbroken}$ symmetry in~\eqref{SSBpat}.} but that there is no analogous excitation with~$\Delta S_f^3 = -1$. Conversely, all states with~$\Delta S_f^3 = 1$ are linear combinations of the~\eqref{AlSta}, and hence the same is true for~$H_C|e^+ e^- ; {q',q} \rangle$.

We can diagonalize the action of~$H_C$ on the states~$|e^+ e^- ; {q',q} \rangle$ by constructing linear combination of definite momentum~$\vec k = (k_x, k_y)$. This results in a formula similar to~\eqref{CurrNew}, except that we must flip the sign of the momentum,\footnote{~The prefactor is chosen so that these states are normalized as
\begin{equation}\label{epmnorm}
    \langle e^+ e^- ; \vec k | e^+ e^- ; \vec k' \rangle = (2 \pi)^2  \delta^{(2)}\left(\vec k - \vec k'\right)~.
\end{equation}}  
 \be \label{epmkdef}
 |e^+e^-; \vec k \rangle \equiv \sqrt{2 \pi \over B} \int { dq dq' \over (2\pi)^2 } \, 2\pi \delta( q-q' + k_y) \, \exp\left({i k_x ( q + q' ) \over 2 B} \right) (\psi^1_{q'})^\dagger \psi^2_{q} |\downarrow~\rangle~.
 \ee 
 These states diagonalize the Coulomb interaction~\eqref{VerForPos},
   \be 
H_C |e^+e^-; \vec k \rangle =   \omega(k) |e^+e^-; \vec k \rangle~.
 \ee 
A simple way to deduce the dispersion relation~$\omega(k)$, which uses~\eqref{CurrNew} and~$j^0(\vec p)|\downarrow~\rangle = 0$, begins with the observation that  
 \begin{equation}\label{jonkstate}
     j^0(\vec p) |e^+e^-; \vec k\rangle =  2 i \exp\left( -{ {\vec p}^{\, 2} \over  4 B}\right)  \sin \left( {k_x p_y - k_y p_x \over 2B} \right) \, |e^+e^-; \vec k - \vec p\rangle~,
 \end{equation}
consistent with the fact that~$j^0(\vec p)$ carries momentum~$-\vec p$. Since~$H_C \sim \int d^2 p \, j^0(- \vec p) j^0(\vec p) / {\vec p}^{\, 2}$ is bilinear in the current, we can immediately apply~\eqref{jonkstate} twice to obtain
 \be\la{GenExp} 
 \omega(k) =    e^2   \int {d^2 p \over (2\pi)^2 }  \,  {\exp\left( -{ {\vec p}^{\, 2} \over  2 B}\right) \over {\vec p}^{\, 2}}  \,    \left(1- \exp\left(  {i ( k_x p_y - k_y p_x) \over B } \right)  \right)~.   
 \ee  
The dispersion relation~$\omega(k)$ only depends on the magnitude~$k = |\vec k|$ of the momentum, as required by rotational invariance. This becomes manifest when we express~\nref{GenExp} in polar coordinates,   
  \be \la{DispRel}
  \omega(k) = { e^2 \over 2 \pi } \int_0^\infty { d y \over y } \, \exp\left( - y^2 {B  \over2  k^2 } \right) \left[  1- J_0( y) \right]  = { e^2 \over 4 \pi } \left[  \log\left( { k^2 \over 2 B}\right) + \Gamma\left( 0, { k^2 \over 2 B } \right) + \gamma_E  \right]~, 
  \ee 
  where~$J_0(y)$ is a Bessel function of the first kind, $\Gamma(0,z) = \int_z^\infty { d t \over t } e^{-t} $ is the incomplete gamma function, and~$\gamma_E$ the Euler-Mascheroni constant.

Let us expand the dispersion relation~\eqref{DispRel} for small and large momenta, and comment on the relevant physics in these regimes. In the small-$k$ regime we find 
\be\la{Smallk}
\omega(k) = \alpha k^2 \,  \left (1 - { k^2 \over 8 B } + \cdots \right) ~, \quad \alpha =  { e^2\over 8 \pi B}~, \quad {\rm for } \; k^2 \ll B~. 
\ee
This displays the expected gapless Nambu-Goldstone bosons as~$k \to 0$, with the quadratic dispersion relation~$\omega \sim k^2$ that is typical of ferromagnets. In section~\ref{LowSigma} we will discuss the low-energy sigma model that describes these gapless excitations, including their interactions with the massless transverse photon that we have neglected above. 

It is worth emphasizing that the coefficient~$\alpha$ of the leading~$O(k^2)$ term in~\nref{Smallk} comes from the~$p$-integral in \nref{GenExp}, expanded at small~$\vec k$, so that most of the contribution is concentrated around~$p \sim \sqrt{B}$. Since~$p$ is the momentum of the exchanged photon, this means that the coefficient~$\alpha$ is set by this short-distance interaction of the lowest Landau Level modes.   In the Wilsonian language of section~\ref{QEDEFT}, this means that we would always capture the full~$\alpha$ as long as the Wilsonian cutoff~$\Lambda \ll \sqrt{B}$.

Now let us consider the large-$k$ regime of the dispersion relation~\eqref{DispRel},  
\be\la{Largek}
\omega(k) = { e^2 \over 4 \pi }  \left[  \log\left( { k^2 \over 2 B }\right)  + \gamma_E  + O \big (e^{ - {k^2 \over 2 B} }\big)    \right] ~, \quad{\rm for} \; k^2\gg B~. 
\ee

  \begin{figure}[t]
   \begin{center}
    \includegraphics[scale=.6]{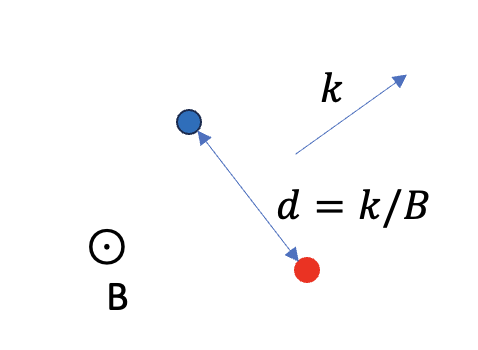}
    \end{center}
    \caption{Classical picture of the~$e^+e^-$ excitation for large momentum~$k \gg \sqrt{B}$: as the positively charged fermion (red) and its negatively charged antiparticle (blue) move in the~$k$-direction, the magnetic Lorentz force pulling them a distance~$d = k/ B$ apart in the transverse direction is exactly balanced by their Coulomb attraction.}
    \label{Pair}
\end{figure}

\noindent The leading logarithmic growth of the energy~$\omega(k)$ can be understood semiclassically, as shown in figure~\ref{Pair}: it is just the Coulomb energy~$V_C \sim {e^2 \over 2\pi} \log d$ of an~$e^+e^-$ pair separated by a distance~$d \sim k$. To understand this relation, consider the attractive Coulomb force~$f_C = {e^2 \over 2 \pi d}$ that acts along~$d$ in figure~\ref{Pair}; it is balanced by the magnetic Lorentz force if both charges move with velocity~$v = f_C / B$ in the direction transverse to~$d$, along~$k$ in figure~\ref{Pair}. Since it follows from~\eqref{Largek} that~$v = { d \omega(k) \over d k } = {e^2 \over 2 \pi k}$, we conclude that~$d \sim k$ as expected, with coefficient
\begin{equation}\la{dkrel}
    d = {k \over B}~.
\end{equation}

\noindent Several comments are in order:
\begin{itemize}
\item[(i)] It follows from~\eqref{dkrel} that~$d \to \infty$ as~$k \to \infty$, so that the charged particles become infinitely separated; for large finite~$k \gg \sqrt{B}$ they are well separated, $d \gg \ell_B$.\footnote{~Recall that the magnetic length~$\ell_B = 1 /\sqrt{B}$ is the smallest distance scale that can be probed in the lowest Landau level.} Thus, large~$k$ is associated with large distances, not high energies, and consequently there is no UV cutoff on~$k$.  We have encountered similar phenomena in sections~\ref{LLLEFT} and~\ref{QEDEFT}. 

Note that the velocity of a well-separated pair is very small,
\begin{equation}\la{nonrelv}
    v = {e^2 \over 2 \pi B d} \ll 1~, \qquad d \gg \ell_B = {1 \over \sqrt{B}}~.
\end{equation}
Thus the particles are effectively non-relativistic, even though they were originally massless.

\item[(ii)] The logarithmically confining Coulomb potential comes from the small-$p$ region of the integral~\nref{GenExp} over photon momenta~$\vec p$. This is consistent with the idea that large~$k$ corresponds to long spatial distances, see figure~\ref{Pair} and point~(i) above. The regime of small photon momentum is sensitive to the Wilsonian cutoff~$\Lambda$ we impose, as discussed in section~\ref{QEDEFT}. To obtain~\eqref{Largek}, we have integrated out all photon momenta. 

It is interesting to reinstate a small non-vanishing Wilsonian cutoff~$0 < \Lambda \ll e^2$, so that we only integrate over photon momenta~$p \gtrsim \Lambda$. In this case, we find that the energy of a well-separated particle-hole pair is finite and given by~${e^2 \over 4 \pi} \log {B / \Lambda^2}$.  Since it follows from~\eqref{nonrelv} that the motion is non-relativistic, we can say that the effective mass of each charged fermion is half of the total energy,  
\be \la{EffMass}
\mu_\text{eff}(\Lambda)  =  { e^2 \over 8 \pi } \log {B \over   \Lambda^2} ~.
\ee 
Of course, this expression diverges as $\Lambda \to 0$.

\item[(iii)] The Coulomb energy of a well-separated pair with~$k \sim d \to \infty$ diverges as~$V_C \sim {e^2 \over 2 \pi} \log d$, and hence the individual charged fermions cannot be isolated -- they are confined (see below for further comments). This is reflected in their infinite effective mass~\eqref{EffMass} in the~$\Lambda \to 0$ limit. 

\end{itemize}

\noindent Since QED is a theory with fundamental charges and no one-form symmetry, the statement of confinement in point (iii) above requires elaboration:

The main point is that the~$e^+ e^-$ two-particle bound states~\eqref{epmkdef} described above, with dispersion relation~\eqref{DispRel}, are exactly stable for all momenta~$k$. This is certainly true within the lowest Landau level approximation, because all gauge-invariant excitations carry positive spin~$\Delta S_f^3  \geq 1$ relative to the ground state~$|\downarrow~\rangle$, and~\eqref{epmkdef} has the smallest such spin $\Delta S_f^3 = +1$. This approximation is unassailable for all but exponentially large momenta, at which the logarithmic term in~\eqref{Largek} is comparable with the Landau-level energy separation. 

In this regime, one might worry that the state could decay into some of the higher Landau levels. To assess whether this is possible,  we have to remember that the momentum~$\vec k$ is conserved, and that it is almost entirely due to the electric and magnetic fields. In particular, $k$ is proportional to the electric dipole moment, i.e.~to the separation~$d$ between the electron and the positron, as reflected in~\eqref{dkrel}. This means that any higher-Landau-level mode would also need to be created at the same locations to conserve momentum, leading to the same Coulomb energy, plus the gap to the higher Landau level. It is important in this discussion that the large energy contained in the electric field is spread over long distances so that we can trust the low-energy approximation when we calculate it. 

Thus, there is always a net energy cost when we try to excite a higher Landau level mode with the same momentum~$\vec k$, so that the bound states~\eqref{epmkdef} cannot decay.

\subsection{Generalization to Other Interactions}

As reviewed in~\cite{Miransky:2015ava}, the phenomenon of symmetry-breaking in the lowest Landau level is not unique to the Coulomb Hamiltonian~\eqref{VerFor} analyzed above. Here we will briefly indicate how to generalize our discussion to other interactions. 

Let us replace the 2+1 dimensional Coulomb potential~${e^2 \over 2\pi} \log r$ in position space by a more general, rotation-invariant potential~$v(r)$. A natural generalization of the Hamiltonian~\eqref{VerFor} is then given by
 \be \la{Hvgen}
 H_v =  { 1 \over 2} \int { d^2 p  \over (2 \pi)^2 } \, \left[ j^0 ({\vec p}\, )\right]^\dagger \,  \tilde v( p)  \, j^0({\vec p })~,
    \ee
where $\tilde v(p)$ is the rotationally-invariant Fourier transform of the potential~$v(r)$. If the potential is suitably repulsive, so that~$\tilde v(p) > 0$ for all~$p$, then it follows from~\eqref{Hvgen} that the vacuum is again annihilated by all operators~$j^0({\vec p})$. This leads to exactly the same vacua as in the case of the Coulomb interaction, which were described in section~\ref{vacua}. 

An important example of a potential that obeys this positivity condition is the 3+1 dimensional Coulomb potential~$v(r) \sim {1 / r}$, with~$\t v(p) \sim {1 / p} > 0$, which describes the quantum-Hall ferromagnets in~\cite{PhysRevB.30.5655,sondhi1993skyrmions,girvin1999quantumhalleffectnovel} (briefly reviewed in appendix~\ref{FMQHE}), as well as those that occur in graphene~\cite{Semenoff:2011ya}. Another commonly studied case (see e.g.~\cite{Gusynin:1994re,Semenoff:1999xv}) is a current-current contact interaction,  for which~$\tilde v(p) = \t v$ is constant.\footnote{~An interaction involving four UV Dirac fermions is not renormalizable in 2+1 dimensions. Here we mean that the theory has been UV completed in a way that leads to the effective Hamiltonian~\eqref{Hvgen} with constant~$\t v > 0$ in the lowest Landau level.}  

We can immediately generalize the analysis of the excitations in section~\ref{excitations}. In particular, the dispersion relation is given by replacing~$e^2/ p^2 \to \t v(p)$ in~\eqref{GenExp},    
 \be \la{DispReloth}
  \omega(\vec k)  =   \int {d^2 p \over (2\pi)^2 } \,   \left(1- e^{ i {( k_x p_y - k_y p_x)/ B } }  \right) {e^{ - {p^2 \over 2 B}} \,  \tilde v(|\vec p|)  }~,
\ee 
whose small-$k$ expansion gives the dispersion relation of the gapless Nambu-Goldstone bosons, 
  \be  \la{GolDi}
  \omega(k \to 0) =  \alpha k^2 ~,~~~~~\alpha =  { 1 \over 4 B^2 }  \int {d^2 p \over (2\pi)^2 }  \,    p^2  {e^{ - {p^2 \over 2 B}} \,  \tilde v(p)  }~.
  \ee

  Formula \nref{DispReloth} has an interesting interpretation: the product of $\tilde v(\vec p) $ and $ e^{ - {p^2 \over 2 B}} $ in momentum space gives a convolution between the potential~$v(\vec x)$ and a Gaussian~$g(\vec x) \sim e^{- B|\vec x|^2/2}$ in position space. We can think of~$g(\vec x)$ as a rotationally-invariant wavefunction in the lowest Landau level. Then~\nref{DispReloth} is nothing but a Fourier-transform to position space, if we take the positions to be~$ x^a_k \equiv \ep^{ab} k_b/B$, as suggested by figure~\ref{Pair}, 
  \be \la{OmPos} 
  \omega(\vec k) =   \hat v(0)-\hat v(\vec x_k)  ~, \qquad x^a_k \equiv {\ep^{ab} k_b \over B} ~, \qquad \hat v(\vec x) \equiv  v(\vec x) * g(\vec x) ~, \qquad g(\vec x) \equiv { B \over 2 \pi } e^{ - B |\vec x|^2/2}~. 
  \ee 
  Here, the star~$*$ indicates the convolution in position space.

Using~\eqref{OmPos}, we can consider the large-momentum limit~$k\gg \sqrt{B}$ of the dispersion relation. If the potential $v(r)$ decays as~$r \to \infty$ in position space, then~$\omega(k \to \infty) = \hat v(0)$ is constant, and thus a well-separated particle-hole pair of fermions has finite energy. By contrast, for the Coulomb potential~$v(r) \sim -{e^2 \over 2\pi} \log r$,  the second term in~\nref{OmPos} dominates at large $k$, so that we recover the logarithmically divergent $\omega(k \to \infty)$ in~\nref{Largek}. A similar conclusion would follow for any confining~$v(r)$ that grows at long distances.

\subsection{The $\t S^3$ Sigma Model at Low Energies} 
   \la{LowSigma}

  We continue our discussion of QED$_3$, where the Coulomb Hamiltonian~\eqref{VerForPos} for the zero modes triggers the symmetry-breaking pattern~\eqref{SSBpat}, resulting in the squashed three-sphere~$\t S^3$ of vacua in~\eqref{s3vacua}. The gapless Nambu-Goldstone Bosons (NBGs) are the transverse photon, and the particle-hole bound states found in section~\ref{excitations}. We will now describe the effective non-linear sigma model, with target space~$\t S^3$, that describes these excitations at suitably low energies and momenta. 

\subsubsection{The~$\C\P^1$ Base of the Hopf Fibration}

As discussed around~\eqref{s3vacua}, the~$\t S^3$ vacuum manifold is the total space of an~$S^1$ circle bundle that is Hopf-fibered over a~$\C\P^1 = S^2$ base. For presentation purposes, we begin with the~$\C\P^1$ base, deferring a discussion of the Hopf fiber until section~\ref{Hfiber}. An important caveat is that some aspects of the physics are not correctly captured by the~$\C\P^1$ base alone, and we will point them out explicitly as we go along.

The~$\C\P^1$ base of the Hopf fibration is described by the vacua~$|n^A\rangle$ labeled by a unit vector~$n^A$, so that~$n^A n^A = 1$. Recall that the all-down state~$|\downarrow~\rangle$ has~$n^A = - \delta^A_3$, so that it corresponds to the south pole of the~$\C\P^1$. In the deep IR, we therefore introduce a dynamical~$SU(2)_f$ unit vector field~$n^A(t, \vec x)$, whose vev dictates which vacuum~$|n^A\rangle$ we are talking about. By comparing with~\eqref{ChargDen}, we see that the triplet fermion bilinear flows in the IR sigma model to
\begin{equation}\label{OAtonA}
    \CO^A =  {B \over 2\pi} n^A + \left(\text{derivative terms}\right)~.
\end{equation}

As usual, the low-energy fluctuations of~$\vec n(x)$ describe gapless NGBs, whose quadratic small-$k$ dispersion relation
\begin{equation}\label{smallkBis}
    \omega(k \ll \sqrt{B}) = \alpha k^2~, \qquad \alpha =  { e^2\over 8 \pi B}~,
\end{equation}
we found exactly in~\eqref{Smallk}. At leading order, the effective action for~$n^A(t, \vec x)$ is therefore a non-linear sigma model with target space~$\C\P^1 = S^2$ that is second-order in the spatial gradients~$\vec \grad$, but first order in time-derivatives~$\d_t$, as is typical of ferromagnets (see e.g.~\cite{Fradkin:2013anc}),\footnote{~Importantly, quantum fluctuations of the order parameter do not destroy the symmetry, because the propagator that follows from~\eqref{smallkBis} is integrable at small frequencies and momenta.}  
\be \la{SMG}
  S_{\mathbb{C P}^1} =  \int dt dx^2 \, \left[ - { B \over 2 \pi }  \, {\CA}_\alpha \d_t \theta^\alpha -   { e^2 \over ( 8 \pi)^2 } \,  \vec \grad n^A \cdot \vec \grad n^A \right]~, \qquad n^A n^A = 1~.
  \ee
Let us explain these two terms, whose form is largely fixed by the symmetries, and how to determine their coefficients:
\begin{itemize}
\item The first term in~\eqref{SMG} is written using standard angular coordinates~$\theta^\alpha(x) = (\theta, \varphi)$ on the~$S^2 = \C\P^1$ target space parametrized by the unit vector~$n^A(x)$.\footnote{~The~$S^2$ metric is~$ds^2_{S^2} = d\theta^2 + \sin^2 \theta \, d \varphi^2$.} Then~$\CA_\alpha$ is a standard~$U(1)$ connection on that~$S^2$, whose curvature~$d \CA$ is proportional to the volume form: it is an~$SU(2)_f$ -invariant unit monopole in target space, 
\begin{equation}\la{ConFor}
    d \CA = \half \sin \theta \, d\theta d\varphi~, \qquad \int_{S^2} d\CA = 2\pi~.
\end{equation}
Since the first term in~\eqref{SMG} is invariant under gauge-transformations of~$\CA$, it is thus also~$SU(2)_f$ invariant.

\item Up to a sign, the coefficient of the first term in~\eqref{SMG} can be determined by placing the whole theory on a large spatial~$S^2$ (not to be confused with the~$S^2 = \C\P^1$ target space of the sigma model). To determine the ground states, we can ignore the spatial gradients in~\eqref{SMG} and focus on the effective quantum mechanics of~$\theta^\alpha(t)$, with action 
  \be \la{phaseQM}
   S_\text{QM on spatial $S^2$} = -N \int dt \, \CA_\alpha(\theta) \d_t \theta^\alpha~, \qquad N \in \Z~,
\ee 
where we have used the flux quantization condition~\eqref{fluxquant}. This is a Wilson line of charge~$-N$ (i.e.~a quantum-mechanical Chern-Simons term) for~$\CA_\alpha$. Quantizing the first-order action~\eqref{phaseQM}, whose phase-space is the round~$S^2 = \C\P^1$ with~$SU(2)_f$ isometry and total phase-space volume $2\pi N$, gives an~$N+1$ dimensional Hilbert spin with total~$SU(2)_f$ spin 
\begin{equation}
    s_f = {N \over 2}~.
\end{equation}
This is the correct ground state on a large spatial~$S^2$, as discussed around~\eqref{TotSpin}.\footnote{~Since all~$SU(2)_f$ representations are (psuedo-) real, changing the sign of~\eqref{phaseQM} leads to a conjugate, and hence equivalent, $SU(2)_f$ representation.}

\item The sign of the first term in~\eqref{SMG} will be fixed below, by examining the flavor charges of the NGB excitations. Another check is discussed in section~\ref{photanom}.

\item The coefficient of the (manifestly~$SU(2)_f$-invariant) second term in~\eqref{SMG} is fixed by computing the excitations of the NGBs and matching with~\eqref{smallkBis}. 
\end{itemize}

The last two points require further discussion of the NGB fluctuations around the all-down vacuum~$|\downarrow~\rangle$ described by~$n^A = - \delta_3^A$, or equivalently~$\theta = \pi$. The Hermitian~$SU(2)_f$ Cartan generator stabilizing this axis is~$S_f^3 = -i \d_\phi$. Let us use complex stereographic coordinates~$z$, with~$z = 0$ corresponding to the south pole,\footnote{~The usual stereographic coordinate adapted to the north pole is~$z_\text{N} = \tan{\theta \over 2} e^{i\varphi}$, and it has~$S_f^3$ charge~$+1$. It is related to our~$z = z_\text{S}$ with~$S_f^3 = -1$, which is adapted to the south pole, via an inversion~$z_\text{N} = 1/z_\text{S}$.}
\begin{equation}\la{zExprzb}
    z = \tan{\pi - \theta \over 2} e^{-i\varphi}~.
\end{equation}
Clearly~$z$ has~$S_f^3$ charge~$-1$. We will also need the K\"ahler connection and the metric,\footnote{~We have chosen a gauge in which~$\CA$ is regular at the south pole~$z = 0$. The gauge-invariant K\"ahler form is~$d\CA = {i dz \wedge d \b z / (1 + |z|^2)^2}$.}
\begin{equation}\la{ConExpl}
    \CA =  - {i \over 2} \cdot { \b z dz - z d \b z \over 1 + |z|^2}~, \qquad ds^2 = {4 dz d\b z \over (1 + |z|^2)^2}~. 
\end{equation}
Expanding near the south pole~$z \simeq 0$, the action~\eqref{SMG} becomes
\be \la{SMGzexp}
  S_{\mathbb{C P}^1} \simeq  \int dt dx^2 \, \left[  { i B \over 2 \pi }  \, \b z \d_t z -   { 4 e^2 \over ( 8 \pi)^2 } \,  \vec \grad \b z \cdot \vec \grad z \right]~.
  \ee
Then expanding in plane waves~$z \sim e^{- i \omega t + i \vec k \cdot \vec x}$ and using the equation of motion gives the correct dispersion relation~\eqref{smallkBis}. 

Canonically quantizing~\eqref{SMGzexp} leads to the following equal-time commutators,
\begin{equation}
    [z(\vec x), \b z(\vec y)] =  {2 \pi \over B} \delta^{(2)}(\vec x - \vec y)~.
\end{equation}
Then
\begin{equation}\label{zmodes}
    z(t, \vec x) = \sqrt{2 \pi \over B} \int {d^2 k \over (2 \pi)^2} \, e^{- i \omega(k) t + i \vec k \cdot \vec x} \, a_{\vec k}~, \qquad [a_{\vec k}, a^\dagger_{\vec k'}] = (2 \pi)^2 \delta^{(2)}(\vec k - \vec k')~.
\end{equation}
Note that~$z(\vec x)$ only contains only contains annihilation operators.\footnote{~The fact that~$z$ only contains annihilation (rather than only creation) operators is dictated by the sign of the first term in~\eqref{SMG}. Note that this situation never arises in a relativistic context, where the equations of motion for bosonic fields are always second-order in the time-derivatives and lead to creation operators (related to anti-particles) in the mode expansion.} The vacuum~$|\downarrow~\rangle$ that we are expanding around is precisely the Fock vacuum annihilated by the~$a_{\vec k}$,
\begin{equation}\label{aonvac}
    a_{\vec k} |\downarrow~\rangle = 0~.
\end{equation}
Since~$\b z$ has~$S_f^3 = +1$, it follows that the creation operators~$a^\dagger_{\vec k}$ produce excitations of that charge on top of the vacuum~$|\downarrow~\rangle$. This is precisely the charge of the NGBs, as discussed around~\eqref{AlSta}.

The fact that~$z(x)$ in~\eqref{zmodes} only contains annihilation operators reflects the famous fact that ferromagnets only possess one Nambu-Goldstone boson (or spin-wave excitation), even though they spontaneously break~$SU(2) \to U(1)$. This in turn is due to the non-vanishing charge density for the~$U(1)$ Cartan in~\eqref{ChargDen}, see e.g.~\cite{Watanabe:2014fva} for a modern discussion with references. Note that, on its own, this charge density does not mean that the~$U(1)$ symmetry is spontaneously broken in an ordinary ferromagnet -- though it will be the case for us, once we take into account the coupling to the transverse photon (section~\ref{photanom}), or equivalently the Hopf fiber (section~\ref{Hfiber}).

\subsubsection{Coupling to the Photon and Anomaly Matching} \label{photanom}

The leading-order effective action~\eqref{SMG} for the~$\C\P^1$ NGBs exhibits a problem we have already encountered several times, see especially the discussions around~\eqref{CSsplit} and below~\eqref{CurrNew}: since QED is a relativistic theory in the UV, the background magnetic field~$B$ breaks Lorentz-invariance (as well as~$\CC$ and~$\CT$) spontaneously, and thus the effective action should only depend on fully relativistic combinations of the total~$U(1)$ gauge field~\eqref{qeda},\footnote{~The spatial gradients in the second term of~\eqref{SMG} can be written in a manifestly covariant way by constructing projection operators using~$dA_\text{total}$, which leads to higher-order interactions with the photon field~$f$, see e.g.~\cite{Golkar:2014wwa}.}
\begin{equation}\label{qedabis}
    A_\text{total}  = A + a~, \qquad (dA)_{xy} = B~.
\end{equation}
This can be achieved by covariantizing the first term in~\eqref{SMG} to a mixed, fully relativistic Chern-Simons term, 
   \be \la{CSte}
  S_\text{mixed CS} = - { 1 \over 2 \pi } \int  dA_\text{total} \wedge { \cal A} _\alpha d \theta^\alpha   =  \int d^3 x \, \left[-{B \over 2 \pi} \CA_\alpha \d_t \theta^\alpha\right] - { 1 \over 2 \pi } \int a \wedge d{\cal A}~.
  \ee
As in the discussion around~\eqref{CSsplit}, we expect this term to arise from carefully integrating out the lowest-Landau-level fermions, which acquire a large effective mass~\eqref{EffMass}. Note that the two terms on the right side of~\eqref{CSte} are separately gauge invariant.\footnote{~Strictly speaking, this is true if~$a$ is a standard~$U(1)$ connection. Since it is actually a Spin$^c$ connection, with half-integer fluxes~\eqref{spinc}, it must be defined more carefully, e.g.~by extending spacetime~$\CM_3$ to a four-dimensional bulk~$\CM_4$. See section~4.1.2 of~\cite{Dumitrescu:2024jko} for a detailed discussion.} They are also invariant under the spontaneously broken~$\CC$ and~$\CT$ symmetries ($A_\text{total}$ and~$\CA$ are both~$\CC$-odd, see~\eqref{Cdef}, but they transform oppositely under~$\CT$, see~\eqref{Tdef}), and the same is true of the second term in the effective action~\eqref{SMG}. 

The Chern-Simons term~\eqref{CSte} couples the~$\C\P^1$ sigma model to the previously decoupled transverse photon. This coupling has several important consequences:
\begin{itemize}
    \item[1.)] It identifies the electric current~$j^\mu$, which couples via~$\Delta S = \int d^3 x \, a_\mu j^\mu$, as the topological (or Skyrmion) current~$j^\mu_\text{top}$ of the~$\C\P^1$ sigma model, which is therefore gauged, 
  \be \la{CoupFi}
  j^\mu \quad \to \quad j^\mu_\text{top} =  - {1 \over 2\pi} \ep^{\mu\nu\rho} (d\CA)_{\alpha\beta} \d_\nu \theta^\alpha \d_\rho \theta^\beta = - { 1 \over 8 \pi }   \ep^{\mu \nu \rho } \, \vec n \cdot  ( \partial_\nu \vec n \times \partial_\rho \vec n)~.  
    \ee 
Here, the arrow denotes the RG flow to the IR. A particle-like Skyrmion configuration with unit winding number around the~$\C\P^1$ has charge~$-1$, just as the microscopic fermions.\footnote{~It follows that the Skyrmions must be fermions, and hence the~$\C\P^1$ sigma model also contains the Hopf term described in~\cite{Wilczek:1983cy, Freed_2018}. As explained in~\cite{Dumitrescu:2024jko}, this is inevitable once the mixed Chern-Simons term~\eqref{CSte} coupling the dynamical Spin$^c$ gauge field~$a$ to the~$\C\P^1$ sigma model is carefully defined.} We know these to be confined by the logarithmic Coulomb force in 2+1 dimensions. As we will see below, this fate is shared by the Skyrmions, thanks to the coupling~\eqref{CoupFi}. Note that this coupling explicitly involves the (previously decoupled) transverse photon~$\vec a$, as well as the spatial electric current~$\vec j$, and both are needed to make the second term in~\eqref{CSte} gauge invariant. 

 \item[2.)] We can perform a sensitive check of~\eqref{CoupFi} by expanding the charge density in the vicinity of the south-pole~$z = 0$ of the~$\C\P^1$ target space using~\eqref{ConExpl}, 
\begin{equation}
    j^0_\text{top} = -{1 \over 2\pi} (d\CA)_{xy} \simeq -{i \over 2\pi} \left(\d_x z \d_y \b z - \d_y z \d_x \b z\right)~.
\end{equation}
Substituting the mode expansion~\eqref{zmodes}, we find
\begin{equation}\label{elcureff}
    j^0_\text{top}(\vec p) = {i \over B} \int {d^2 k \over (2 \pi)^2} \,  (k_x p_y - k_y p_x) \, a^\dagger_{\vec k -\vec p} a_{\vec k}~.
\end{equation}
Note that we are free to reorder this expression at no cost. It follows immediately that the topological current~\eqref{elcureff} annihilates the vacuum thanks to~\eqref{aonvac}, 
\begin{equation}
    j^0_\text{top}(\vec p)|\downarrow~\rangle = 0~,
\end{equation}
exactly as the electric charge density~$j^0(\vec p)$ in the microscopic description~\eqref{CondOk}.  

The action of~\eqref{elcureff} on single Nambu-Goldstone excitations $|\text{NGB}; \vec k\rangle \equiv a^\dagger_{\vec k}|\downarrow~\rangle$ is\footnote{~For small momenta~$\vec k$, the states~$|\text{NGB}; \vec k\rangle \equiv a^\dagger_{\vec k}|\downarrow~\rangle$ should be identified with the particle-hole bound states~$|e^+ e^-; \vec k\rangle$ in~\eqref{epmkdef}, including their normalization~\eqref{epmnorm}.} 
\begin{equation}
j^0_\text{top}(\vec p) |\text{NGB}; \vec k\rangle = {i \over B} \left(k_x p_y - k_y p_x\right) |\text{NGB}; \vec k - \vec p\rangle~.
\end{equation}
This matches the microscopic relation~\eqref{jonkstate} to leading order in the small momenta~$\vec k, \vec p$, as is appropriate for our low-energy sigma model.

 \item[3.)] As shown explicitly in section~4.1.3 of~\cite{Dumitrescu:2024jko}, the Chern-Simons term~\eqref{CSte} matches all 't Hooft anomalies of QED, with inflow action~\eqref{U2infl}, once we coupled to background gauge fields for the~$U(2)$ symmetry~\eqref{u2symm}. 

\item[4.)]  The Chern-Simons term modifies the symmetry-breaking pattern from that of an ordinary antiferromagnet (described below~\eqref{aonvac}) with a decoupled photon, to the diagonal breaking pattern~\eqref{SSBpat}. This is easiest to see once we dualize the photon -- a task to which we now turn.
\end{itemize}

\subsubsection{Hopf Fiber from the Dual Photon}\label{Hfiber}   

Together with the Maxwell kinetic term, the action for the photon~$a_\mu$ that follows from~\eqref{CSte} is given by
\begin{equation}\label{photonac}
    S_\text{photon} = \int \, \left( {1 \over 2 e^2 } f \wedge * f - {1 \over 2\pi} f \wedge \CA\right)~.
\end{equation}
It is useful to dualize the photon to a compact scalar~$\chi  \sim \chi + 4\pi$. (The reason for the non-standard periodicity is explained below.)  How to this carefully in the presence of the Chern-Simons term in~\eqref{photonac} was worked out in section 4.3 of~\cite{Dumitrescu:2024jko}. Locally, the duality is expressed by the relation\footnote{~The compact scalar~$\sigma$ used in~\cite{Dumitrescu:2024jko} has~$2\pi$-periodicity, so that~$\chi = 2 \sigma$. Note that a~$4\pi$ winding of~$\chi$ leads to unit electric flux in~\eqref{eflux}} 
\be \label{eflux}
    { 1 \over e^2 } * f=  {1 \over 4\pi} \left(d\chi + \cos \theta  d \varphi\right)~, \qquad \chi \sim \chi + 4 \pi~. 
\ee 
Substituting into~\eqref{photonac}, and adding the first-order term and the spatial gradients in~\eqref{SMG}, we obtain the following low-energy effective action up to second order in derivatives,
\be \la{KinPhot}
   S_{\t S^3}=  - { e^2 \over 32 \pi^2  } \int d^3 x \,  \left[ (\partial_\mu \chi + \cos \theta \, \partial_\mu  \varphi )^2  + \half  \Big( (\vec \grad \theta)^2 + \sin^2 \theta (\vec \grad  \varphi)^2 \Big)   \right] + \int d^3 x \, { B \over 4 \pi } \, \cos \theta \, \d_t \varphi~. 
    \ee 
Here~$\mu = 0, 1, 2$ is a spacetime index, while~$\vec \grad$ denotes a spatial gradient. 

Focusing only on the spatial derivatives, we recognize~\eqref{KinPhot} as a non-linear sigma model, whose target space is a squashed three-sphere~$\t S^3$ in Hopf coordinates (hence the~$\chi \sim \chi + 4 \pi$ periodicity). This~$\t S^3$ has~$U(2)$ symmetry, i.e.~it is not round, and precisely agrees with the space of vacua we found in~\eqref{s3vacua} after taking into account the mixing of the~$SU(2)_f$ and the~$U(1)_m$ symmetries. The squashed~$\t S^3$ metric only differs from a round~$S^3$ by a relative factor of two between the base and the fiber. Note that the last term in~\eqref{KinPhot} can be absorbed into the relativistic Hopf-fiber kinetic term $\sim (\d_\mu \chi + \cdots)^2$, if we replace
\begin{equation}\label{chiBfluc}
    \chi \quad \to \quad \chi_\text{total} = { 4\pi B \over  e^2 } t + \chi~.
\end{equation}
Thanks to~\eqref{eflux} this describes a constant magnetic field~$f_{xy} = B$, with fluctuations~$\chi$. We will interpret the linear time-dependence of~$\chi_\text{total}$ below.

It is instructive to compute the $SU(2)_f$ charge density from~\nref{KinPhot}. Using the Noether procedure, we find\footnote{~This is consistent with~\eqref{OAtonA}, and the fact that~$j^{0, A} = \half \CO^A$ to leading order in~$e^2$, as shown in~\eqref{vOisvJ}. In other words, the~$O(e^2)$ derivative terms in~\eqref{OAtonA} and~\eqref{su2genS3} need not obviously agree.}
\be \label{su2genS3}
    j^{0,A} = { B \over 4 \pi } n^A  +   { e^2 \over 16 \pi^2 } ( \d_t \chi + \cos \theta \d_t  \varphi) \, n^A~.
\ee 
Similarly, the~$U(1)_m$ charge density can be obtained by noting that~$e^{i \chi}$ has~$U(1)_m$ charge~$+2$, while the other variables are neutral, so that 
\begin{equation}\label{jmnonrel}
    j_m^0 = {e^2 \over 8 \pi^2}  (\d_t \chi_\text{total} + \cos\theta\, \d_t \varphi) = {B \over 2\pi} +  {e^2 \over 8 \pi^2}  (\d_t \chi + \cos\theta\, \d_t \varphi)~.
\end{equation}
If we now expand~\eqref{su2genS3} and~\eqref{jmnonrel} around the south pole~$n^A = -\delta^A_3$, we find that~$j^0_m + 2 j^{0, A = 3}$ contains no background charge, nor any terms linear in the NGBs. This confirms that the (properly normalized) unbroken~$U(1)$ symmetry is given by~\eqref{SSBpat}, 
\begin{equation}
    U(1)_\text{unbroken} = \half \left(U(1)_m + U(1)_f\right)~,
\end{equation}
where~$U(1)_f$ has integer charges~$2 S_f^3$. The linear combination orthogonal to~\eqref{SSBpat} is spontaneously broken,
\begin{equation}
    U(1)_\text{broken} = \half \left(U(1)_m - U(1)_f\right)~.
\end{equation}
The corresponding charge density (again expanded around the south pole) is given by
\be\label{brokencurr}
\half j^0_m - j^{0, A = 3} = {B \over 2\pi} + {e^2 \over 8 \pi^2} (\d_t \chi - \d_t\varphi) + (\text{terms with at least 2 NGBs})~. 
\ee
Note that it contains terms linear in the NGBs, as befits a spontaneously broken symmetry, as well as a background charge~$\sim B$. Thus, we are dealing with a superfluid for the~$U(1)_\text{broken}$ symmetry. This can already be gleaned from the linear time-dependence of~$\chi_\text{total}$ in~\eqref{chiBfluc}, which shows that ordinary time translations, generated by the Hamiltonian~$H$, are spontaneously broken. An unbroken time-reversal symmetry can be defined by suitably mixing~$H$ with the~$U(1)_\text{broken}$ charge obtained by integrating~\eqref{brokencurr} over all of space. The order parameter that triggers this symmetry-breaking pattern is the~$i = 2$ component~$\CM^{i = 2} \sim e^{i \chi_\text{total}/2}$ of the minimal~$N=1$ monopole operator.\footnote{~By contrast, $\CM^{i = 1}$ vanishes at the south pole. See~\cite{Dumitrescu:2024jko} for more details on how the monopole operator~$\CM^i$ is represented in the IR sigma model.}  

The fact that the target-space of the sigma model~\eqref{KinPhot} is~$\t S^3$ has two immediate, important consequences:
\begin{itemize}
\item The target space has no two-cycles, $\pi_2(\t S^3) = 0$. Consequently, there is no topology to support Skyrmions: they are confined because they carry electric charge and couple to the massless (dual) photon, as discussed below~\eqref{CoupFi}. 

\item The target space is three-dimensional, so that we can add a topological~$\theta$-term with coefficient~$\theta = \pi$ to the action,
\be \label{thetaterm}
    S_{\theta\text{-term}} = \pi \int d^3 x \, \left[ { 1 \over 16  \pi^2 } \sin \theta \, d\chi  d\theta  d\varphi  \right]~.
    \ee 
    The expression in square brackets is the unit volume form on~$\t S^3$, so that~$\theta \sim \theta + 2\pi$ and~$\theta = \pi$ is compatible with time-reversal. In fact, this term is not optional, because it is responsible for matching the 't Hooft anomalies of the theory in the~$\t S^3$ sigma model presentation~\cite{Chester:2024waw,Dumitrescu:2024jko}. As explained in~\cite{Dumitrescu:2024jko}, it arises automatically when we dualize~\eqref{photonac} while being careful about global issues. 
\end{itemize}

\subsubsection{Further Comments}\label{furthercommentssec}

Here we make some further remarks about the low-energy effective action~\eqref{KinPhot}:
\begin{itemize}
    \item In the deep IR, it is weakly coupled, as expected for Nambu-Goldstone bosons. However, it becomes strongly coupled at energies $E \sim e^2$ and momenta $p \sim \sqrt{B}$, as can be seen by making the following rescalings in the~$\C\P^1$ part~\nref{SMG} of the effective action, 
   \be 
   \la{NewSpaTi} 
   t e^2 = \hat t~, \qquad x \sqrt{B} = \hat x~,
   \ee 
which has coefficients of order one when expressed in terms of the dimensionless quantities~$\hat t , ~\hat x $. Of course, in that regime \nref{SMG} is not a good description and we should return to the four-fermion Coulomb Hamiltonian~\nref{VerFor}. 

If we perform the rescalings~\eqref{NewSpaTi} in the full sigma-model~\nref{KinPhot}, which also includes the Hopf fiber coming from the dual photon~$\chi$, we see that all terms with two time derivatives in \nref{KinPhot} are multiplied by~${e^4 \over B} \ll 1$. The upshot is that we should keep the leading kinetic term~$\sim (\d_t \chi)^2$ of the dual photon, but that the terms~$\sim \d_t \chi \d_t \varphi, (\d_t \varphi)^2$ are subleading and can be dropped relative to the spatial gradient terms.\footnote{~A related comment is that we can set the energy to zero in intermediate photon lines. This is reminiscent of the discussion below~\nref{VerFor}, where we explained why the energy of the exchanged photon vanishes.} This can be seen directly from~\eqref{KinPhot}, because the leading equation of motion~$\d_t \varphi \sim \vec \grad^2 \varphi$ shows that the terms~$\sim \d_t \chi \d_t \varphi, (\d_t \varphi)^2$ are effectively higher-order in the gradient expansion.

\item Let us consider adding small mass terms of the form~\eqref{masses},
\begin{equation}
    \Delta S_\text{masses} = \int d^x \, \left( m \CO + m_A \CO^A\right)~.
\end{equation}
It follows from~$\CO^A \sim n^A$ in~\eqref{OAtonA} that~$n^A$ is pinned to be parallel with~$m_A$, giving the~$\C\P^1$ NGBs a mass-squared~$\sim |m_A|$, and leaving only the (dual) photon massless. By contrast, the singlet mass~$m$ (which breaks~$\CC\CT$ but preserves all continuous symmetries) leaves all NGBs massless. Since~$\CO \sim j^0 \sim j^0_\text{top}$, it follows from~\eqref{CoupFi} that the leading~$m$-deformation is a total derivative, consistent with the discussion below~\eqref{massH}. 

\item Results very similar to ours were obtained in the context of quantum hall ferromagnets~\cite{sondhi1993skyrmions}, where 2+1 dimensional fermions interact with a 3+1 dimensional Coulomb potential~$v(r) \sim 1/r$. (See  appendix~\ref{FMQHE} for slightly more detail.) 
 
An important difference is that our purely 2+1 dimensional problem involves the dual photon~$\chi$, which couples to the~$\C\P^1$ degrees of freedom at leading order in the (spatial) derivatives. This deforms the target space of the sigma model to an~$\t S^3$, which does not possess any two-cycles or Skyrmions.

By contrast, $\C\P^1$ Skyrmions play an important role in quantum hall ferromagnets~\cite{sondhi1993skyrmions}, because the 3+1 dimensional Coulomb interaction is subleading in the derivative expansion: the Coulomb energy of a large Skyrmion of size~$L$ scales as~$1 / L$, compared to its~$O(1)$ gradient energy coming from the spatial terms in the~$\C\P^1$ sigma model~\eqref{SMG}. 

\end{itemize}

\subsection{Comparing with the Weak Magnetic Field Regime~$B \ll e^4 $. } \label{weakfield}

In this subsection,  we compare the preceding discussion, valid for strong magnetic field~$B\gg e^4$,  with the weak-field case $B\ll e^4$. In this regime, the gauge interactions dominate over the effects of the magnetic field and must be treated first. This is a strongly-coupled problem that cannot be solved exactly, and the fate of QED$_3$ with $N_f=2$ flavors in the deep IR is a hotly debated topic (see~\cite{Dumitrescu:2024jko} for a recent summary with references). 

Here we will focus on the picture put forward in~\cite{Chester:2024waw, Dumitrescu:2024jko}. These papers proposed that the theory spontaneously breaks~$U(2) \to U(1)_\text{unbroken}$ via the condensation of the minimal~$N =1$, $SU(2)_f$ doublet monopole operator~$\CM^i$ discussed above~\eqref{nfluxjm}. In the deep IR, this leads to a fully relativistic sigma model whose target space is a squashed~$\t S^3$,   
\be \la{RelDum}
S_{\text{relativistic } \t S^3} =  -e^2  \int d^3 x \, \left[  c_1 (\partial_\mu  \chi + \cos \theta \partial_\mu \varphi)^2 + c_2\Big( (\partial_\mu \theta)^2 + \sin^2 \theta ( \partial_\mu \varphi)^2  \Big)  \right] + S_{\theta\text{-term}}~,
\ee 
with~$\theta$-term~\eqref{thetaterm} to match the 't Hooft anomalies. The unknown dimensionless constants~$c_1, c_2 >0$ determine the~$U(2)$ symmetric sigma-model metric. Since the problem contains no small parameters (because~$e^2$ is the only scale), we expect both of them to be~$O(1)$ numbers. As in our discussion above, $\chi \sim \chi + 4 \pi$ is the dual photon, so that~$c_1 = c_2$ corresponds to a round~$S^3$. 

To introduce a small magnetic field~$B \ll e^4$, we use the fact that a constant magnetic field corresponds to a uniform~$U(1)_m$ charge density, as in~\eqref{nfluxjm},\footnote{~This is consistent with~\eqref{jmnonrel}, since~$c_1 = {1 / 32\pi^2}$ in~\eqref{KinPhot}.}
\be \la{MagDens}
 j^0_m  = 4  e^2   c_1 \left( \partial_t \chi_\text{total} + \cos \theta \partial_t \varphi \right) = { B \over 2 \pi }~, \qquad \chi_\text{total} = {B t \over 8 e^2 c_1 \pi} + \chi~.
 \ee 
Substituting~$\chi \to \chi_\text{total}$ in~\nref{RelDum}, and expanding in the fluctuations, we obtain an effective action very similar to~\nref{KinPhot},\footnote{~To obtain this expression, we have dropped some of the subleading terms with two time derivatives, as discussed below~\eqref{NewSpaTi}.} 
\be \la{LoAct}
   S_{\t S^3}=  - { e^2  } \int d^3 x \,  \left[ c_1 (\partial_\mu \chi + \cos \theta \, \partial_\mu  \varphi )^2  + c_2  \Big( (\vec \grad \theta)^2 + \sin^2 \theta (\vec \grad  \varphi)^2 \Big)   \right] + \int d^3 x \, { B \over 4 \pi } \, \cos \theta \, \d_t \varphi~,
    \ee 
where we have dropped the~$\theta$-term. 

This means that we get very similar behavior for $B\ll e^4$ as we got for $B\gg e^4$. In particular, the symmetry-breaking patterns and the low-energy effective actions are identical, with different~$O(1)$ coefficients~$c_{1,2}$ in~\eqref{LoAct}. This prompts us to conjecture that a similar low-energy description is valid for all values of~$B$, with~$c_1, c_2$ replaced by functions of~$e^4/B$ that interpolate between~\nref{LoAct} and \nref{KinPhot}.

\section{Generalization to~$N_f \geq 4$ Flavors} 
\la{LargeNsec}

In this section, we generalize the preceding discussion to even~$N_f \geq 4$. A new feature will be that we can extend our results to all magnetic fields (not just very strong ones) in the large-$N_f$ limit.
  
  \subsection{Strong Magnetic Field~$B \gg (e^2 N_f)^2$} 
  \la{SmallBSec}

   This discussion is very similar to the one above for $N_f=2$.  As we go to 
   energy scales less than $\sqrt{B}$,  we can integrate out the higher Landau levels, and the theory is still weakly coupled at that scale. For energies lower than $\sqrt{B}$, we basically have the same effective theory we had previously, except that there are now~$N_f$ fermion zero-modes~$\psi_q^i~(i = 1, \ldots, N_f)$ and a background charge $- N_f/2$. In other words, the bare Chern Simons term for the $U(1)$ gauge field in~\eqref{LagraNor} is now~$k_\text{bare} = - N_f/2$. This implies that we need to -- on average -- fill $N_f/2$ of the fermions in each Landau orbital~$q$. As before, it turns out that the lowest-energy state is translationally invariant, with exactly~$N_f/2$ filled fermions at every~$q$; moreover they must be the same flavors at every~$q$. 
   
   At the level of Lie algebras (indicated by lowercase letters), the flavor symmetry acting on the fermions is~$su(N_f)$, and the half-filling pattern described above spontaneously breaks $su(N_f) \to u(1)_f \times su(N_f/2) \times su(N_f/2)$.\footnote{~For a complete discussion of the symmetries of QED with general~$N_f$, including all global issues, see~\cite{Benini:2017dus, Cordova:2017kue, Dumitrescu:2024jko}.}  As in the~$N_f = 2$ case, the magnetic~$U(1)_m$ symmetry associated with the massless photon is also spontaneously broken, in a way that is intertwined with the~$su(N_f)$ symmetry; the order parameter that captures the full spontaneous breaking pattern is again the minimal~$N = 1$ monopole operator in the theory with~$N_f$ flavors, which has~$N_f/2$ totally antisymmetrized~$su(N_f)$ indices,\footnote{~\label{fn:diracmasses}One way to see this is to give suitably time-reversal invariant masses to all but~$N_f = 2$ of the fermions, e.g.~we can give~$N_f/2 - 1$ of them mass~$+m$ and~$N_f/2-1$ of them mass~$-m$. This returns us to the~$N_f = 2$ problem, where the~$N = 1$ monopole acquires a time-dependent vev, as discussed below~\eqref{brokencurr}.}   \begin{equation}\label{nfmonopolevev}
       \langle \CM^{[i_1 \ldots i_{N_f/2}]}\rangle \neq 0~.
   \end{equation}
This is precisely the symmetry-breaking pattern considered in section 5 of~\cite{Dumitrescu:2024jko}.  As was shown there, the resulting vacuum manifold is given by
\begin{equation}\label{nfvacua}
  \text{vacuum manifold } \; \CV \; = \;    {U(N_f) \over SU(N_f/2) \times U(N_f/2)}~.
\end{equation}
In this formula, we are using Lie groups (not Lie algebras) and all discrete quotients have been accounted for. As a sanity check, substituting~$N_f =2$ leads to~$U(2)/ U(1) = {\t S^3}$. We will say more about the non-linear sigma model with target space~\eqref{nfvacua} in section~\ref{largeNfaction} below. 

The vacuum manifold~\eqref{nfvacua} is a circle bundle (described by the dual photon) over a base space~$\CG$,\footnote{~Up to discrete quotients, the base~$\CG$ of the fibration coincides with the Grassmannian
\begin{equation}\label{grass}
    Gr(N_f, N_f/2) = {U(N_f) \over   U(N_f/2) \times U(N_f/2)}~.
\end{equation}}
\begin{equation}\label{s1fib}
    S^1 \to \CV \to \CG~.
\end{equation}
The NGBs described by the base~$\CG$ generalize the~$\C\P^1$ excitations of the~$N_f = 2$ case we found in section~\ref{excitations}. One difference is that these excitations now transform under the unbroken~$u(1)_f \times su(N_f/2) \times su(N_f/2)$ symmetry in the   $(\Box , \bar \Box)_{1}$ representation. Their dispersion relation is identical to~\nref{GenExp} and~\nref{DispRel}, with no $N_f$ dependence. In particular, at small momenta we find~\eqref{Smallk},
\begin{equation}\label{smalkgenNf}
    \omega(k) = \alpha k^2~, \qquad \alpha = {e^2 \over 8 \pi B}~, \qquad k^2 \ll B~.
\end{equation}

  \subsection{Interlude: Large $N_f$ Without a Magnetic Field} 
  \la{ZMF}

 \begin{figure}[t]
    \begin{center}
    \includegraphics[scale=.5]{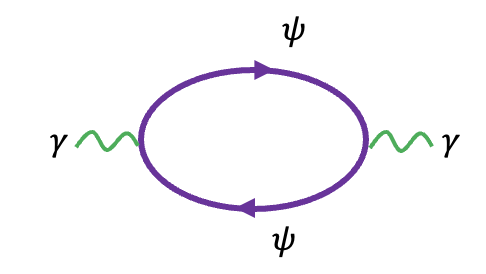}
    \end{center}
    \caption{One-loop fermion bubble correcting the photon propagator. }
    \label{1Loop}
\end{figure}

  The theory for vanishing magnetic field is solvable for large $N_f$. Let us recall some of its features: at leading order in~$N_f$, the only relevant one-loop diagram is the fermion bubble shown in figure~\ref{1Loop}. In Euclidean signature, it leads to the following one-loop vacuum polarization (see appendix~\ref{OneLoopDiag} for more details), 
  \be \la{OLDI}
  \Pi_{\mu \nu} (p) = - N_f \int { d^3 k \over (2\pi)^3} \,    \text{Tr} \left[ \gamma^\mu { 1 \over \not \! k  } \gamma^\nu { 1 \over \not \! k + \not \! p } \right] =-  { N_f \over 16} (p^2 \delta_{\mu \nu } - p_\mu p_\nu ) { 1 \over |p | }~.  
  \ee 
This means that the full photon kinetic term effectively has the following momentum-dependent gauge coupling,  
  \be 
{1 \over e^2_\text{eff}(|p|)} \equiv { 1 \over   e^2 } + { N_f \over 16 } { 1 \over |p | } 
  \ee 
  We see that the second term dominates at low energies. It gives a scale-invariant propagator for the vector field with an effective coupling of order $1/N_f$. Thus the theory remains weakly-coupled at large $N_f$, and we can reliably conclude that it flows to a CFT~\cite{Appelquist:1988sr} at energies smaller than~$e^2 N_f$.

 \subsection{Large $N_f$ with any Magnetic Field} 
 
 For large $N_f$ and any magnetic field~$B$, the theory remains solvable via a combination of large-$N_f$ techniques and an analysis similar to the one we performed above in the strong-field regime. 

 \subsubsection{Fermion Loop Corrections to the Photon Effective Action}
 
 One important element in this discussion is a computation of the correction to the effective action for the photon due to the fermion one loop diagram shown in figure \ref{1Loop}, but now in the presence of a magnetic field. Of course we expect to get~\nref{OLDI} for large momenta~$p \gg \sqrt{B}$.  For very small momenta $p\ll \sqrt{B}$ we would expect to receive a contribution only from the zero-modes in the lowest Landau level. However, the one-loop diagram with fermions exclusively in the lowest Landau level actually vanishes,  because its only non-vanishing component in Euclidean signature is proportional to a sum of terms of the form\footnote{~Here we use the Euclidean zero-mode propagator~\eqref{fermpropE}, and~$\omega$, $\omega'$ are Euclidean momenta in the~$\tau = x^3$ direction.}
   \be \la{OneLLLL}
 \Pi_{33}(\vec p, \omega) \sim  \sum_{i = 1}^{N_f} \int_{-\infty}^\infty d \omega' \, { 1 \over (i \omega  + i \omega' - \mu_i )}  \,  { 1 \over ( i \omega' - \mu_i ) }  =0 ~.
    \ee 
This vanishes separately for each flavor, because both poles for $\omega'$ are on the same side of the complex plane. Here we have assumed that the~$N_f$ fermion flavors get different effective masses~$\mu_i$ that we should sum over. For now these masses can just viewed as IR regulators (we are after all starting with the massless theory in the UV), but we already seen in the~$N_f = 2$ case that the residual Coulomb interactions do give rise to such a mass, as we discussed around \nref{EffMass}. Since these masses must respect the unbroken symmetries, they are effectively Dirac masses aligned with the monopole vev~\eqref{nfmonopolevev}. For instance, if~$\CM^{1 \cdots N_f/2}$ gets a vev, then the first~$N_f/2$ flavors get the same positive effective mass~$\mu_1 = \cdots = \mu_{N_f/2} = \mu > 0$, and the remaining~$N_f/2$ flavors the same negative mass~$\mu_{N_f/2+1} = \cdots = \mu_{N_f} = -\mu$. (see section~5 of~\cite{Dumitrescu:2024jko} for more detail). 

Of course the full one-loop diagram in figure \ref{1Loop} is non-zero at low momenta, but since only the higher Landau levels contribute, they give rise to a local term in the photon effective action, which scales as $N_f/\sqrt{B}$. There is a simple way to determine this contribution without evaluating the full diagram -- as we explain in appendix \ref{App1Loop} -- using the conformal symmetry of the problem that is spontaneously broken by the magnetic field. 
     The final result can be summarized by the following extra term in the effective action for the photon, 
   \be \la{EffCou}
    S_{\rm photon, \, one-loop} =   { \tilde c N_f \over \sqrt{B} } \int d^3 x \,  { 1 \over 2  } \,  \left ( f_{0a}^2 - \half f_{xy}^2 \right) ~,~~~~~
    {\rm with } 
    ~~~~\tilde c \equiv  { 3 \zeta( 3/2) \over 8  \pi^2 \sqrt{2}   } ~.
       \ee
    Here $f_{\mu\nu}$ parametrizes the fluctuations of the photon around the background magnetic field~$B$. To obtain~\eqref{EffCou} one first computes the energy for a constant background $B$ and then suitably replaces $B \to B +f$ to extract \nref{EffCou}; see appendix \ref{App1Loop} for details, and \cite{Herzog:2025ddq} for a recent discussion with further references.  
 
 The full effective kinetic term for the photon is obtained by adding~\nref{EffCou} to the tree-level kinetic term coming from the classical action~\eqref{LagraNor},
 \be \la{FullKin}
 S_\text{eff, photon} = \half \int d^3 x \, \left[    { 1 \over e^2} ( f_{0a}^2 - f_{xy}^2 )  +  { \tilde c N_f \over \sqrt{B} } \left ( f_{0i}^2 - \half f_{xy}^2 \right)
 \right]~.
 \ee 
 Let us make some comments about this result: 
 \begin{itemize} 
 \item 
 	When $\sqrt{B} \ll e^2 N_f$, the induced kinetic term \nref{EffCou} dominates over the original tree-level kinetic term. This means that we can set $e^2=\infty$, so that the discussion applies to the addition of a magnetic field to the low energy CFT. 
  \item 
 	 The relative factor of two between the $f_{0a}^2$ and $f_{xy}^2$ in the induced kinetic term~ \nref{EffCou} is the usual one expected for a relativistic conformal (super)fluid in 2+1 dimensions. In other words, the magnetic field represents a large background $U(1)_m$ charge and the conformal symmetry constrains the form of the effective action for the fluctuations~\cite{Hellerman:2015nra}, see appendix \ref{App1Loop}. 
 	 	\item 
 	 At large $N_f$, the overall coefficient of the action \nref{FullKin} is always large at the scale $\sqrt{B}$. This means that,  after integrating out the higher Landau levels, we end up with a theory that contains fermions only in the lowest Landau level, and these fermions are weakly coupled at scale $\sqrt{B}$.   	
 	  \item 
 	  While \nref{EffCou} gives the leading effective action for low-momentum photons, we saw in section~\ref{qed3sec} that photons of momenta~$\lesssim \sqrt{B}$ play a role in determining the symmetry-breaking vacuum, and the low-energy effective theory of the Nambu-Goldstone Bosons. We must thus analyze the one-loop diagram in figure~\ref{1Loop} in more detail, which we will do presently.   
 	  \end{itemize}

 \subsubsection{The Effective Coulomb Potential, Symmetry-Breaking, and Excitations} \label{effpot}

 In appendix~\ref{TevenLoop} we determine the full effective action for~$a_0$ at coincident times, but arbitrary spatial momenta~$\vec p$, which is sufficient to determine the instantaneous Coulomb interaction between the zero modes that generalizes~\eqref{VerFor}. In Lorentzian signature,
 \begin{equation}\label{statica0}
     S_{O(a_0^2), \text{ static}} = \half \int d t \, \int {d^2 p \over (2 \pi)^2} \, a_0(-\vec p, t) \, |\vec p|^2 \left({1 \over e^2} + {N_f \over \sqrt{B}}  \Pi\left(\t p \right)\right) \, a_0(\vec p, t)~, \quad \t p \equiv {|\vec p| \over \sqrt{B}}~.
 \end{equation}
Here~$\Pi(\t p)$ is defined as the static limit of a particular component of the Euclidean vacuum polarization tensor~\eqref{OLDI}, which is computed by the fermion one-loop diagram in figure~\ref{1Loop}, but now in a non-zero magnetic field,
\begin{equation}\label{Pidefmain}
    \Pi_{33}(\vec p, p_3 = 0) \equiv - |\vec p|^2 \, {N_f \over \sqrt{B}} \, \Pi(\t p )~, \qquad \t p = {|\vec p| \over \sqrt{B}}~.
\end{equation}
See appendix~\ref{TevenLoop} for more details on the definition and computation of~\eqref{Pidefmain}. The final integral representation for~$\Pi(\t p)$ appears in~\eqref{Piofqfinal}.\footnote{~In appendix~\ref{TevenLoop} we set~$B = 1$, so that~$\t p = |\vec p|$ there.} The numerical result is plotted in figure~\ref{LoopDiaFig} above. Let us examine this function in two limits:

  \begin{figure}[t]
\begin{center}
\includegraphics[scale=0.4]{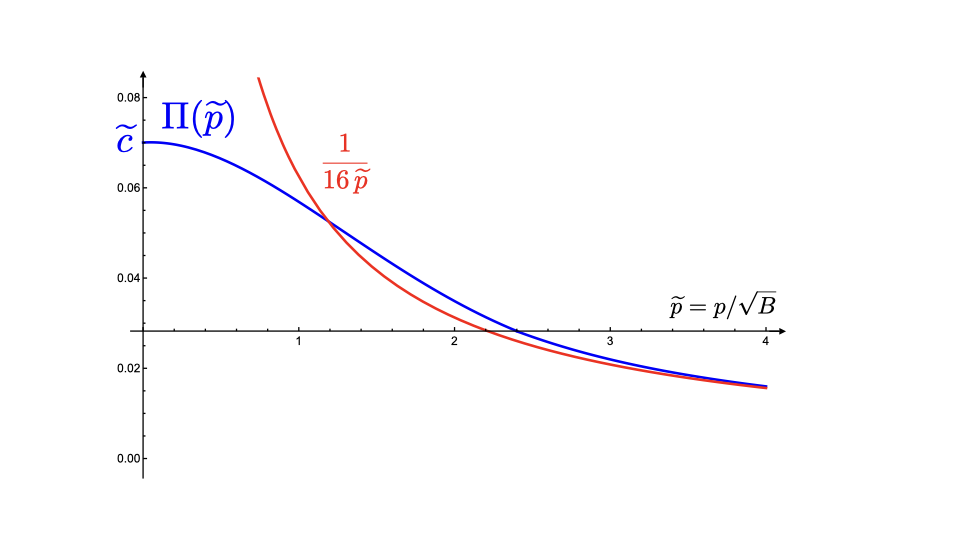} 
\caption{The function~$\Pi(\t p)$ defined in~\eqref{Pidefmain} (see also appendix~\ref{TevenLoop}) as a function of~$\t p = p/\sqrt{B}$ is shown in blue. The long-distance value~$\Pi(0) = \t c = {3 \zeta(3/2) / 8 \pi^2 \sqrt{2}} \simeq 0.070$ is shown on the vertical axis; the short-distance asymptotics~$\Pi(\t p \to \infty) = {1 / 16 \t p}$ are shown in red. }
\label{LoopDiaFig}
\end{center}
\end{figure}

\begin{itemize}
    \item At long distances, we find that
 \be \la{LowP}
 \Pi(\tilde p \to 0) = { \tilde c } = {3 \zeta(3/2) \over 8 \pi^2 \sqrt{2}}~,
 \ee 
  in accordance with~\nref{EffCou}. 

  \item At short distances, we find that
  \begin{equation}
      \Pi(\t p \to \infty) = {1 \over 16 \t p}~,
  \end{equation}
  in accord with~\nref{OLDI}. 
\end{itemize}

We can now determine the effective Coulomb potential in momentum space, which is simply given by static~$a_0$ propagator obtained by inverting~\eqref{statica0}, 
\begin{equation}\label{effCoul}
\t v_{\text{eff.} \, C}(\t p) = {1 \over \displaystyle B \, {\t p}^{\,2} \left({1 \over e^2} + {N_f \over \sqrt{B}} \, \Pi(\t p)\right)}~.
\end{equation}
Since this is positive for all momenta, $\t v_{\text{eff.} \, C}(\t p) > 0$, it follows from the discussion below~\eqref{Hvgen} that the vacua are identical to the ones we found in section~\ref{SmallBSec}, where we only considered large~$B$-fields and tree-level Coulomb exchange. Similarly, the quantum numbers of the excitations are also the same.

However, the dispersion relation~$\omega(k)$ of these excitations {\it is} affected by the modification of the Coulomb potential in~\eqref{effCoul}, and we must substitute~$\t v_{\text{eff.} \, C}(\t p)$ into~\nref{DispReloth}, which applies for a general potential. In particular, we find in the small-$k$ limit~\eqref{GolDi} that 
 \be \la{alphaGen}
 \omega(k\to 0) = \alpha k^{2} ~, \qquad \alpha = { 1 \over 8 \pi \sqrt{B} N_f } \int_0^\infty \t p \, d \t p \;\;    \frac{ \displaystyle e^{ -   \tilde p^{\,2}/2}   }{  \displaystyle \left( \frac{ \sqrt{B} }{e^2 N_f } +  \Pi(\tilde p)     \right) }~.  
    \ee 
This formula is valid for any value of $e^2 N_f/\sqrt{B}$. In the strong-field limit~$\sqrt{B} \gg e^2 N_f$, it reduces to~\eqref{smalkgenNf}.  By contrast, in the weak-field limit~$\sqrt{B} \ll e^2 N_f$ we have a magnetic field in the conformal regime of the theory, and the coefficient~$\alpha$ evaluates to
    \be \la{ALNf}
    \alpha = { 1 \over 8 \pi \sqrt{B} N_f } \int_0^\infty \t p \, d \t p \; \;   { e^{ -   \tilde p^{\,2}/2}   \over   \Pi(\tilde p) } \simeq {0.85 \pm 0.01 \over \sqrt{B} N_f }~,
    \ee 
    where we have done the integral numerically (with some estimate of the error) using the function~$\Pi(\t p)$ in equation~\eqref{Piofqfinal} of appendix \ref{OneLoopDiag}. Note that, in the conformal regime, the only scale in the problem is $\sqrt{B}$, and that is what sets the units of~$\alpha$ in~\eqref{ALNf}. Moreover, $\alpha$ is~$1/N_f$ suppressed. 
     
   \subsubsection{The Low-Energy Effective Action for the Nambu-Goldstone Bosons} \label{largeNfaction}

   We would like to write down the analogue of the effective action~\eqref{KinPhot} for general~$N_f$. To do this, we again use the dual photon~$\chi \sim \chi + 4 \pi$ to describe the~$S^1$ fiber of the vacuum manifold~$\CV$ in~\eqref{s1fib}, and coordinates~$\theta^\alpha$ on the base~$\CG$,\footnote{~Here~$\alpha = 1, \ldots, \text{ dim } \CG = \half N_f^2$.}  
\be \la{InterC}
\begin{split}
   S_{\CV}=~ & { 1 \over 32 \pi^2 } \int d^3 x \,    \left( { \displaystyle e^2 ( \partial_t \chi- 2 {\cal A}_{\alpha } \d_t \theta^\alpha )^2 \over \displaystyle  1 + { \tilde c N_f e^2 \over 2 \sqrt{B} } }    -  { \displaystyle  e^2 ( \vec \grad \chi - 2 {\cal A}_{\alpha } \vec \grad   \theta^\alpha )^2 \over \displaystyle  1 + { \tilde c N_f e^2 \over   \sqrt{B} } }    - 4\pi \alpha B \, g_{\alpha\beta}(\theta) \vec \grad \theta^\alpha \cdot \vec \grad  \theta^\beta  \right) \\
   & - \int {d^3 x} \, {B \over 2 \pi} \CA_\alpha \d_t \theta^\alpha + S_\text{top}~.
\end{split}
\ee 
In more detail:
\begin{itemize}
\item $\CA$ is a~$U(1)$ connection on the base space~$\CG$ of the circle bundle~\eqref{s1fib}, whose field strength determines the first Chern class of the bundle, 
\begin{equation}
    c_1(\CV) = \left[{d \CA \over 2\pi}\right]~.
\end{equation}
Moreover, $\CA$ is invariant under the~$SU(N_f)$ symmetry acting on~$\CG$, up to gauge transformations. This generalizes~\eqref{ConFor} to arbitrary~$N_f$.  Note that~$\CA$ appears in the covariant derivative of the dual photon~$\chi$ with a factor of~$2$ because~$\chi \sim \chi + 4\pi$, and that the first-order term involving~$\CA$ takes the same form as in~\eqref{SMG}.  

\item The terms involving the dual photon are no longer Lorentz invariant, and they depend on the two couplings in the photon kinetic term~\nref{FullKin}. In the small-$B$ regime appropriate to the CFT, they take the form of a conformal superfluid, as discussed in~\cite{Hellerman:2015nra}, which is fibered over the base~$\CG$. 

\item The spatial gradients along the base~$\CG$ involve the coefficient $\alpha$ computed in \nref{alphaGen}, which depends on~$e^2$, $B$, and~$N_f$. Recall that~$\alpha$ is~$\sim e^2 /B$ in the strong-field regime, and~$\sim 1 / N_f \sqrt{B}$ in the conformal weak-field regime. Thus, the entire first line of~\eqref{InterC} is~$O(e^2)$ and~$O(\sqrt{B}/N_f)$ in these two regimes, respectively. By contrast, the first-order term on the second line is~$O(B)$ in both regimes. 

\item The metric~$g_{\alpha\beta}(\theta)$ that appears in the spatial gradient terms is the~$SU(N_f)$ invariant metric on the base~$\CG$, which is locally identical to the homogeneous metric on the Grassmannian~\eqref{grass}. It is therefore fixed by the symmetries, up to an overall scale. 

We fix this scale in the following way: the base space~$\CG$ has an embedded~$\C\P^1$ sub-manifold, and restricting the sigma model maps~$\theta^\alpha(x)$ to that~$\C\P^1$ reduces the symmetry-breaking pattern and the massless NGBs (though not necessarily the coefficients in the effective action) to the~$N_f = 2$ case considered previously.\footnote{~This can be justified by giving suitably time-reversal invariant masses to all but~$N_f = 2$ of the fermions, e.g.~we can give~$N_f/2 - 1$ of them mass~$+m$ and~$N_f/2-1$ of them mass~$-m$.} Note, however, that the coefficients do agree with~\eqref{KinPhot} if we take~$B$ to be very large, in which case~$\alpha = {e^2 \over 8 \pi B}$ and the radius of the~$\C\P^1$ base is half the radius of the fiber. 

\item Finally, there must be terms -- collectively referred to as~$S_\text{top}$ in~\eqref{InterC} -- that are associated with the topology of the manifold~$\CV$ and whose job it is to match the~'t~Hooft anomalies of the theory. These could be generalizations of the~$\theta$-term~\eqref{thetaterm}, which are constructed using classes in~$H^3(\CV, \Z)$, but there could also be Wess-Zumino terms associated with~$H^4(\CV, \Z)$, which are necessarily absent in the~$N_f = 2$ case.\footnote{~This would naturally reflect the fact that the 't Hooft anomalies of QED for~$N_f \geq 4$, which were determined in~\cite{Benini:2017dus, Hsin:2024abm}, contain terms that are absent for~$N_f = 2$.} We leave the detailed investigation of these terms to the future. 

\end{itemize}

  \subsection{Application to Monopole Operators } \label{monOps}
  
  In this section,  we point out the connection between the preceding discussion and the study of monopole operators in the conformal field theory that arises in the IR of QED$_3$ with large $N_f$. In a CFT, the spectrum of conformal dimensions is the same as the spectrum of states on $S^2 \times (\text{time})$. In the discussion below, we set the radius of the~$S^2$ to unity. Following~\cite{Borokhov:2002ib}, operators with~$U(1)_m$ charge~$N$ are states with magnetic flux~$2 \pi N$ on~$S^2$.  Such operators were discussed, for example,  in  \cite{Pufu:2013vpa,Dyer:2013fja,Chester:2017vdh,Dupuis:2021flq}. 
   
  The discussion in this paper applies mostly directly to the case of operators with large magnetic charge~$N \gg 1$, for which there is no significant difference between the spatial~$S^2$ and a plane. The leading contribution to the scaling dimension of these operators is given by the Casimir-like vacuum energy in a magnetic field computed explicitly in appendix~\ref{App1Loop},
  \be 
  \Delta = 4\pi \epsilon = 4 \pi \gamma B^{3/2} N_f =    
 {\zeta( {3\over 2} ) \over 4 \pi } N^{3/2} N_f ~,~~~~~~~N , N_f \gg 1~, 
 \ee 
 where we have used the fact that the flux on a unit sphere is~$ 4\pi B = 2 \pi N$. Results for finite values of the magnetic flux $N$ were obtained in~\cite{Pufu:2013vpa,Dyer:2013fja,Chester:2017vdh,Dupuis:2021flq}. General aspects of the large-$N$ charge expansion were discussed in~\cite{Hellerman:2015nra}, but here we have some new features due to the interplay of the magnetic and the flavor symmetries. 
 
 In the leading large-$N_f$ approximation,  all the states in the lowest Landau level are degenerate. Our discussion enables us to characterize the further splittings which were indirectly seen in section 3.1.3 of \cite{Chester:2017vdh}.
  In infinite volume, we found spontaneous symmetry breaking. In finite volume, the symmetry is restored, but the analogous statement is that the ground states form a particular representation of the $SU(N_f)$ flavor symmetry.

  \begin{figure}[t] 
     \begin{center}
    \includegraphics[scale=.3]{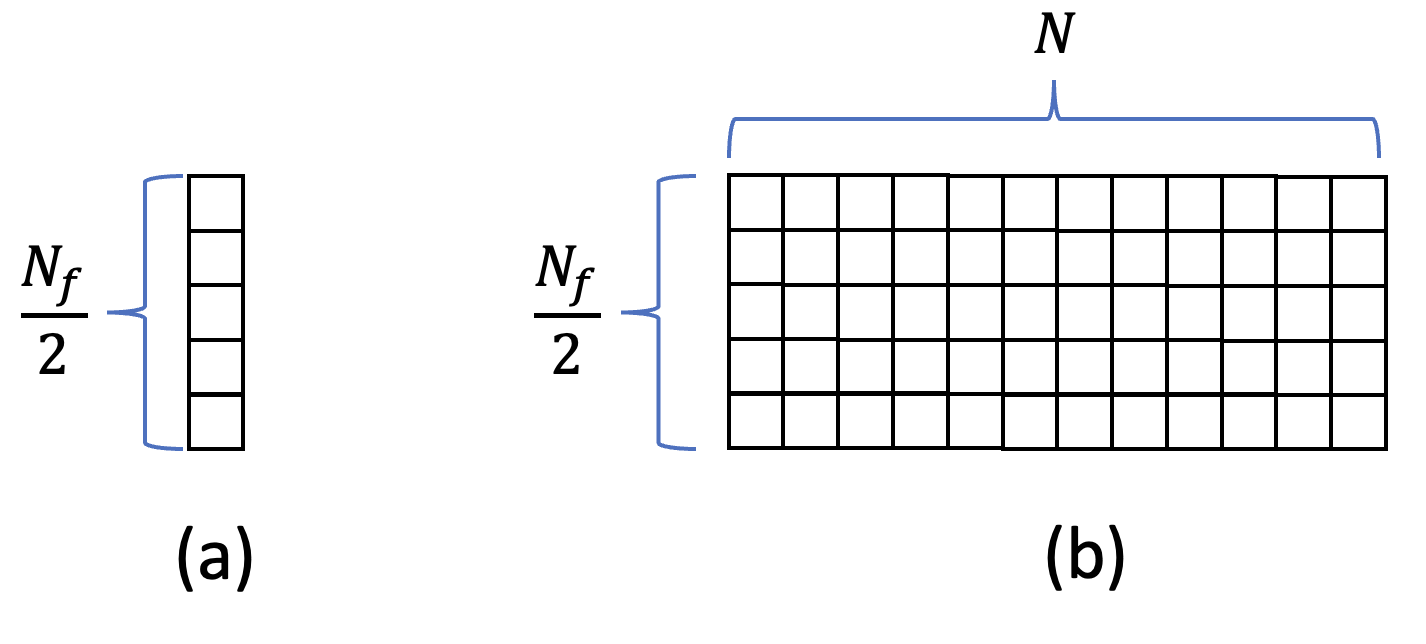}
    \end{center}
    \caption{ Young diagrams for various~$SU(N_f)$ representations: (a) totally antisymmetric representation describing a single orbital with $N_f/2$ fermions; (b) representation describing the lowest energy state on~$S^2$ after we combine all $N $ orbitals. }
    \label{Young}
\end{figure}
  
This representation can be described as follows: 
  for each orbital, we are filling  $N_f/2$ fermions, leading to an antisymmetric combination of $N_f/2$ fundamentals. This has a Young diagram that is a vertical column of $N_f/2$  boxes, see figure \ref{Young}a. We now have  $N$ orbitals, where $N$ is the quantized magnetic flux.  We then expect a symmetric combination of all the fermions for each orbital. This gives a Young diagram that contains $N_f/2$ rows, each of length $N$, as show in figure \ref{Young}b. This is the representation of the monopole operator. The same conclusion was reached in~\cite{Dyer:2013fja}, where the condition of vanishing charge density (our~\eqref{CondOk}) was also imposed. 
  
  To see this in slightly more detail, note that a single orbital with label $m$  leads to a state transforming in the totally anti-symmetric~$SU(N_f)$ representation in figure~\ref{Young}a,
  \be \label{oneorbital}
   \prod_{i=1}^{N_f/2} \left(\psi^{i}_m\right)^\dagger |0 \rangle~,
   \ee 
    up to~$SU(N_f)$ transformations. A generalization of the discussion around~\eqref{CondVa} shows that the ground state that minimizes the Coulomb Hamiltonian~\eqref{VerForPos} (and thus has vanishing charge-density~\eqref{CondOk}) is given by a copy of~\eqref{oneorbital} for every orbital, 
   \be \la{Cus}
   \prod_{m= - j}^{ j} \prod_{i=1}^{N_f/2} \psi^{i\dagger}_m |0 \rangle ~,~~~~~~~~2 j +1 = N~.
   \ee 
   This state is unique, up to $SU(N_f)$ transformations, and thus the ground states form an irreducible $SU(N_f)$ representation -- precisely the one whose Young diagram appears in figure~\ref{Young}b. Note that the state  \nref{Cus} has vanishing~$SU(2)_R$ angular momentum on the sphere, i.e.~it has spacetime spin~$j  = 0$, because it can be reordered so that each flavor contributes a manifestly anti-symmetric combination~$\prod_{m = -j}^j \psi_m^i$, which is therefore a singlet under the~$SU(N) \supset SU(2)_R$ symmetry under which the~$\psi_m^i$ transform in the~$N$-dimensional fundamental representation.

One could further discuss excited states on top of the lightest monopole. The simplest ones correspond to adding a fermion-hole pair, i.e.~they are the Nambu-Goldstone bosons whose effective action in flat space is~\eqref{InterC}. In principle, the spectrum of these excitations can be worked out on the sphere, and even for finite~$N$, using methods similar to the ones in \cite{NAKAJIMA1994327}.  Of course, in the large-$N$ limit we can use the flat-space formulas, as explicitly checked for the case in \cite{NAKAJIMA1994327}.  In particular,  for the lowest-energy excitations we can use~\nref{alphaGen} with $k^2 \to \ell(\ell+1)$ and $e^2 \to \infty$, where $\ell$ is the~$SU(2)_R$ angular momentum on the sphere.

\section{Discussion} 

In this paper, we have discussed the dynamics of 2+1 dimensional QED with massless fermions in the presence of a constant and uniform magnetic field. We considered only the time-reversal symmetric theory. 

For large magnetic fields, $B\gg (e^2 N_f)^2$, at leading order in the $e^4/B$ expansion,  the theory has a large set of low energy states, with a degeneracy that is exponential in the area. At first sight, one might therefore expect a rather complicated problem in degenerate perturbation theory at the next order. However, the physics is rather simple due to the form of the leading interaction, which can be written as a sum of squares \nref{VerFor}. This then leads to a set of equations~\nref{CondOk} for the vacuum. Furthermore, these equations can all be solved with a simple choice of vacuum, where the state in all individual orbitals is the same. 

In the $N_f =2$ case, this is just a state where all~$SU(2)$ flavor spins are pointing in the same direction, just as in an ordinary ferromagnet, for the same reasons originally articulated in the early days of quantum mechanics \cite{Heisenberg}. This ground state is very simple, a product state with a trivial entanglement structure. The simplicity of the leading interaction also permits an analysis of the spectrum, where we have magnons that are reminiscent of the ones for the ordinary Heisenberg ferromagnet. 
For the same reasons, the low-energy theory is somewhat similar to the one encountered in ferromagnets. The new feature is that we have another massless field, the dual photon.  

It is curious that in the $N_f =2$ case, we also get a similar description for small magnetic fields, given the conjectured low-energy theory in \cite{Chester:2024waw, Dumitrescu:2024jko}. 
It is natural to conjecture that, for any non-zero magnetic field, the low-energy dynamics is continuously connected. Of course, the coefficients of the low-energy theory could depend on $e^2/\sqrt{B}$. 

For large $N_f$, in the absence of a magnetic field, we have a CFT at low energies.  We can consider a small magnetic field that lies within the conformal regime. This field breaks the conformal symmetry and leads to a vacuum structure that is similar to that encountered for any $N_f$ and large magnetic fields. At large~$N_f$, we can compute the coefficients of the low-energy theory for any value of $e^2N_f/\sqrt{B}$.  

Throughout, we have described the low-energy dynamics in terms of a sigma model. It is tempting to wonder whether, at higher energies, we might have a non-commutative version, since non-commutative theories have made an appearance in quantum Hall physics. 

Some of our discussion would continue to hold in the presence of a four-fermion interaction, such as in the Nambu-Jona-Lasinio or Gross-Neveu models.\footnote{~Since these interactions are not renormalizable, we can imagine inducing them by integrating out a massive boson with a suitable Yukawa coupling.} In these cases, there is no dynamical photon, and the magnetic field would be a purely external background, but symmetry breaking can nevertheless be triggered by a similar mechanism \cite{Gusynin:1994re,Semenoff:1999xv}.

A natural generalization of the problem studied above would involve considering theories with Chern-Simons terms. In this case, we expect a somewhat similar picture: the Chern-Simons term dictates how many Landau levels we are filling. When all Landau levels are completely filled or empty, we do not expect any flavor-symmetry breaking; when they are partially filled, we do expect some symmetry breaking because the Coulomb repulsion prefers filling the last Landau level using fermions with identical quantum numbers. 

Another interesting problem is the case of QCD$_3$ in a constant magnetic field for baryon-number symmetry. In this case, we expect a similar pattern of symmetry breaking, but now the theory becomes strongly coupled and confining at low energies. Nevertheless, the phenomenon of symmetry breaking may be calculable if it happens at sufficiently high scales.

Finally, it would be interesting to better understand the magnetic catalysis phenomenon in 3+1 dimensions.

\bigskip

\subsection*{Acknowledgments}

\noindent  We are grateful to S.~Pufu and S.~Sondhi for discussions and for pointing out several references. We also thank I.~Klebanov, N. Seiberg, and D.~Son for discussions, and~S.~Pufu for comments on the manuscript. The work of TD was supported in part by U.S. Department of Energy award DE-SC0025534, as well as the Simons Collaboration on Global Categorical Symmetries. The work of JM was supported in part by U.S. Department of Energy grant DE-SC0009988.

\appendix

\section{Brief Review of Quantum Hall Ferromagnets}
 \la{FMQHE} 
  
 In \cite{PhysRevB.30.5655,sondhi1993skyrmions,girvin1999quantumhalleffectnovel}, the problem of electrons in a magnetic field, confined to a two dimensional plane,  at filling fraction $\nu=1$ was considered. In addition, these papers considered the limit where the coupling to the magnetic dipole of the electron vanishes (or can be neglected). In this problem, each Landau level orbital can be occupied by no electron, a spin up or down electron, or both. The $SU(2)$ rotation symmetry is playing a role similar to our $SU(2)_f$ flavor symmetry. We briefly review here some aspects of their discussion with the goal of emphasizing where some of the differences arise.

   One difference is that the Coulomb potential is 
  \be 
   v(r) = { e^2 \over 4 \pi } { 1 \over r }  \quad \Longleftrightarrow \quad 
   \t v (\vec p) = { e^2 } \int_{-\infty}^{\infty} { d p_3 \over 2 \pi }  { 1\over {\vec p}^{\, 2} + p_3^2 } = { e^2 \over 2 | \vec p \,|}~, 
   \ee 
  in position or Fourier space.
  With this minor difference, the dispersion relation takes the form obtained in~\cite{Bychkov1981,PhysRevB.30.5655},\footnote{~The difference with equation (4.12) of \cite{PhysRevB.30.5655} is that ${e^2_{\rm here} \over 4 \pi } = e^2_{\rm there}$. }
  \be 
  \omega(k)  = { e^2 \over 4 \pi }{ B \over k  } \int_0^\infty dy [ 1- J_0(y) ] e^{- y^2 B/(2k^2) } 
  = { e^2 \sqrt{B} \over 4 \sqrt{ 2 \pi}  } [ 1 - e^{ - \nu } I_0(\nu) ]~, \qquad \nu \equiv   { k^2 \over 4 B } 
  \ee 
  This behaves as $\omega \sim  k^2/\sqrt{B}$ at small $k^2$,  displaying the Nambu-Goldstone bosons. This leads to a sigma model description which has soliton (or Skyrmion) solutions \cite{sondhi1993skyrmions, girvin1999quantumhalleffectnovel}. 
  
  We can find these solutions easily as follows. If we focus on time-independent solutions, we see that the action is conformally invariant. So we are interested in maps from compactified space, say $S^2$, to the $S^2$ target space. One simple solution (with one unit of topological charge) would be the identity map between the two spheres. We can generate a full family of solutions by acting with $SL(2,\C)$ transformations of the sphere. Explictly, 
  \be 
  \cos \theta = { r^2 - |\lambda|^2 \over r^2 + |\lambda|^2 } ~, \qquad \sin \theta e^{i \varphi } = { 2 \lambda (x + i y)  \over r^2 + |\lambda|^2 }  ~,~~~~~r^2 \equiv x^2 + y^2~, 
  \ee 
  where  $\lambda $ is complex and $(x,y)$ are coordinates in the transverse plane.  The phase of $\lambda$ controls an orientation inside the target sphere. 
  
  These Skyrmions have a center of mass coordinate $(x_0, y_0)$ and a complex size parameter~$\lambda$. The effective action for the center of mass is first-order in time~\cite{PhysRevB.54.R2331},
  \be \la{KinLa}
   S =   B \int dt  \, x_0 \dot y_0~.   
  \ee 
As discussed in \cite{sondhi1993skyrmions}, the electrostatic energy of a single Skyrmion decreases as $e^2/|\lambda |$ when $|\lambda|  \to \infty$. This means that for large $|\lambda|$ we can neglect the electrostatic forces when we compute the Skyrmion solution.
     As we remarked at the end of section \ref{furthercommentssec},  this is not the case in QED$_3$, where the photon is 2+1 dimensional and there is no limit in which we can approximately discuss a Skyrmion solution.

\section{Quantum Mechanics of a Complex Fermion}  
\label{FermApp}
In this appendix, we will review some aspects of a quantum mechanical two-level system, or qubit, in the free fermion presentation.

\subsection{Lorentzian Action}  

We start in Lorentzian signature with real time $t$ and a single complex fermion $\psi(t)$ with the following action, Hamiltonian, and equal time anti-commutation relations,
\begin{equation}\la{Sferm}
S = \int dt \, \left(i \psi^\dagger \d_t \psi + \mu \psi^\dagger \psi - E_0\right)~, \qquad H = -\mu \psi^\dagger \psi + E_0~, \qquad \big\{\psi^\dagger, \psi\big\} = 1~. 
\end{equation}
Some comments:
\begin{itemize}
\item[(i)] The fermion mass $\mu \in \R$ can have either sign;\footnote{~In quantum mechanics, the fermion mass is typically~$m_\text{QM} = -\mu$. This should not be confused with the 2+1 dimensional Dirac mass~$m$ in~\eqref{mDirac}, which gives rise to quantum mechanical zero modes with~$\mu = +m$.} it can be thought of as a chemical potential for the~$U(1)$ symmetry under which $\psi$ has charge~$-1$ (see appendix~\ref{sec:gsferm} below). As discussed around~\eqref{SzmQM}, a~2+1 dimensional Dirac fermion~\eqref{mDirac} in a magnetic field~$F_{xy}  = B > 0$ gives rise to a Lagrangian of the form~\eqref{Sferm}, with~$\mu = m$, for every zero mode~$\psi_q(t)$ in the lowest Landau level.  
\item[(ii)] The c-number counterterm $E_0 \in \R$ is the vacuum energy; it fixes the operator ordering in the Hamiltonian $H$. 
\end{itemize}

The Hilbert space~$\CH$ is two-dimensional and spanned by the unit-norm states $|0\rangle$ and~$|1\rangle$, which satisfy
\begin{equation}
\psi |0\rangle = 0~, \qquad \psi^\dagger|0\rangle = |1\rangle~, \qquad \psi^\dagger |1 \rangle = 0~, \qquad \psi |1\rangle = |0\rangle~.
\end{equation}
Thus~$|0\rangle$ is the Fock vacuum annihilated by $\psi$, which we will view as a lowering operator. The ordering of the spectrum depends on the sign of~$\mu$: the minimum energy ground state is $|\Omega \rangle = |0\rangle$ when~$\mu < 0$ and~$|\Omega \rangle = |1\rangle$ when~$\mu > 0$. As long as~$\mu \neq 0$, the propagator is
\begin{equation}\la{fermpropL}
\langle \Omega| T \{ \psi(t) \psi^\dagger(0) \} | \Omega \rangle = \int {d \omega \over 2\pi} \, e^{-i \omega t} \, {i \over \omega + \mu - i \, \text{sign}(\mu) \, \ep}~.
\end{equation}
When~$\mu = 0$, there are two degenerate ground states (see appendix~\ref{sec:degqm} below for further details), which is reflected in the fact that the~$\mu \to 0$ limit of~\eqref{fermpropL} is ambiguous. 

\subsection{Global Symmetries} \la{sec:gsferm}

Let us examine the global symmetries of~\eqref{Sferm}: 
\begin{itemize}
\item[1.)] For all~$\mu \in \R$, there is a~$U(1)$ symmetry under which $\psi$ has charge~$-1$ and~$\psi^\dagger$ has charge~$+1$.  The Hermitian charge operator is
\begin{equation}\label{Qdef}
Q \equiv j^0 = \psi^\dagger \psi + k_\text{bare}~, \qquad k_\text{bare} \in \Z~, \qquad [Q, \psi] = - \psi~, \qquad [Q, \psi^\dagger] = \psi^\dagger~.
\end{equation}
Here~$k_\text{bare}$ is a c-number ordering counterterm that determines the~$U(1)$ charge of the Fock vacuum; it must be quantized to ensure~$U(1)$ gauge invariance (see below).

\item[2.)] A unitary charge-conjugation symmetry $\sfC$ that acts as $\psi \to \psi^\dagger$ and $\mu \to - \mu$. Thus~$\sfC$ is only a symmetry when $\mu = 0$, and it does have (mixed) 't Hooft anomalies, as discussed in appendix~\ref{sec:degqm} below. 

\item[3.)] For all~$\mu \in \R$, an anti-unitary time-reversal symmetry $\sfT$, which acts as $\psi(t) \to \psi(-t)$. 
\end{itemize}
Note that $\sfC\sfT$ only differs from the action of $\CC\CT$ on Dirac zero modes in the lowest Landau level~\eqref{CTonzm} by a~$U(1)$ phase rotation. This means that the implications of these two symmetries are the same, as long as the~$U(1)$ symmetry is present. 

The~$U(1)$ symmetry enables us to couple~$\psi$ to a~$U(1)$ background gauge field~$A_0(t)$ via
\be\la{DSferm}
\Delta S[A_0] = \int dt \, A_0 j^0 = \int dt \, A_0 \left(\psi^\dagger \psi +k_\text{bare}\right)~.
\ee
This amounts to replacing~$\d_t \to D_t = \d_t - iA_0$ in~\eqref{Sferm}, and also adding a quantum-mechanical Chern-Simons term with level~$k_\text{bare}$ for the~$U(1)$ background gauge field~$A_0$. The latter specifies a background Wilson line, whose $U(1)$ charge $k_\text{bare} \in \Z$ must be integrally quantized as in~\eqref{Qdef}.  Note that~$A_0$ transforms under~$\sfC$ and~$\sfT$ in the same way as~$\mu$, i.e.~$\sfC: A_0 \to -A_0$ and~$\sfT : A_0(t) \to A_0(-t)$, so that the bare Chern-Simons term in~\eqref{DSferm} always preserves~$\sfT$, but is incompatible with~$\sfC$. 

\subsection{Euclidean Action} 

The continuation to Euclidean signature proceeds via
\be
t = - i \tau~, \qquad A_0 = i A_\tau~, \qquad i S = - S_E~,
\ee
where~$\tau$ denotes Euclidean time and~$S_E$ is the Euclidean action. Applying this to~\eqref{Sferm} and~\eqref{DSferm}, we obtain
\begin{equation}\label{sfermE}
S_E[\mu, A_\tau] = \int d\tau \, \left(\psi^\dagger \d_\tau \psi - (\mu + i A_\tau) \psi^\dagger \psi - i k_\text{bare} A_\tau+ E_0 \right)
\end{equation}
The Euclidean~$A_\tau$ is the imaginary part of a complexified chemical potential. In the the absence of~$A_\tau$, the Euclidean propagator is given by the Wick rotation of~\eqref{fermpropL},
\begin{equation}\la{fermpropE}
\langle \psi(\tau) \psi^\dagger(0)\rangle = \int {d p_\tau \over 2\pi} \, e^{i p_\tau \tau} \, {1 \over i p_\tau - \mu}~.
\end{equation}

\subsection{Partition Functions and Counterterms} \label{appZ}

Let us compute the thermal partition function~$Z[\beta, \mu, A_\tau]$ of the theory in the presence of~$A_\tau$, i.e.~the Euclidean partition function on a circle of circumference~$\beta$, with anti-periodic boundary conditions for the fermions. This will allow us to detect 't Hooft anomalies for the global symmetries. Note that both~$U(1)$ and~$\sfT$ are compatible with the fermion mass~$\mu$, and hence we can regulate the theory in such a way that they are free of anomalies (e.g.~by introducing suitable Pauli-Villars regulator fields). By contrast, insisting on~$\sfC$-symmetry pins~$\mu = 0$, and thus~$\sfC$ can participate in 't Hooft anomalies.

Using non-anomalous~$U(1)$ background gauge transformations, we can make~$A_\tau$ time-independent; in this gauge~$A_\tau$ is completely determined by its holonomy~$\alpha$,
\begin{equation}
A_\tau = {\alpha\over \beta}~, \qquad \alpha = \int_0^\beta d\tau \, A_\tau~.
\end{equation}
Note that large gauge transformations of~$A_\tau$ on the thermal circle shift~$\alpha$ by an element of~$2 \pi \Z$, so that it is an angle, and~$Z(\beta, \mu, \alpha)$ must be a gauge-invariant function of this angle. It is straightforward to compute the thermal partition function via a trace over Hilbert space,
\begin{equation}\la{Ztherm}
Z(\beta, \mu, \alpha) = \tr_{\CH} e^{- \beta\left(  - (\mu + i {\alpha \over \beta}) \psi^\dagger \psi - i k_\text{bare} {\alpha \over \beta} + E_0  \right) } = e^{- \beta E_0 + i k_\text{bare} \alpha } \left(1 + e^{\beta \mu + i \alpha}\right)~.
\end{equation}
Note that this is indeed a gauge-invariant function of the holonomy~$\alpha \sim \alpha + 2\pi$. 

We can now determine the effective Euclidean action at zero temperature by extracting the~$\beta \to \infty$ limit of~$- \log Z(\beta, \mu, \alpha)$ from~\eqref{Ztherm},
\begin{equation}\la{seffE}
S_{E, \, \text{eff}}[\mu, A_\tau] =  \int d\tau \,  \left(-{1 + \text{sign}(\mu) \over 2}(\mu + i A_\tau)   - i k_\text{bare} A_\tau + E_0\right)~. 
\end{equation}
Here we have assumed a non-vanishing fermion mass~$\mu \neq 0$. (The massless case is discussed in appendix~\ref{sec:degqm} below.) Several comments are in order: 
\begin{itemize}
\item The effective action is local, as expected of a gapped theory. 
\item The effective Chern-Simons level~$k_\text{eff} $, defined as the the coefficient of~$-iA_\tau$ in the effective action, is an integer, 
\begin{equation}
k_\text{eff} = k_\text{bare} + \half (1 + \text{sign}(\mu)) = \begin{cases} k_\text{bare} \quad\quad \;\; (\mu < 0) \\ k_\text{bare} + 1\quad (\mu > 0) \end{cases}~.
\end{equation}
In particular, the effective action is indeed~$U(1)$ gauge invariant.
\item The effective action trivially respects~$\sfT$ symmetry, which does not act on the holonomy~$\alpha$.\footnote{~\la{noTa}This is because the gauge field~$A_\tau(\tau) \to A_\tau(-\tau)$ transforms as a twisted one-form under the orientation-reversing diffeomorphism~$\sfT$. By contrast, we refer to the conventional, geometric orientation-reversing transformation~$A_\tau(\tau) \to -A_\tau(-\tau)$ as~$\sfC\sfT$. Thus~$\sfT$ leaves~$\alpha$ invariant, while~$\sfC$ and~$\sfC\sfT$ change its sign. } 

\item Charge-conjugation~$\sfC$ is not a symmetry unless~$\mu = 0$, but if we view~$\mu$ as a spurion with transformation rule~$\sfC : \mu \to -\mu$, then the effective action~\eqref{seffE} should be charge-conjugation invariant. Instead, we find that it shifts by a local c-number term under the action of~$\sfC$,
\begin{equation}\la{muanom}
\quad S_{E,\,\text{eff}}[-\mu, -A_\tau] = S_{E,\,\text{eff}}[\mu, A_\tau] + \int d\tau \, \big( - \mu + i (2 k_\text{bare} + 1) A_\tau\big)~.
\end{equation}
Any part of this shift that cannot be removed by adjusting the counterterms $E_0$, $k_\text{bare}$ in~\eqref{seffE} reflects an 't Hooft anomaly of the massless~$\mu = 0$ theory (discussed in appendix~\ref{sec:degqm} below). The real part of the effective action can be rendered~$\sfC$-invariant by tuning
\begin{equation}
E_0 =  \half \mu + \cdots~,
\end{equation}
where the ellipsis indicates the possibility of a Taylor series in even powers of~$\mu$.\footnote{~Note that odd powers of~$|\mu|$ are~$\sfC$-invariant but not analytic at~$\mu = 0$. They are thus not valid local counterterms and cannot arise from integrating out heavy states. By contrast, the non-analyticity of the term~$\sim -\half |\mu|$ in~\eqref{seffE} at~$\mu =0$ arises from integrating out massless fermions there.} We could similarly restore~$\sfC$-invariance of the imaginary part by taking~$k_\text{bare} = -\half$, but only at the cost of ruining~$U(1)$ gauge invariance. This shows that the~$\mu = 0$ theory has a mixed anomaly between~$\sfC$ and~$U(1)$ (see below), which is reflected at~$\mu \neq 0$ in the non-invariance of~$\Im S_{E, \, \text{eff}}$ under spurious~$\sfC$ transformations~\eqref{muanom}.
\end{itemize}

\subsection{Symmetries and 't Hooft Anomalies in the Massless Case}\la{sec:degqm}

We will now discuss the massless theory, where~$\mu = 0$. At this point the states~$|0\rangle$ and~$|1\rangle$ are degenerate, and they transform under an~$SO(3)$ global symmetry generated by
\begin{equation}\la{Jdef}
J_3 = \psi^\dagger \psi - \half~, \qquad J_+ = J_1 + i J_2 = \psi^\dagger~, \qquad J_- = J_1 - i J_2 = \psi~,
\end{equation}
which satisfy the standard~$SO(3)$ commutation relations,
\be
[J_3, J_\pm] = \pm J_\pm~, \qquad [J_+, J_-] = 2 J_3~.
\ee
Note that the~$SO(3)$ Cartan~$J_3$ in~\eqref{Jdef} precisely agrees with the~$U(1)$ charge~$Q = \psi^\dagger \psi + k_\text{bare}$ in~\eqref{Qdef}, if we choose~$k_\text{bare} = -\half$. Similarly, the charge-conjugation symmetry~$\sfC$ is an~$SO(3)$ rotation sending~$J_3 \to - J_3$ (i.e.~it is a Weyl reflection).

The Hilbert space, which is two dimensional, transforms as an~$SU(2)$ doublet, and hence it realizes the~$SO(3)$ symmetry projectively.\footnote{~Since the normalization of~$J_3$, which is completely fixed by the~$SO(3)$ Lie algebra, requires~$k_\text{eff} = - \half$, all states in the Hilbert space have half-integer~$U(1)$ charges: $J_3|0\rangle = -\half |0\rangle$, so that~$|0\rangle$ is the spin-down state of the~$SU(2)$ doublet, and~$J_3 |1\rangle = \half |1\rangle$, so that~$|1\rangle$ is the spin-up state. By contrast, all operators (e.g.~$\psi$) transform in faithful~$SO(3)$ representations and thus have integer~$U(1)$ charges.} This constitutes an~$SO(3)$ 't Hooft anomaly, characterized by the following invertible anomaly inflow action (or SPT),
\begin{equation}\la{so3infl}
S_{E, \, SO(3)~\text{inflow}} = i \pi \int_{\CM_2} w_2(SO(3))~,
\end{equation}
Here~$\CM_2$ is an Riemannian two-manifold, on whose boundary~$\d \CM_2$ the quantum mechanical system under discussion resides. The second Stiefel-Whiteny class~$w_2(SO(3)) \in H^2(\CM_2, \Z_2)$ of the~$SO(3)$ background gauge bundle (suitably extended to~$\CM_2$) vanishes if and only if the bundle can be lifted to~$SU(2)$. When we restrict from~$SO(3)$ background gauge fields to the~$U(1)$ Cartan background gauge field~$A$ already discussed above, then~\eqref{so3infl} reduces to
\begin{equation}\la{u1infl}
S_{E, \, U(1)~\text{inflow}} = {i \pi } \int_{\CM_2} {dA \over  2\pi}~.
\end{equation}
This is a conventional~$U(1)$~$\theta$-angle, with~$\theta = \pi$, and it constitutes an 't Hooft anomaly if there is a symmetry that pins~$\theta = \pi$. One example is charge conjugation~$\sfC : A \to -A$, but in fact any symmetry -- for instance $\sfC\sfT$ -- that pins~$\theta = \pi$ has a mixed anomaly with~$U(1)$. 

This mixed anomaly can be seen explicitly by examining the partition function~\eqref{Ztherm} in the massless limit,
\be\la{muzeroZ}
Z(\beta, \mu = 0, \alpha) = e^{- \beta E_0 } \left(e^{i k_\text{bare} \alpha}  + e^{i (k_\text{bare} + 1) \alpha}\right)~.
\ee
Let us make some observations: 
\begin{itemize}
\item[(i)] When~$k_\text{bare} = -\half$, as required by the~$SO(3)$ symmetry~\eqref{Jdef}, then~\eqref{muzeroZ} is not a gauge-invariant function of the~$U(1)$ holonomy~$\alpha \sim \alpha + 2\pi$. It is, however, an even and real function of~$\alpha$, compatible with the~$\sfC$ and~$\sfT$ symmetries.
\item[(ii)] If instead we insist on~$U(1)$ gauge invariance, then~$k_\text{eff} \in \Z$ must be quantized and~\eqref{muzeroZ} is not~$\sfC$-invariant (though it is still compatible with~$\sfT$, see footnote~\ref{noTa}).
\item[(iii)]  The partition function vanishes when the~$U(1)$ holonomy is~$\alpha =  \pi$ (modulo~$2 \pi \Z$). This is because the fermion~$\psi$ effectively has periodic (rather than thermal) boundary conditions at this point, so that the Dirac operator~$i D_\tau = i \d_\tau + A_\tau$ that appears in the action~\eqref{sfermE} has a zero eigenvalue.\footnote{~\label{fn:Dspec}The eigenvalues of~$i D_\tau$ on the thermal circle are~$\lambda_ n = {2\pi  \over \beta}\left(n - \half + {\alpha \over 2 \pi} \right)$, with~$n \in \Z$.} 
\end{itemize}

In more detail, the real part of the effective action~$S_{E, \, \text{eff}} = - \log Z(\beta, \mu = 0, \alpha)$ is a non-local function of~$\alpha$ that preserves all symmetries,
\begin{equation}\la{reSe}
\Re S_{E, \, \text{eff}} = \beta E_0 - \half \log \left(2 + 2 \cos \alpha\right)~,
\end{equation}
while its imaginary part is more subtle,
\begin{equation}\la{imSe}
\Im S_{E, \, \text{eff}} = - k_\text{bare} \alpha - \Im \log \left(1 + e^{i\alpha}\right) = - k_\text{bare} \alpha - \half [\alpha]~.
\end{equation}
Here the gauge-invariant quantity~$[\alpha]$ is defined in the following way: because of point (iii) above, the imaginary part~\eqref{imSe} is only well defined when~$\alpha \notin \pi + 2 \pi \Z$. Such values of~$\alpha$ have a unique gauge representative~$[\alpha]$ satisfying~$- \pi < [\alpha] < \pi$. Note, however, that~$[\alpha]$ jumps discontinuously by~$2 \pi$ whenever~$\alpha$ crosses~$\pi + 2 \pi \Z$, which correctly accounts for the sign change of the partition function at these values of~$\alpha$. In fact, it can be checked that
\begin{equation}\label{eq:etainv}
\eta(i D_\tau) = - {[\alpha] \over \pi}~, \qquad \alpha \notin \pi + 2 \pi \Z~,
\end{equation}
where~$\eta(i D_\tau)$ is the~$\eta$-invariant of Atiyah, Patodi, and Singer~\cite{Atiyah:1975jf}, computed for the Dirac operator~$iD_\tau$,\footnote{~Here the~$\eta$-invariant can be defined as~$\eta(i D_\tau) = \lim_{\ep \to 0+} \sum_{n \in \Z} e^{-|\lambda_n|} \text{sign}(\lambda_n)$, where~$\lambda_n$ are the eigenvalues of~$i D_\tau$. Using the explicit formula in footnote~\ref{fn:Dspec}, we find~\eqref{eq:etainv}.} which furnishes a uniform way to define~$\Im S_{E, \, \text{eff}}$ for Dirac fermions in odd spacetime dimensions~\cite{Alvarez-Gaume:1984zst} (see also~\cite{Witten:2015aba,Seiberg:2016rsg}).

The upshot is that~\eqref{imSe} is gauge invariant (whenever there are no zero modes, so that it is well defined modulo~$2 \pi \Z$), but it is not~$\sfC$-invariant for any quantized Chern-Simons level~$k_\text{bare} \in \Z$, due to the half-integer coefficient of~$[\alpha]$.\footnote{~Note that the anomalous shift~$\Delta \Im S_{E, \, \text{eff}} = [\alpha] = \int d\tau\,  A_\tau \, (\text{mod} \, 2 \pi \Z)$ under the action of~$\sfC$ is precisely canceled by the corresponding shift of the inflow action~\eqref{u1infl}.} It is standard to refer to this coefficient as contributing to the effective Chern-Simons level~$k_\text{eff} = k_\text{bare} + \half$, but this is imprecise: a Chern-Simons term with half-integer level is not gauge invariant, while~\eqref{imSe} is gauge invariant (whenever there are no zero modes). However, since~$\alpha - [\alpha] \in 2 \pi \Z$, it follows that small variations~$\delta [\alpha] = \delta \alpha$ agree. For the purpose of varying the effective action with respect to~$\alpha$, e.g.~to determine the effective~$U(1)$ charge, it is thus acceptable to treat~$\Im S_{E, \, \text{eff}}$ as an effective Chern-Simons term with half-integer level,
\begin{equation}\la{effCSferm}
\delta S_{E, \, \text{eff}} = - i k_\text{eff} \int d \tau \, \delta A_\tau~, \qquad k_\text{eff} = k_\text{bare} + \half~.
\end{equation}

 \section{Induced Kinetic Terms for the Gauge Field}
 \la{App1Loop}

 Here we derive the leading (in a gradient expansion) effective action~\eqref{Fconfkin} for the gauge field that arises when we integrate out one massless two-component Dirac fermion~$\Psi$ of unit charge. For a recent discussion and further references, see  \cite{Herzog:2025ddq}.  This answer will also be used in~\nref{EffCou}. A simple way to do this calculation is the following: first, we compute the energy of a configuration with a constant magnetic field~$F_{xy} = B$. This can be done by summing over energy of all the Landau levels in~\eqref{LLenergy}
 \be \la{EnLL}
 E_n = \pm\sqrt{2 n B}~, \qquad n \in \Z_{\geq 0}.
 \ee 
As discussed below~\eqref{LLenergy}, the degeneracy of each level is~${B A \over 2\pi}$, where~$A$ is the area. We can now obtain the ground state energy by summing~$- \half |E_n|$ over all states~\cite{Cangemi:1994by}. 
Note that we must sum over all Landau levels (including those in the filled Dirac sea); equivalently, we must sum over both electrons and positrons. This leads to the overall factor of $2$ in   
\be \la{GSec}
 { E } = 2 \times \left ( { B A  \over    2 \pi } \right) \sum_{n=1}^\infty \left(-\half \sqrt{2 n B }  \right) 
  \quad \Longrightarrow \quad \epsilon = { E \over A } = B^{3/2} \gamma ~, \quad \gamma \equiv { \zeta( 3/2) \over 4  \pi^2 \sqrt{2}   }~. 
  \ee 
Here we have used~$\zeta$-function regularization to carry out the sum over~$n$, and~$\zeta(-1/2) = - {\zeta(3/2) \over 4 \pi }$. Note that this regularization method automatically subtracts the $B$-independent UV divergence in~$\epsilon$. 
    
Note that the~$\epsilon \sim B^{3/2}$ scaling in~\eqref{GSec} is expected based on scaling symmetry. When we consider small, long distance,  fluctuations~$F_\text{total} = F + f$ around the constant magnetic field~$F_{xy} = B$, then we expect to find a Lorentzian effective action that is invariant under the Lorentz and conformal symmetries of the massless 2+1 dimensional Dirac equation coupled to~$A_\text{total}$,
  \be 
  S_\text{eff}= - \gamma \int d^3x \,  \left( \half F_\text{total}^2 \right)^{3/4} ~,~~~~~~{\rm with } ~~~~~ \half F_\text{total}^2 = B^2 + 2 B f_{xy} + f_{xy}^2 - f_{0a}^2~.
  \ee 
Here the coefficient is determined by comparing with~\nref{GSec} for the case~$f = 0$. As discussed in \cite{Hellerman:2015nra}, 
to quadratic order leads to a conformally-invariant kinetic term for~$f$,\footnote{~Note that there is no linear~$\CO(f)$ term, since~$\int f_{xy} = 0$.}
  \be \la{FinAc}
  S_\text{eff} \quad \supset  \quad { \t c \over 2  \sqrt{B} } \int d^3 x \,  \left(  f_{0a}^2  - \half f_{xy}^2 \right)~, \qquad \tilde c =  { 3 \gamma \over 2 } = {3 \zeta( 3/2) \over 8  \pi^2 \sqrt{2}   }~.
  \ee 
  This is the result for~$N_f = 1$ flavor of two-component Dirac fermion. In general, we must multiply~$\gamma$ by~$N_f$. 
  
  We can also compute the first coefficient~$\sim f_{0a}^2$ in \nref{FinAc} directly by looking at the small-momentum limit of the one loop fermion diagram analyzed in appendix~\ref{TevenLoop}. The second term~$\sim f_{xy}^2$ can also be computed this way, using a similar diagram. We have computed these integrals numerically and verified~\nref{FinAc}.

  \section{Fermion One-Loop Diagrams in a Magnetic Field} 
  \la{OneLoopDiag}

Here we discuss the one-loop fermion bubble diagram, with two external photons, depicted in figure~\ref{1Loop}. 

\subsection{Assorted Propagators}

The Dirac propagator in Lorentzian signature arising from the action~\eqref{mDirac} is given by
\begin{equation}
    G(x-x') \equiv \langle 0| T{\Psi(x) \b \Psi(x')}|0\rangle = \int {d^3 p \over (2 \pi)^3} \, e^{i p \cdot (x-x')} \t G(p)~, \qquad \t G(p) = {i \gamma^\mu p_\mu + m \over p^2  + m^2}~.
\end{equation}
Wick rotating~\eqref{mDirac} leads to the Euclidean action
\begin{equation}\label{eq:EDprop}
S_{E,\,  \text{Dirac}}(m) = \int d^3 x \left( i \b \Psi \gamma^\mu \d_\mu  \Psi    - i m \b \Psi \Psi\right)~.
\end{equation}
Here~$\gamma_\mu$ are the Euclidean gamma matrices; they are Hermitian and satisfy~$\{\gamma_\mu, \gamma_\nu\} = 2\delta_{\mu\nu}$. We will take them to be the Pauli sigma-matrices,
\begin{equation}
    \gamma_\mu = \sigma_\mu~, \qquad \mu = 1, 2, 3~.
\end{equation}
The Euclidean Dirac propagator is given by
\begin{equation} \la{EuclProA}
    G_E(x-x') = \langle \Psi(x) \b \Psi(x')\rangle = \int {d^3 p \over (2 \pi)^3} \, e^{i p \cdot (x-x')} \t G_E(p)~, \qquad \t G_E(p) = {-  \gamma^\mu p_\mu + i m \over p^2  + m^2}~.
\end{equation}
As usual, $x^0 = -i x^3$, $p^0 = - i p^3$, and~$\gamma^0 = - i \gamma^3$, so that~$\t G (p^0 = - i p^3) = -i \t G_E(p^3)$.

We will also need the Dirac propagator~$G_E^B(x-x')$ in a constant magnetic background $F_{xy} = B$. We will only need this in Euclidean signature. It should satisfy
\begin{equation}
    \big(i \gamma^\mu (\d_\mu - i A_\mu) - i m \big)G_E^B(x-x') = \delta^{(3)}(x-x')~.
\end{equation}
Let us make the following ansatz,
\begin{equation}\label{ansatz}
    G_E^B(x-x') \equiv e^{i \int_{x'}^x A} \int {d^3 p \over (2 \pi)^3} \, e^{i p \cdot (x-x')} \hat G(p)~.
\end{equation}
Here~$A$ is integrated on a straight line connecting~$x$ and~$x'$, defining an open Wilson line that captures the (background) gauge dependence of the propagator. The remainder~$\hat G(p)$ is gauge invariant; following Schwinger~\cite{PhysRev.82.664}, it can be represented as an integral over the Euclidean length~$s$ of the particle's worldline.
In Landau gauge, where~$A_2 = Bx^1$ is the only non-zero component of~$A$, the phase 
$\int_{x'}^x A = {B \over 2} (x^1 +{x'}^1  )(x^2-{x'}^2)$. 
Then
\begin{equation}
    \left(- \gamma^\mu p_\mu - im - {i B \over 2} \left(\gamma^1 \d_{ p_2} - \gamma^2 \d_{p_1}\right)\right) \hat G(p) = \mathds{1}_{2 \times 2}~.
\end{equation}
It is straightforward to check that this equation is satisfied by
\begin{equation}\label{proptimeS}
\begin{split}
    & \hat G(p) = \int_0^\infty ds \, \hat G(p, m, s)~, \\
    & \hat G(p_\mu, m, s) = 
    \exp\left(-s(m^2 + p_3^2) - {{\vec p}^{\, 2} \over B} \tanh{Bs}\right) \times \\
    & \qquad \qquad \qquad \times \left(- \gamma^\mu p_\mu + i m + i \left(\gamma_1 p_2 - \gamma_2 p_1\right) \tanh{Bs} \right) \left(\mathds{1} - i \gamma_1 \gamma_2 \tanh{Bs}\right)~.
\end{split}
\end{equation}
Note that this reduces to~\eqref{EuclProA}  in the~$B \to 0$ limit. The~$s \to \infty$ limit corresponds to low energies and projects onto the zero modes~\eqref{DirSol} in the lowest Landau level, which have~$\gamma_1\gamma_2$ eigenvalue~$+i$.

We can also compute the pure zero mode contribution, directly from the mode expansion~\eqref{Pmodexp}, after stripping off the Wilson line phase factor,
\begin{equation}\label{zeromodeprop}
    \hat G_{0}(p) =  -  \left(\mathds{1} - i \gamma^1 \gamma^2 \right) {\exp\left(- {{\vec p}^{\, 2} \over B}\right) \over  p_3 + im} = \left(\mathds{1} - i \gamma^1 \gamma^2 \right) \int_0^\infty ds \, e^{- s (m^2 + p_3^2)- {\vec p}^{\,2}/B} (i m - p_3) ~.
\end{equation}
The integrand in the second expression precisely agrees with the~$s \to \infty$ limit of the integrand in the full propagator~\eqref{proptimeS}.

In this appendix, we use~\eqref{proptimeS} to compute fermion bubble diagram with one and two external photons. Thus, the gauge-dependent phase in~\eqref{ansatz} drops out. We set~$B = 1$ below when it is convenient (it can be restored by dimensional analysis), and we only work in Euclidean signature. 
  
\subsection{Current One-Point Function and~$\hat k_\text{eff}$}

Consider a single Dirac fermion of mass~$m$, with Euclidean action~\eqref{eq:EDprop}. We turn on a magnetic field~$B = (dA)_{xy}$ and fluctuations~$a_\mu$ by substituting~$\d_\mu \to \d_\mu - i A_{\text{total}, \mu}$, with~$A_\text{total} = A + a$, as in \eqref{qeda}. Then the Dirac current couples to~$a_\mu$ via 
\begin{equation}
    \Delta S_E = - \int d^3 x\, a_\mu j^\mu~, \qquad j_\mu = - \b \Psi \gamma_\mu \Psi~.
\end{equation}
Thus~$j_\mu = {\delta / \delta a^\mu(x)}$ when acting on the Euclidean partition function~$Z[a]$. 

By symmetry, the only non-vanishing one-point function of the current is
\begin{equation}
    \langle j_3(x)\rangle = \int {d^3 p \over (2 \pi)^3} \, \int_0^\infty ds \, \text{Tr}\left(\gamma_3 \hat G(p, m, s)\right) = {i B \over 4 \pi} \text{sign}(m)~.
\end{equation}
Here~$\hat G(p, m, s)$ is the (gauge-invariant part of) the fermion propagator in~\eqref{proptimeS}. This contributes the following term in the effective action,
\begin{equation}
    \Delta S_{E, \text{eff}} = - {i B \text{sign}(m) \over 4 \pi} \int d^3 x \, a_3~.
\end{equation}
If we integrate over space, this precisely corresponds to a Chern-Simons term in one Euclidean dimension, with level
\begin{equation}
    \Delta \hat k_\text{eff} = \half \text{sign}(m) {B \over 2 \pi} (\text{Area}) = {N \over 2} \text{sign}(m)~,
\end{equation}
where we have used the flux-quantization~$N = {B \over 2 \pi} (\text{Area})$. Thus each one of the~$N$ zero modes in the lowest Landau level contributes~$\half \text{sign}(m)$ to the effective Chern-Simons level~$\hat k_\text{eff}$.

\subsection{Current Two-Point Function}

We are interested in the~$O(a^2)$ terms in the effective action, which at one-loop come from the connected current two-point function, in Euclidean momentum space
\begin{equation}
    \Pi_{\mu\nu}(q) = \int d^3 x \, e^{- i q \cdot x} \, \langle j_\mu(x) j_\nu(0)\rangle_\text{connected}~,
\end{equation}
so that the~$O(a^2)$ term in the one-loop effective action reads\footnote{~Here~$a_\mu(q) \equiv \int d^3 x \, e^{- i q \cdot x} a_\mu(x)$.}
\begin{equation}\label{Sa2eff}
    S_{E, \, O(a^2), \text{ one loop}} = - \half \int {d^3 q \over (2 \pi)^3} \, a_\mu(-q) \Pi_{\mu\nu}(q) a_\nu(q)~.
\end{equation}

\subsubsection{Time-Reversal-Even Effective Coulomb Potential}\la{TevenLoop}

Let us consider the static part of 
\begin{equation}
    \Pi_{33}(\vec q, q_3 = 0) \equiv - |{\vec q}\,|^{ 2} \Pi(|\vec q|)~,
\end{equation} 
which contributes to the effective, instantaneous Coulomb potential via
\begin{equation}
S_{E, \, O(a_3^2) , \text{ static, one loop}} = {1 \over 2} \int d\tau \int {d^2 q \over (2 \pi)^2} \, a_3(-\vec q, \tau) {\vec q}^{\, 2} \Pi(|\vec q|) a_3(\vec q, \tau)~,
\end{equation}
where~$\tau = x_3$ is Euclidean time. We can resum this one-particle irreducible (1PI) contribution coming from the fermion bubble using intermediate tree-level propagators coming from the classical action~$S_{E} \supset {1 \over 2 e^2} \int d^3 x\, (\vec \grad a_3)^2$, to obtain the effective Coulomb potential in momentum space,
\begin{equation}
    \t v_{\text{eff.} \, C}(|\vec q|) = {1 \over |\vec q|^2 \left({1 \over e^2} + \Pi(|\vec q|)\right)}~.
\end{equation}
We could evaluate~$\Pi(|\vec q|)$ directly; this leads to some  divergences at intermediate stages that ultimately cancel due to gauge invariance.

This can be avoided by relating~$\Pi(|\vec q|)$ to a different component of the polarization tensor~$\Pi_{\mu\nu}$, using the Ward identity that follows from conservation of the current~$j_\mu$,
\begin{equation} \la{ConsEqu}
    q^\mu \Pi_{\mu\nu}(q) = 0~.
\end{equation}
We can now use rotational invariance, which implies that $\Pi_{a3 } $ is proportional to $q_a$, together with the~$\CT$-even part of the~$\nu = 3$ component of \nref{ConsEqu} to conclude that 
\begin{equation}\label{Pitrick}
\Pi^\text{even}_{a3}(\vec q, q_3 \to 0) = q_a q_3   \Pi(|\vec q|)~.
\end{equation}
Thus we must compute one component of~$\Pi_{a3}^\text{even}$ to leading, linear order in small~$q_3$.\footnote{~It can be checked explicitly that there is no~$q_3$-independent term.}
We choose to evaluate
\begin{equation}\label{Pi13int}
    \Pi_{a = 1, 3}^\text{even}(q) = -  \int{d^3 p \over (2 \pi)^3} \, \int_0^\infty ds ds' \, \text{Tr} \left(\gamma_1 \hat G(p, m = 0, s) \gamma_3 \hat G(p-q, m = 0, s'\right)~,
\end{equation}
directly in the massless limit.\footnote{~When~$m = 0$ the theory has time-reversal symmetry, so that~$\Pi_{a, 3} = \Pi^\text{even}_{a, 3}$ has no~$\CT$-odd part.} The loop integral over~$\vec p$ is Gaussian and thus easily done, leaving an absolutely convergent integral over the proper time parameters~$s, s'$. It is convenient to change integration variables to 
\be \la{VarChas}
 s = { u + r \over 2} ~, \qquad s' = {u - r \over 2} ~, \qquad \int_0^\infty ds ds' \to \half \int_0^\infty d u \int_{-u} ^u dr~.
 \ee 
Expanding the integral~\eqref{Pi13int} at small~$q_3$, we indeed find that it is linear in~$q_3$, so~\eqref{Pitrick} allows us to extract
\begin{equation}\label{Piofqfinal}
    \Pi(|\vec q|) = \int_0^\infty du \int_{-u}^u dr \, \frac{ e^{- \frac{|\vec q|^2}{2}   (\cosh u -\cosh r)/\sinh u} (u \cosh r \sinh u -r \sinh r \cosh u)}{16 \pi ^{3/2} u^{3/2} \sinh^2 u}~,
\end{equation}
in units where~$B = 1$. This can now be integrated numerically. If there is more than one flavor we should multiply~$\Pi(|\vec q|)$ by~$N_f$.

 \subsubsection{Time-Reversal-Odd Effective Chern-Simons Level~$k_\text{eff}$}\la{Toddapp}

 Current conservation, together with the unbroken symmetries, implies that the most general $\CT$-odd two-point function function in momentum space must take the form
 \begin{equation}\label{keffdefApp}
     \Pi_{\mu\nu}^\text{odd}(q) = {1 \over 2 \pi} \ep_{\mu\nu\rho} q^\rho k_\text{eff}(|\vec q|, q_3)~.
 \end{equation}
 When~$k_\text{eff}$ is  a constant, substituting into~\eqref{Sa2eff} gives a conventional, relativistic Chern-Simons term with level~$k_\text{eff}$ for the gauge field~$a_\mu$. 

We will evaluate the static limit~$k_\text{eff}(|\vec q|, 0)$, in the presence of a non-zero mass~$m \neq 0$ for the Dirac fermion. We will compute
\begin{equation}\label{Pi13Odd}
    \Pi^\text{odd}_{a = 1, 3}(q_1 = 0, q_2,q_3 = 0) = -  \int{d^3 p \over (2 \pi)^3} \, \int_0^\infty ds ds' \, \text{Tr} \left(\gamma_1 \hat G(p, m, s) \gamma_3 \hat G(p-q, m, s'\right)~,
\end{equation}
which only has a~$\CT$-odd part, as can be seen by comparing with~\eqref{Pitrick}. This integral can be evaluated the same way as~\eqref{Pi13int}, by first evaluating the gaussian~$\vec p$-integrals, before changing variables as in~\eqref{VarChas}. Together with~\eqref{keffdefApp}, this lets us extract
\begin{equation}\label{keffqint}
    k_\text{eff}(|\vec q|, 0) = {m \over 4 \sqrt{\pi}}  \int_0^\infty du \int_{-u}^u dr \, \frac{ \cosh r \;  e^{-m^2 u - \frac{|\vec q|^2}{2}   (\cosh u -\cosh r)/ \sinh u}}{\sqrt{u} \, \sinh u }~.
\end{equation}
Two comments are in order:
\begin{itemize}
    \item[(i)] When~$\vec q = 0$, at long distances, we find
    \begin{equation}\label{zeromomCS}
        k_\text{eff}(q = 0) = {m \over 4 \sqrt{\pi}}  \int_0^\infty du \int_{-u}^u dr \, \frac{ \cosh r \;  e^{-m^2 u}} {\sqrt{u} \, \sinh u } = \half \text{sign}(m)~.
    \end{equation}
   An interesting point is that for small~$m$, the integral~\eqref{zeromomCS} is dominated by large~$u$, which means that the lowest Landau level is contributing (more on this below). 
    \item[(ii)] When~$\vec q \to \infty$, the integral~\eqref{keffqint} decays, and we find 
    \begin{equation}\label{largemomCS}
        k_\text{eff}(|\vec q| \to \infty, 0) = 0~.
    \end{equation}
\end{itemize}
Note that~\eqref{zeromomCS} and~\eqref{largemomCS} above are separately scheme-dependent -- they can be shifted by the same amount via a Chern-Simons counterterm in the action. (Indeed, precisely such a joint shift by~$\half$ is needed to make the computations above compatible with~\eqref{keffdir}.) However, their scheme-independent difference is the expected half-integral shift~$\Delta k_\text{eff} = \half \text{sign}(m)$ between UV and IR, very similar to what happens in the absence of a magnetic field~\cite{Closset:2012vp}.  

Let us elaborate on point (i) above and verify which purple fermion modes running in the loop depicted in figure~\ref{1Loop} are contributing to~\eqref{zeromomCS}. We claim that the dominant contribution at small masses~$m \ll 1$ (i.e.~$m \ll \sqrt{B}$) comes from mixed diagrams, where one propagator is a zero mode, and the other one a higher Landau level. To see this, let us replace one of the propagators in~\eqref{Pi13Odd} by the pure zero-mode propagator in~\eqref{zeromodeprop}. It follows from~\eqref{OneLLLL} that the zero modes by themselves do not contribute, i.e.~this is a mixed diagram of the type described above. Evaluating it gives
\begin{equation}
    \Delta k_\text{eff}(q = 0)\big|_\text{mixed} = \half \text{sign}(m) \int_0^\infty ds' \, \frac{1 - \text{erf}\left(\sqrt{s'} | m| \right)}{ \cosh^2{ s'} (\tanh s' +1)^2}~,
\end{equation}
where~$\text{erf}(z) = {2 \over \sqrt{\pi}} \int_0^z dt \, e^{-t^2}$ is the error function. Since the integral is exponentially convergent, we can expand~$\text{erf}(z)$ around~$z = 0$ and integrate term by term, giving
\begin{equation}\label{mixedCSsmallm}
    \Delta k_\text{eff}(q = 0)\big|_\text{mixed} = {1 \over 4} \text{sign}(m) + O(m)~, \qquad m \ll 1~.
\end{equation}
We see that the leading~$O(1)$ term contributes half of the answer~\eqref{zeromomCS}. Adding the other mixed diagram that is obtained by exchanging zero-mode and higher-Landau-level propagators, completely saturates~\eqref{zeromomCS}. This shows that loops with only higher-Landau-level fermions only contribute at subleading~$O(m)$ order, canceling the~$O(m)$ terms in~\eqref{mixedCSsmallm}.

\newpage

\bibliographystyle{utphys}
\bibliography{GeneralBibliography}

\end{document}